\documentclass[12pt]{article}
\usepackage{graphicx}
\usepackage{latexsym,amsmath,amsfonts,amssymb,color}
\usepackage[latin1]{inputenc}
\usepackage{cite}
\usepackage[american]{babel}
\usepackage{hyperref}
\newcommand{\be}{\begin{equation}}
\newcommand{\ee}{\end{equation}}
\newcommand{\beq}{\begin{equation}}
\newcommand{\eeq}{\end{equation}}
\newcommand{\bea}{\begin{eqnarray}}
\newcommand{\eea}{\end{eqnarray}}

\newcommand{\ba}{\begin{eqnarray}}
\newcommand{\ea}{\end{eqnarray}}
\begin{document}
\baselineskip=15.5pt
\pagestyle{plain}
\setcounter{page}{1}
%--------+---------+---------+---------+---------+---------+---------+
%Body

\def\ie{{\em i.e.},}
\def\eg{{\em e.g.},}
\newcommand{\rc}{\nonumber\\}
\newcommand{\bear}{\begin{eqnarray}}
\newcommand{\eear}{\end{eqnarray}}
\newcommand{\Tr}{\mbox{Tr}}    % trace over gauge indices
\newcommand{\ack}[1]{{\color{red}{\bf Pfft!! #1}}}

\def\a{\alpha}
\def\b{\beta}
\def\c{\gamma}
\def\d{\delta}
\def\eps{\epsilon}           % Also, \varepsilon
\def\f{\phi}               %      \varphi
\def\vf{\varphi}  \def\tvf{\hat{\varphi}}
\def\vp{\varphi}
\def\g{\gamma}
\def\h{\eta}
\def\j{\psi}
\def\k{\kappa}                    % Also, \varkappa (see below)
\def\l{\lambda}
\def\m{\mu}
\def\n{\nu}
\def\o{\omega}  \def\w{\omega}
\def\p{\pi}
\def\q{\theta}  \def\th{\theta}                  %     \vartheta
\def\r{\rho}                                     %     \varrho
\def\s{\sigma}                                   %     \varsigma
\def\t{\tau}
\def\u{\upsilon}
\def\x{\xi}
\def\z{\zeta}
\def\pt{\hat{\varphi}}
\def\tt{\hat{\theta}}
\def\lab{\label}
\def\6{\partial}
\def\wg{\wedge}
\def\atanh{{\rm arctanh}}
\def\bpsi{\bar{\psi}}
\def\bt{\bar{\theta}}
\def\bvf{\bar{\varphi}}
\def\W{\Omega}
\def\tr{{\rm tr}}

\numberwithin{equation}{section}

\renewcommand{\theequation}{{\rm\thesection.\arabic{equation}}}

%%%%%%%%%%%%%%%%%%%%%%%%%%%%%%%%%%%%%%%%%%%%%%%%%%%%%%%%%%%%%%%%%%%%%%%%%%
%%%%%%%%%%%%%%%%        AND THERE WE GO!           %%%%%%%%%%%%%%%%%%%%%%%%
%%%%%%%%%%%%%%%%%%%%%%%%%%%%%%%%%%%%%%%%%%%%%%%%%%%%%%%%%%%%%%%%%%%%%%%%%%

%\begin{flushright}
%HIP-2019-01/TH
%\end{flushright}

\begin{center}

\centerline{\Large {\bf   Supersymmetric probes in warped $AdS_6$}}

\vspace{8mm}

\renewcommand\thefootnote{\mbox{$\fnsymbol{footnote}$}}
Jos\'e Manuel Pen\'\i n${}^{1,2}$\footnote{jmanpen@gmail.com},
Alfonso V. Ramallo${}^{1,2}$\footnote{alfonso@fpaxp1.usc.es} and 
Diego Rodr\'\i guez-G\'omez ${}^{3}$\footnote{d.rodriguez.gomez@uniovi.es}

\vspace{4mm}

${}^1${\small \sl Departamento de  F\'\i sica de Part\'\i  culas} 
{\small \sl and} \\
${}^2${\small \sl Instituto Galego de F\'\i sica de Altas Enerx\'\i as (IGFAE)} \\
{\small \sl Universidade de Santiago de Compostela} \\
{\small \sl E-15782 Santiago de Compostela, Spain} 
\vskip 0.2cm
\vskip 0.2cm

${}^3${\small \sl Department of Physics, Universidad de Oviedo\\ C. Federico Garc\'\i a Lorca 18, 33007, Oviedo, Spain
}

\end{center}

\vspace{8mm}
\numberwithin{equation}{section}
\setcounter{footnote}{0}
\renewcommand\thefootnote{\mbox{\arabic{footnote}}}

\begin{abstract}
We consider defects in 5d field theories corresponding to higher-rank generalizations of the $E_{N_f+1}$ theories; holographically dual to the Brandhuber-Oz background in type I' String Theory. We concentrate on codimension 2 and 1 defects, corresponding, respectively, to 3d and 4d defect Quantum Field Theories. We study holographically such defect theories by considering supersymmetric probe D4- and D6-branes in the $AdS_6$, whose fluctuations allow us to study the spectrum of mesonic operators of the defect theories. In the case of D4-branes, we also consider wrappings in the internal space which can be regarded as generalizations of the configurations capturing the antisymmetric Wilson loop.

\end{abstract}

\newpage
\tableofcontents

\section{Introduction}

Quantum Field Theories (QFT's) in dimensions larger than four have become a very important ingredient in the modern approach to supersymmetric QFT, where they play a central role as building blocks which, upon compactification on appropriate surfaces, give rise to virtually all other lower-dimensional QFT's. This provides a powerful and deep approach to SUSY QFT which, in particular, allows to understand relations among theories, most notably dualities. 

The mere existence of QFT's in d$>$4 is remarkable \textit{per se}. For instance, gauge theories in 5d seem \textit{a priori} uninteresting. Since their gauge coupling is irrelevant, they are not power-counting renormalizable and thus they seem not to be consistent QFT's by themselves. Nevertheless, as first argued in \cite{Seiberg:1996bd,Intriligator:1997pq}, under some favorable circumstances, 5d gauge theories may be regarded as the endpoint of RG flows triggered from UV fixed points where a 5d Superconformal Field Theory (SCFT) realizing the unique 5d superalgebra $F(4)$ lives. More generically, combinations of String Theory techniques (in particular 5-brane webs in type IIB String Theory  \cite{Aharony:1997ju,Aharony:1997bh}) as well as exact computations relying on localization (such as the index in \cite{Kim:2012gu}), have shown the existence of a wide class of 5d SCFT's which are intrinsically strongly coupled, some of which admit a mass-deformation triggering a flow into an IR 5d gauge theory. In parallel, the $AdS/CFT$ correspondence provides complementary means to test the properties of some such SCFT's. Very recently, a large class of $AdS_6$ backgrounds which can be regarded as the backreacted geometry of certain families of 5-brane webs have been constructed in type IIB String Theory in \cite{DHoker:2016ujz,DHoker:2017mds,DHoker:2017zwj} (see also \cite{Apruzzi:2014qva,Kim:2015hya}). These geometries contain the (T-dualized) version \cite{Lozano:2012au} of the celebrated Brandhuber-Oz solution (and its orbifold generalizations) in type I' String Theory \cite{Brandhuber:1999np,Bergman:2012kr}, as well as the non-abelian T-dual solutions of \cite{Lozano:2013oma,Lozano:2012au,Lozano:2018pcp}. Using these backgrounds it had been possible to perform very detailed analysis of the dual 5d theories \textit{e.g.} \cite{Ferrara:1998gv,Jafferis:2012iv,Bergman:2012qh,Assel:2012nf,Bergman:2013koa,Pini:2014bea,Gutperle:2018vdd,Passias:2018swc,Gutperle:2018wuk,Bergman:2018hin,Fluder:2018chf,Kaidi:2018zkx,Chaney:2018gjc,Fluder:2019szh,Chang:2017mxc,Choi:2018fdc,Choi:2019miv}.

One natural way to probe a theory is to consider its dynamics in the presence of boundaries. Such boundaries must carry the appropriate degrees of freedom and thus host a lower-dimensional QFT which is coupled with the ambient QFT. These defect QFT's can be very interesting by themselves and lead to new insights in QFT (\textit{e.g.} \cite{Gaiotto:2008sd,Gaiotto:2008ak}). The story becomes particularly interesting in the case of ambient CFT's with a gravity dual, where it has been shown that, by considering the appropriate branes in the dual geometry, it is possible to find a gravity dual to the defect QFT \cite{Karch:2001cw,Karch:2000gx,DeWolfe:2001pq}. This gravity dual can be either in terms of probe branes in the geometry dual to the ambient QFT or, if the number of probe branes is large (\textit{i.e.} if the number of degrees of freedom of the defect is large), in terms of the fully backreacted geometry corresponding to the brane intersection which gives rise to the defect QFT (see \textit{e.g.} \cite{DHoker:2007zhm,DHoker:2007hhe}). As a by-product of these constructions, since in particular the defect theories typically host degrees of freedom in the fundamental representation of the gauge group, defect QFT's with a gravity dual provide a way to find a holographic description for more realistic QFT's. This has been explored in the past to study the operator content and particle masses in these theories (see \textit{e.g.} \cite{Constable:2002xt,Erdmenger:2003kn,Erdmenger:2005bj,Arean:2006pk,Arean:2006vg,Arean:2007nh}).

Motivated by these considerations, in this paper we study, from a holographic point of view, SUSY-preserving defects in 5d SCFT's. We will concentrate on the higher-rank version of the $E_{N_f+1}$ theories of \cite{Seiberg:1996bd}, whose gravitational dual is the Brandhuber-Oz (BO) solution \cite{Brandhuber:1999np}. The latter corresponds to the near-brane region of a configuration of $N$ D4-branes near an O8$^-$ orientifold with $N_f<8$ D8-branes. Upon backreaction, the type I' SUGRA solution is the 10d uplift \cite{Cvetic:1999un} of the $AdS_6$ solution of Romans'  $F(4)$ SUGRA \cite{Romans:1985tw}. The full 10d geometry is a warped $AdS_6\times \mathbb{S}^4$ with non-zero Romans' mass. The geometry is singular at the equator of the sphere, which is really a half-$\mathbb{S}^4$, reflecting the presence of the O8-D8's. In turn, the dual 5d fixed point theory is a rank $N$ SCFT with global symmetry $E_{N_f+1}$ which admits a mass-deformation, by turning on a Yang-Mills coupling, into a conventional 5d gauge theory with gauge group $USp(2N)$, a hypermultiplet in the 2-index antisymmetric representation and $N_f$ hypermultiplets in the fundamental representation. 

In view of the String Theory construction one may, in principle, classify all possible supersymmetric defects by adding supersymmetric branes to the D4-O8-D8 configuration. Depending on the number of added branes the most convenient description will be either as probe branes in the BO background or as fully backreacted backgrounds. Indeed, some such defects have been studied in the past. In \cite{Dibitetto:2018iar} (see also \cite{Nunez:2001pt}), codimension 3 defects leading to defect QFT's with conformal invariance were studied upon backreaction from a supergravity point of view, finding the corresponding $AdS_3$ backgrounds. This was extended in \cite{Suh:2018tul,Hosseini:2018usu,Suh:2018szn,Dibitetto:2018gtk} to codimension 4 defects, leading to $AdS_2$ backgrounds. Instead, in this paper we will study codimension 1 and 2 defects. These correspond, respectively, to D6- and (certain) D4-branes, which we will treat as probes. Hence we will be able to study the open string degrees of freedom associated to the defect theory, allowing us to study the operator content of the defect theories. 

Codimension 1 defects corresponding to D6-branes lead to 4d conformal defect QFT's. In turn, the case of D4-branes is richer in the sense that they can be arranged in two different ways preserving supersymmetry in both cases. One such arrangement engineers codimension 2 defects which correspond to non-conformal 4d defect theories. The other corresponds to codimension 4 defects, which can actually be regarded as the Wilson loops studied in \cite{Assel:2012nf}. Our analysis shows that there is actually a wider class of solutions out of which those in \cite{Assel:2012nf} are a particular case.

The structure of this paper is as follows: we start in section \ref{background} with a lightning review of the BO background and its dual CFT, introducing in particular appropriate coordinate choices for our future purposes. In section  \ref{D4_defects} we turn to codimension 2 defects engineered as probe D4-branes in the BO background, discussing their supersymmetry as well as their fluctuations, proposing an operator-fluctuation correspondence. In section \ref{D4_wrapped} we study a different configuration of D4-branes, where they wrap the internal space and correspond to generalizations of the configuration capturing the antisymmetric Wilson loop. In section \ref{D6} we turn to codimension 1 defects, constructed through probe D6-branes, whose supersymmetry and fluctuations (including the operator-fluctuation map) we discuss. Finally, we summarize our main results and conclusions in section \ref{conclusions}. Since the technical aspects are a bit lengthy, we compile the details of the manipulations leading to the results in the main text in a number of appendices. In appendix \ref{Killing} we review the relevant Killing spinors. In appendices \ref{D4_defects_fluct} and \ref{fluc_warpped_D4} we compute the fluctations for the defect D4's and wrapped D4's respectively. Finally, in appendix \ref{Fluct_D6} we compute the fluctuations for the defect D6's.

\section{The D4-D8 background}\label{background}

The BO background corresponds to the near-brane geometry of the D4-O8-$N_f$ D8 system. Since the constituents are mutually supersymmetric, it can be constructed modularly starting with the O8-D8's geometry and later adding the D4-branes. Since it will be useful for latter purposes, let us briefly review the construction. Assuming the D8-branes along $\{x^1,\,\cdots,\,x^8\}$, so that $x^9$ is the transverse coordinate to the D8-brane stack, the string frame 10d metric corresponding to the D8-branes is 

\begin{equation}
ds^2=\frac{dx_{1,8}^2}{\sqrt{\frac{x^9}{m}}}+\sqrt{\frac{x^9}{m}}\,(dx^9)^2\,.
\end{equation}
The parameter $m\sim 8-N_f$ is the Romans' mass. Upon introducing $x^9\sim z^{\frac{2}{3}}$ the background can be brought to a conformally flat form \cite{Polchinski:1995df}

\begin{equation}
\label{ansatz}
ds^2=\Omega(z)^2\,\Big[ dx_{1,8}^2+dz^2\Big]\,,\qquad \Omega(z)=H^{-\frac{1}{4}}\,,\qquad H=\Big(\frac{2}{3}mz\Big)^{\frac{2}{3}}\,.
\end{equation}
We now add a stack of N D4-branes, which we will assume to be coincident with the D8-branes. In order for them to be SUSY, they must be completely inside the D8-O8. Thus, let us split the $\mathbb{R}^8$ into the $\mathbb{R}_{\rm wv}^4$ wrapped by both the D4-branes and the D8-branes parameterized by $\{x^1,\,\cdots,\,x^4\}$, and the $\mathbb{R}_{X}^4$ transverse to the D4-branes inside the D8-branes parameterized by $\{X^1,\,\cdots,\,X^4\}$. The Freund-Rubin ansatz for such a brane system is  

\begin{equation}
\label{FR}
ds^2=\Omega(z)^2\,\Big[ h^{-\frac{1}{2}}dx_{1,4}^2+h^{\frac{1}{2}}\,\big(d\vec{X}^2+dz^2\big)^2\Big]\,,\qquad F_6=d(h^{-1})\wedge d{\rm Vol}_{{\rm Min}_{1,4}}\, , \qquad e^{-\Phi}=H^{\frac{5}{4}}\,h^{\frac{1}{4}}\,.
\end{equation}
Using the equations of motion, one easily finds

\begin{equation}
h=\frac{C}{(\vec{X}^2+z^2)^{\frac{5}{3}}}\,,
\end{equation}
where $C$ a constant, to be fixed by flux quantization, proportional to $N$. One can now introduce polar coordinates in $\mathbb{R}_X^4$, denoting the radial coordinate by $R$:\
\beq
R\,=\,\sqrt{\vec X^{\,2}}\,\,
\qquad\leadsto\qquad
d\vec{X}^2\,=\,dR^2\,+\,R^2\,d\Omega_3^2\,\,.
\eeq
Upon doing the further change of coordinates

\begin{equation}
\label{change}
R=r^{\frac{3}{2}}\,\sin\alpha\,,\qquad z=r^{\frac{3}{2}}\cos\alpha\, ,
\end{equation}
the geometry becomes the Brandhuber-Oz $AdS_6$ background\footnote{We use Poincare coordinates for $AdS_6$ where $ds^2_{AdS_6}\,=\,r^2\,dx^2_{1,4}\,+\,{dr^2\over r^2}$. } 
\beq
ds^2_{10}\,=\,\frac{Q^3}{K^8(\alpha)}\,\Big[{9\over 4}\,ds^2_{AdS_6}\,+\,d\Omega_4^2\Big]\,,\qquad e^{-\Phi}\,=\,{\big[K(\alpha)\big]^{20}\over Q^6}\,,\qquad \qquad K(\alpha)\,\equiv\, Q^{{5\over 16}}\,\Big(3m\,\cos\alpha\Big)^{{1\over 24}}\,\,,
\label{10d_metric_string}
\eeq
where $d\Omega_4^2\,=\,d\alpha^2\,+\,\sin^2\,\alpha\,d
\Omega_3^2$. 
In addition there is a RR 4-form $F_4$ given by:
\beq
F_4\,=\,-{10\over 3}\,Q^{{9\over 10}}\,e^{-{2\Phi\over 5}}\,{\rm Vol}({\mathbb S}^4)\,=\,
-{10\over 3}\,Q^{-{3\over 2}}\, K^8(\alpha)\,{\rm Vol}({\mathbb S}^4)\,\quad \leadsto\quad C_5\,=\,-\Big({3\over 2}\Big)^5\,Q^{{3\over 2}}\,r^5\,d^5 x\,\,.
\label{F_4_expression}
\eeq
Finally, the charge quantization conditions require

\beq
m\,=\,{8-N_f\over 4\pi}\,\,,
\qquad\qquad
Q\,=\,\Bigg({2^{11}\,\pi^4\over 3^4\,(8-N_f)}\Bigg)^{{1\over 3}}\,N\,\,.
\label{m_Q}
\eeq

The effect of the change of coordinates in \eqref{change} is two-fold. On the one hand, it produces the $AdS_6$. On the other hand, it allows to combine the $\mathbb{S}^3$ inside the $\mathbb{R}_X^4$ with the extra angular coordinate into an $\mathbb{S}^4$. Note however that the range of $\alpha$ is $0\le \alpha\le \pi/2$. Thus we really have not a full ${\mathbb S}^4$ but an ${\mathbb S}^4$-hemisphere, at whose $\mathbb{S}^3$ boundary, located at $\alpha=\frac{\pi}{2}$, the dilaton diverges. Such boundary is naturally interpreted as the position of the orientifold \cite{Bergman:2012kr}.

The background, as written in \eqref{FR}, keeps track of the $\mathbb{R}^4_X$, which corresponds to the directions inside the D8-branes transverse to the D4-branes; as well as the $z$, which is essentially the coordinate transverse to all branes. Since it will be very useful for our later purposes, let us explicitly quote the string frame background in these coordinates

\beq
ds^2_{10}\,=\,{Q^{{1\over 2}}\over (3m)^{{1\over 3}}}\,
{(\vec X^2+z^2)^{{1\over 6}}\over z^{{1\over 3}}}
\Bigg[\,{9\over 4}\,\big(\vec X^2+z^2\big)^{{2\over 3}}\,dx^2_{1,4}\,+\,
{1\over \vec X^2+z^2}\,(d\vec X^2+dz^2)\Bigg]\,\,.
\label{10dmetric_cartesian}
\eeq
The dilaton is:
\beq
e^{-\Phi}\,=\,{K^{20}\over Q^6}\,=\,Q^{{1\over 4}}\,(3m)^{{5\over 6}}\,
{z^{{5\over 6}}\over (\vec X^2+z^2)^{{5\over 12}}}\,\,.
\label{dilaton_cartesian}
\eeq
This background has a RR 5-form potential $C_5$ given by:
\beq
C_5\,=\,-\Big({3\over 2}\Big)^5\,Q^{{3\over 2}}\,
\big(\vec X^2+z^2\big)^{{5\over 3}}\,d^5x\,\,.
\label{C5_cartesian}
\eeq

\subsection{The dual CFT}

The CFT dual to the BO background above can be read-off from the brane construction as the worldvolume theory on the stack of $N$ D4-branes on top of the O8$^-$+$N_f$ D8-branes. The String Theory construction shows that this is a strongly coupled rank $N$ CFT with global symmetry $E_{N_f+1}$. This symmetry arises as the gauge symmetry on $N_f$ D8-branes on top of an O8$^-$ upon tuning the dilaton to diverge on the D8-brane stack (\textit{cf.} \eqref{10d_metric_string}, which shows that the dilaton diverges at $\alpha=\frac{\pi}{2}$). These theories are the higher-rank generalization of \cite{Seiberg:1996bd,Intriligator:1997pq}.

The 5d CFT's admits a mass-deformation by turning on a Yang-Mills coupling. The String Theory counterpart is to turn on an integration constant in the dilaton which makes it not to diverge on top of the orientifold. At any rate, this deformation triggers an RG flow towards a conventional gauge theory, which can be easily read-off from the brane construction: it is a $USp(2N)$ gauge theory with one antisymmetric hypermultiplet and $N_f$ fundamental hypermultiplets. From this perspective, the global $SO(2N_f)$ symmetry of the hypermultiplets combines with the $U(1)_I$ topological symmetry\footnote{In 5d any gauge theory automatically contains a topologically conserved $U(1)$ current $j_I=\star {\rm Tr}F\wedge F$. The states electrically charged under this symmetry are instantonic particles.} enhancing it to $E_{N_f+1}$ in the UV fixed point. Moreover, the gauge theory provides intuition for the dynamics of the CFT. For instance, the 5d vector multiplet contains a real scalar which is naturally identified with the direction transverse to all branes, \textit{i.e.} $x^9$ (or its $z$ avatar). In turn, the antisymmetric hypermultiplet is naturally identified with the $\mathbb{R}^4_X$ \cite{Bergman:2012qh}. More specifically, denoting, in 4d $\mathcal{N}=1$ language, by $A_{1,\,2}$ the two antisymmetric chirals making up for the antisymmetric hypermultiplet, the meson-like operators constructed out of them are in one-to-one correspondence with the holomorphic functions in $\mathbb{C}^2_X\sim \mathbb{R}^4_X$. Consistently, note that $X^i\sim r^{\frac{3}{2}}$, which (since $A_1\sim X^1+i\,X^2$, $A_2\sim X^3+i\,X^4$) suggests that the corresponding fields have scaling dimension $\frac{3}{2}$, just as expected for scalar fields in 5d. In turn $x^9\sim z^{\frac{2}{3}}\sim (r^{\frac{3}{2}})^{\frac{2}{3}}\sim r$, which is consistent with the identification of $x^9$ with the (real) scalar in the vector multiplet.\footnote{SUSY requires the kinetic term for a scalar in the vector multiplet to be $g_{YM}^{-2}\,\partial\phi^2$. Since in 5d $g_{YM}^2$ has mass-dimension one, $\phi$ has mass-dimension one. Note that scalars in a hypermultiplet have a standard kinetic term, and thus they have the expected mass-dimension $\frac{3}{2}$.}

\section{D4-brane defects}
\label{D4_defects}
The first example of a supersymmetric probe brane configuration in the BO background  we are going to consider is the one corresponding to a D4$'$-brane probe that creates a codimension 2 defect on the worldvolume of the D4-brane of the background. This setup can be represented by the following array:
\beq
\label{D4D4'intersection}
\begin{array}{ccccccccccl}
 &x^1&x^2&x^3& x^4& X^1&X^2 &X^3&X^4&z &  \\
D4: & \times &\times &\times &\times &\_ & \_&\_ &\_ &\_ &      \\ 
D4': &\times&\times&\_&\_&\times&\times&\_&\_&\_ &
\end{array}
\eeq
where the coordinates are those used in (\ref{10dmetric_cartesian})-(\ref{C5_cartesian}). 
To analyze the dynamics of the probe we will use the following set of worldvolume coordiantes:
\beq
\zeta^{a}\,=\,(x^0, x^1,x^2, X^1, X^2)\,\,,
\label{vw_coordinates_D4defects}
\eeq
and we will treat $x^3$, $x^4$,  $X^3$, $X^4$ and $z$ as scalar fields parameterizing the embedding. Actually, we will denote:
\beq
W^1=x^3\,\,,
\qquad\qquad
W^2=x^4\,\,,
\eeq
and we shall adopt the following embedding ansatz:
\beq
W^1\,=\,W^1(X^1, X^2)\,\,,
\qquad
W^2\,=\,W^1(X^1, X^2)\,\,,
\qquad
 X^3, X^4,z={\rm constant}\,\,.
 \eeq
With no loss of generality we can take
\beq
X^3\,=\,L\,\,,
\qquad\qquad
X^4\,=\,0\,\,,
\qquad\qquad
z\,=\,a\,\,.
\eeq
Moreover, we define the variable $\sigma$ as:
\beq
\sigma\,=\,\sqrt{(X^1)^2+(X^2)^2}\,\,.
\label{sigma_def}
\eeq
For such an ansatz, the induced metric on the worldvolume of the probe D4-brane can be obtained by computing the pullback of the line element (\ref{10dmetric_cartesian}):
\beq
ds^2_5\,=\,{Q^{{1\over 2}}\over (3\,m\,a)^{{1\over 3}}}\,
\Bigg[{9\over 4}\,(\sigma^2+L^2+a^2)^{{5\over 6}}\,dx^2_{1,2}\,+\,
h_{ij}\,dX^i\,dX^j\Bigg]\,\,,
\label{induced_metric_D4defects_general}
\eeq
where the indices $i,j$ can take the values $1,2$,  $\partial_i\equiv \partial_{X^i}$ and $h_{ij}$ is the matrix:
\beq
h_{ij}\,=\,{\delta_{ij}\over (\sigma^2+L^2+a^2)^{{5\over 6}}}\,+\,
{9\over 4}\,(\sigma^2+L^2+a^2)^{{5\over 6}}
\big(\,\partial_iW^1\,\partial_jW^1+\partial_iW^2\,\partial_jW^2\,\big)\,\,.
\eeq
The action of the probe is the sum of a Dirac-Born-Infeld (DBI) and Wess-Zumino (WZ) term:
\beq
S\,=\,S_{DBI}\,+\,S_{WZ}\,\,.
\eeq
If we do not excite the woldvolume gauge field, the DBI action can be written as:
\beq
S_{DBI}\,=\,-T_4\,\int d^5\zeta\,e^{-\Phi}\,\sqrt{-\det g_5}\,\,,
\eeq
where $T_4$ is the tension of the D4-brane and $g_5$ is the induced metric (\ref{induced_metric_D4defects_general}). More explicitly we have:
\beq
S_{DBI}\,=\,-\Big({3\over 2}\Big)^3\,Q^{{3\over 2}}\,T_4\,
\int d^5\zeta\,
(\sigma^2+L^2+a^2)^{{5\over 6}}\sqrt{\det h_{ij}}\,\,.
\label{S_DBI_D4defect_general}
\eeq
The WZ term of the action for our setup is given by:
\beq
S_{WZ}\,=\,-T_4\,\int \hat C_5\,\,,
\eeq
where $\hat C_5$ denotes the pullback of the RR 5-form potential to the worlvolume. For our system of coordinates  this pullback is given by:
\beq
\hat C_5\,=\,-\Big({3\over 2}\Big)^5\,Q^{{3\over 2}}\,
(\sigma^2+L^2+a^2)^{{5\over 3}}\,
\big(\partial_1\,W^1\partial_2\,W^2-\partial_1\,W^2\partial_2\,W^1\big) d^3x\wedge dX^1\wedge dX^2\,\,,
\eeq
with $d^3x=dx^0\wedge dx^1 \wedge dx^2$.  It follows from this last expression  that $S_{WZ}$ is given by:
\beq
S_{WZ}\,=\,\Big({3\over 2}\Big)^5\,Q^{{3\over 2}}\,
T_4\,\int d^5\zeta\, 
(\sigma^2+L^2+a^2)^{{5\over 3}}\,
\big(\partial_1\,W^1\partial_2\,W^2-\partial_1\,W^2\partial_2\,W^1\big)\,\,.
\label{S_WZ_D4defect_general}
\eeq

\subsection{Kappa symmetry}
\label{kappa_D4_defects}

We now study SUSY configurations of a D4-brane probe  embedded in the warped  $AdS_6\times {\mathbb S}^4$ background as described in the previous section. These configurations should satisfy the kappa symmetry condition:
\beq
\Gamma_{\kappa}\,\epsilon\,=\,\pm\epsilon\,\,,
\label{kappa_sym_cond}
\eeq
where $\epsilon$ is a Killing spinor and, if there are no excited gauge fields on the worldvolume, the matrix $\Gamma_{\kappa}$ is:
\beq
\Gamma_{\kappa}\,=\,{1\over \sqrt{-\det g_5}}\,
{1\over 5!}\,\epsilon^{a_1\cdots a_5}\,\Gamma_{11}\,
\gamma_{a_1\cdots a_5}\,=\,{1\over \sqrt{-\det g_5}}\,\Gamma_{11}\,
\gamma_{x^0\,x^1\,x^2\,X^1\,X^2}\,\,,
\label{Gamma_kappa_D4}
\eeq
where, in the second step,  we have written the form of $\Gamma_{\kappa}$ for the system (\ref{vw_coordinates_D4defects}) of worldvolume coordinates.  In (\ref{Gamma_kappa_D4}) the $\gamma$'s are induced Dirac matrices along the different worldvolume coordinates and $\gamma_{\alpha_1\cdots \alpha_5}$ denotes their antisymmetrized product. The $\gamma$'s  are related to the flat Dirac matrices of the background (denoted by $\Gamma$) as:
\bear
&&\gamma_{x^{\mu}}\,=\,{3\over 2}\,
{Q^{{1\over 4}}\over (3\,m\,a)^{{1\over 6}}}\big(\sigma^2\,+\,L^2\,+a^2\big)^{{5\over 12}}\,
\Gamma_{x^{\mu}}\,\,,\qquad (\mu=0,1,2)\,\,,\qquad\qquad
\rc\rc
&&\gamma_{X^{i}}\,=\,{Q^{{1\over 4}}\over (3\,m\,a)^{{1\over 6}}}\Bigg[
{3\over 2}\,\big(\sigma^2\,+\,L^2\,+a^2\big)^{{5\over 12}}
\big(\partial_i\,W^1\,\Gamma_{x^4}+\partial_i\,W^2\,\Gamma_{x^5}\big)\,+\,\rc\rc
&&\qquad\qquad\qquad\qquad\qquad
+\,{\Gamma_{X^{i}}\over \big(\sigma^2\,+\,L^2\,+a^2\big)^{{5\over 12}}}\Bigg]\,\,,
\qquad (i=1,2)\,\,.
\eear
After a simple calculation one can check that $\Gamma_{\kappa}$ can be written as:
\beq
\Gamma_{\kappa}\,=\,\Big({3\over 2}\Big)^3\,{Q^{{3\over 4}}\over (3ma)^{{1\over 2}}}
{\big(\sigma^2\,+\,L^2\,+a^2\big)^{{5\over 4}}\over \sqrt{-\det g_5}}\,
\Gamma_{11}\,\Gamma_{x^0 x^1 x^2}\,\gamma_{X^1\,X^2}\,\,,
\eeq
where $\gamma_{X^1\,X^2}$ is given by:
\bear
&&\gamma_{X^1\,X^2}\,=\,
{Q^{{1\over 2}}\over (3\,m\,a)^{{1\over 3}}}\Bigg[
{\Gamma_{X^1 X^2}\over \big(\sigma^2+L^2+a^2\big)^{{5\over 6}}}\,+\,{9\over 4}\,
\big(\sigma^2+L^2+a^2\big)^{{5\over 6}}\,
(\partial_1 W^1\partial_2W^2-\partial_1 W^2\partial_2 W^1)\Gamma_{x^4x^5}\,+\,\rc\rc
&&\qquad\qquad\qquad
+{3\over 2}\,\Big(\partial_2 W^1\Gamma_{X^1x^4}\,+\,\partial_2 W^2\Gamma_{X^1x^5}\,-\,
\partial_1 W^1\Gamma_{X^2x^4}\,-\,\partial_1 W^2\Gamma_{X^2x^5}\,
\Big)\Bigg]\,\,.
\eear
Let us assume that the spinor $\epsilon$ satisfies the projection corresponding to the D4-branes of the background, namely:
\beq
\Gamma_{11}\,\Gamma_{x^0x^1x^2x^3 x^4}\epsilon\,=\,\epsilon\,\,.
\label{D4_projection}
\eeq
Moreover, we impose the following additional projection:
\beq
\Gamma_{11}\,\Gamma_{x^0x^1x^2X^1 X^2}\epsilon\,=\,\epsilon\,\,,
\label{D4prime_projection}
\eeq
which is the one corresponding to having D4$'$-branes extended as in the array (\ref{D4D4'intersection}).  Combined together, (\ref{D4_projection}) and (\ref{D4prime_projection}) lead to:
\beq
\Gamma_{X^1 X^2}\epsilon\,=\,\Gamma_{x^4 x^5}\epsilon\,\,.
\eeq
Then, after a simple calculation one can demonstrate that:
\bear
&&{(3\,m\,a)^{{1\over 3}}\over Q^{{1\over 2}}}
\gamma_{X^1\,X^2}\,\epsilon\,=\,\Bigg[{1\over \big(\sigma^2+L^2+a^2\big)^{{5\over 6}}}+\rc\rc
&&\qquad\qquad
+{9\over 4}\,
\big(\sigma^2+L^2+a^2\big)^{{5\over 6}}\,
(\partial_1 W^1\partial_2W^2-\partial_1 W^2\partial_2 W^1)\Bigg]\Gamma_{X^1 X^2}\epsilon\,+\,\rc\rc
&&\qquad\qquad
+{3\over 2}\,(\partial_2\,W^1+\partial_1 W^2)\,\Gamma_{X^1 x^4}\,\epsilon\,+\,
{3\over 2}\,(\partial_2\,W^2-\partial_1 W^1)\,\Gamma_{X^1 x^5}\,\epsilon\,\,.
\eear
We can now use this result, as well as the projection (\ref{D4prime_projection}), to compute $\Gamma_{\kappa}\,\epsilon$.  One can show that, in order to cancel the terms that do not contain the unit matrix one should impose the  following conditions to the embedding functions $W^1$ and $W^2$:
\beq
\partial_1 W^1\,=\,\partial_2\,W^2\,\,,
\qquad\qquad
\partial_2\,W^1\,=\,-\partial_1 W^2\,\,.
\label{CR_D4}
\eeq
The two equations in  (\ref{CR_D4}) are nothing but the Cauchy-Riemann equations. Indeed, if we define the complex variables $Z$ and $W$ as
\beq
Z\,=\,X^1+i\,X^2\,\,,
\qquad\qquad
W\,=\,W^1+i\,W^2\,\,,
\eeq
as well as the holomorphic and antiholomorphic derivatives $\partial$ and $\bar\partial$ as:
\beq
\partial\,=\,{1\over 2}\big(\partial_1\,-\,i\partial_2)\,\,,
\qquad\qquad
\bar\partial\,=\,{1\over 2}\big(\partial_1\,+\,i\partial_2)\,\,,
\eeq
then, the  BPS conditions (\ref{CR_D4}) become simply:
\beq
\bar\partial\, W\,=\,0\,\,.
\label{holo_BSP_D4}
\eeq
Eq. (\ref{holo_BSP_D4}) is solved by an arbitrary holomorphic function of the type:
\beq
W\,=\,W(Z)\,\,,
\eeq
\ie\ by a function $W$ that depends on $Z$ and not on $\bar Z$.  Moreover, if  (\ref{holo_BSP_D4}) holds one can show that the action of $\Gamma_{\kappa}$ on $\epsilon$ is:
\beq
\sqrt{-\det g_5}\,\Gamma_{\kappa}\epsilon\Big|_{BPS}=
\Big({3\over 2}\Big)^3 {Q^{{5\over 4}}\over (3ma)^{{5\over 6}}}
\big(\sigma^2+L^2+a^2\big)^{{5\over 12}}
\Bigg[1+{9\over 4}\big(\sigma^2+L^2+a^2\big)^{{5\over 3}}
\partial W\bar\partial\bar W\Bigg]\epsilon\,\,.
\label{Gamma_k_epsilon_BPS}
\eeq
Taking into account that:
\beq
\sqrt{\det h_{ij}}\Big|_{BPS}={1\over \big(\sigma^2+L^2+a^2\big)^{{5\over 6}}}
\Bigg[1+{9\over 4}\big(\sigma^2+L^2+a^2\big)^{{5\over 3}}
\partial W\bar\partial\bar W\Bigg]\,\,,
\eeq
one can straightforwardly demonstrate that $\sqrt{-\det g_5}$ for a BPS configuration equals the function multiplying the spinor on the right-hand side of (\ref{Gamma_k_epsilon_BPS}), which proves that $\Gamma_{\kappa}\,\epsilon=\epsilon$ when the holomorphic condition (\ref{holo_BSP_D4}) is satisfied.  

Let us now study the action of the probe for the holomorphic embeddings. By using (\ref{CR_D4}) in (\ref{S_DBI_D4defect_general}) we get that the DBI term for a BPS configuration is:
\beq
S_{DBI}\Big|_{BPS}\,=\,-\Big({3\over 2}\Big)^3\, Q^{{3\over 2}}\,T_4\,
\int d^5\zeta\,\Bigg[1+{9\over 4}\big(\sigma^2+L^2+a^2\big)^{{5\over 3}}
\partial W\bar\partial\bar W\Bigg]\,\,.
\label{S_DBI_D4defect_BPS}
\eeq
Moreover, from (\ref{S_WZ_D4defect_general}) we get the form of the WZ term:
\beq
S_{WZ}\Big|_{BPS}\,=\,\Big({3\over 2}\Big)^5\, Q^{{3\over 2}}\,T_4\,
\int d^5\zeta\,\big(\sigma^2+L^2+a^2\big)^{{5\over 3}}
\partial W\bar\partial\bar W\,\,.
\label{S_WZ_D4defect_BPS}
\eeq
We notice that  (\ref{S_WZ_D4defect_BPS}) cancels against the second term in (\ref{S_DBI_D4defect_BPS}), and thus for any holomorphic embedding we find the no-force condition characteristic of supersymmetry. 

Let us now consider the induced metric for the BPS embeddings. First of all, it is quite convenient to parameterize the $X^1\,X^2$ plane by the radial variable $\sigma$ introduced in (\ref{sigma_def}) and by an angle $\vartheta$ in such a way that:
\beq
(dX^1)^2+(dX^2)^2\,=\,d\sigma^2+\sigma^2\,d\vartheta^2\,\,.
\label{sigma_vartheta_def}
\eeq
Moreover, let us change the radial variable $\sigma$ and  use instead the new coordinate  $\varrho$, related to $\sigma$ as:
\beq
\sigma\,=\,\varrho^{{3\over 2}}\,\sqrt{1\,-\,{L^2+a^2\over \varrho^3}}\,\,.
\label{sigma_varrho}
\eeq
Then, for an holomorphic embedding, one can check that the induced metric on the D4-brane worldvolume  (\ref{induced_metric_D4defects_general}) becomes:
\bear
&&ds^2_{5}\,=\,{Q^{{1\over 2}}\over (3ma)^{{1\over 3}}}\,\varrho^{{1\over 2}}
\Bigg[{9\over 4}\,\varrho^2\,dx^2_{1,2}\,+\,\qquad\qquad\rc\rc
&&\qquad\qquad
+\Big(1+{9\over 4}\,\varrho^5 \partial W\bar\partial\bar W\Big)\,
\Bigg({9\over 4}\,{d\varrho^2\over \varrho^2\big(1-{L^2+a^2\over \varrho^3}\big)}+
\Big(1-{L^2+a^2\over \varrho^3}\Big)\,d\vartheta^2\Bigg)\Bigg]\,\,.
\qquad\qquad
\label{induced_metric_D4defects_BPS}
\eear
Note that in \eqref{induced_metric_D4defects_BPS} $\varrho^3\geq L^2+a^2$. Moreover, note that while $L$ may be taken to zero, $a$ must be non-vanishing. Thus, the space naturally ends at a $\varrho^3\geq a^2$ when $L$ is zero.

When the embedding is trivial, \ie\ when $W$ is constant, the metric (\ref{induced_metric_D4defects_BPS}) in the UV becomes $AdS_4\times {\mathbb S}^1$, up to a warp factor proportional to $\varrho^{{1\over 2}}$. Let us now determine the non-trivial embeddings that preserve this UV limit. Let us suppose that $W(Z)$ behaves in the UV as:
\beq
W\sim Z^{-\lambda}\,\,,
\eeq
where $\lambda$ is a real exponent. Taking into account that, for large $\varrho$, one has
\beq
\varrho^5\, \partial W\bar\partial\bar W\,\sim\,\varrho^{2-3\lambda}\,\,,
\qquad\qquad
(\varrho\to\infty)\,\,,
\eeq
it follows that the exponent $\lambda$ should be bounded from below by:
\beq
\lambda\ge {2\over 3}\,\,.
\eeq
For $\lambda>2/3$ the term depending on $W$ does not contribute to the UV behavior of the induced metric, whereas in the critical case $\lambda=2/3$ the non-trivial bending of the probe changes the radius of the warped 
$AdS_4\times {\mathbb S}^1$. Notice that these marginal embeddings can also be written as 
$Z\sim W^{-{3\over 2}}$. Note that the transverse scalars, corresponding to scalar fields in hypermultiplets in the dual theory, have precisely mass-dimension $\frac{3}{2}$, thus supporting the identification of this configuration with the Higgs branch of the defect field theory.

\subsection{Fluctuations}
 Let us now explore the fluctuations of the probe D4-brane around the simplest of the configurations found above, namely that with $L=W=0$.  The  detailed derivation of the equations governing these fluctuations and the analysis of the different modes is left for appendix \ref{D4_defects_fluct}, summarizing and interpreting here the results. In the unperturbed configuration the scalars $x^3$, $x^4$, $X^3$ and $X^4$ have vanishing value, whereas $z=a$ (see (\ref{zero_order_emb_D4})).  We will allow the probe D4-brane to oscillate around this configuration and we will denote the fluctuations as:
\beq
(\delta x^3, \delta x^4)\,=\,(U^1, U^2)\,\,,
\qquad\qquad
(\delta X^3, \delta X^4, \delta z)\,=\,(Y^1, Y^2, Y^3)\,\,.
\eeq
(see eq. (\ref{D4_fluct_scalars})). Moreover, we  switch on a worldvolume gauge field $A_{\mu}$ (the corresponding fluctuation mode is denoted by $V_{\mu}$). The corresponding second-order action  is written in (\ref{D4_defect_fluct_action}). The metric ${\cal G}_{ab}$ entering the action has been written in (\ref{induced_metric_D4defects}) and it corresponds to a geometry that approaches a (radially warped) $AdS_4\times {\mathbb S}^1$ space in the UV to which we can apply holography in the generalized sense of \cite{Kanitscheider:2008kd}. The radial warping suggests that the dual theory is a 3d non-conformal  defect QFT preserving (3d) $\mathcal{N}=2$,  containing the restriction of the rank $N$ $E_{N_f+1}$ theory on the color D4-branes down to the (3+1)d defect, in addition to $N_{D4}$ hypermultiplets. Since $a^{{2\over 3}}\sim x_9^{D4}$ is the distance between the two types of D4-branes, it sets the flavor mass scale. Thus $x_9^{D4}\sim M$ or, equivalently $a\sim M^{{3\over 2}}$. We stress that the brane configuration becomes singular if $a\to 0$ and, therefore, it is not possible to remove the mass scale $M$. 

To further understand the role of $M$, let us consider the worldvolume action of the probe D4-brane. Let us suppose that we turn on the gauge field $F_{\mu\nu}$ along the Minkowski directions. It is easy to extract the dependence on $a$ of the action (\ref{D4_defect_fluct_action}) by rescaling the holographic coordinate $\varrho$ as $\varrho=a^{{2\over 3}}\,\hat\varrho$. Doing this the dependence on $a$ of the action appears as a global coefficient and one schematically gets:
\beq
S\sim \int d^3x\,{1\over a^{{2\over 3}}}\,F^2\,\sim  \int d^3x\,{1\over M}\,F^2\,\,.
\eeq
Thus, we see that the Yang-Mills coupling for the flavor symmetry is $e^2\sim a^{\frac{2}{3}}\sim x_9^{D4}$.  Therefore 
$e\sim M^{{1\over 2}}$ and the power of $M$ above is naturally understood as coming from the flavor symmetry gauge coupling.  

The dimensions of the operators dual to the $U$, $Y$ and $V_{\mu}$ modes are written in (\ref{Delta_U}), (\ref{Delta_Y}) and (\ref{Delta_V}) respectively.  They depend on the number $n$ that fixes their dependence on the angular coordinate $\vartheta$ (\ie\ the winding number along the ${\mathbb S}^1$ of the metric). This dependence is of the type:
\beq
\Delta\,=\,\Delta(n=0)+{3n\over 2}\,\,,
\eeq
and it is easily understood as due to the insertion of $n$ adjoint scalars represented by the coordinates $(X^1, X^2)$. 
 Focusing on the $Y$ and $V_{\mu}$ modes, the lowest dimension fluctuations have, respectively, $\Delta_Y=\frac{3}{2}$ and $\Delta_V=\frac{3}{2}$.  Given the mass dependence of the  terms in the fluctuation action, it is quite natural to think that the dual operators are rescaled by a factor of the type $M^{\alpha}$ for some $\alpha$,  this rescaling being a multiplicative renormalization of the operator. In the cases of the lowest-lying $Y$ and $V_{\mu}$ modes we conjecture that they are dual to the flavor current conserved multiplet, with the rough identification:
\begin{equation}
\begin{array}{c |  c cc}
 {\rm Fluctuation} & \Delta && {\rm Dual\,\,operator}\\ \hline\\
 Y & \frac{3}{2} && M^{-\frac{1}{2}}\,\bar\psi\,\psi\\ \\
V_{\mu} & \frac{3}{2} && M^{-\frac{1}{2}}\,\Big(\bar\psi\gamma_{\mu}\psi+
q^{\dagger}D_{\mu}q-\tilde{q}^{\dagger}D_{\mu}\tilde{q}\Big)\\ 
 \end{array}
 \label{Fluct_field_D4_defects}
 \end{equation}
where $q$ and $\tilde q$ are  3d scalar fields (with canonical dimension $1/2$) and $\psi$ and $\bar\psi$ are fermionic fields in 3d with dimension $1$.  Notice that, in the $Y$-fluctuations, the distance between color and flavor D4-branes in the space orthogonal to both of them changes. In the holographic setup this distance is related to the quark mass, that sources the meson operator $\bar\psi\,\psi$, which we have identified in (\ref{Fluct_field_D4_defects}) as the dual of the $Y$-fluctuation.  Moreover, it is clear that the fluctuation of the vector field on the probe should couple to a vector current, as in (\ref{Fluct_field_D4_defects}).

In the $U$-mode the probe D4-brane expands into the worldvolume directions of the color D4-branes. This kind of brane recombination is naturally associated with the Higgs branch of the field theory and the corresponding dual operator is a bilinear in the scalar fields \cite{Arean:2006vg,Arean:2007nh}.   Thus, we are led to identify the lowest-lying $U$-mode with an operator of the type  $M^3(q^{\dagger} q-\tilde q^{\dagger}\tilde q)$, where  the power of $M$ has been adjusted to have $\Delta_U=4$ (see (\ref{Delta_U})).

\section{Wrapped D4-branes}\label{D4_wrapped}
In this section we will analyze a second setup with probe D4-branes, in which the probes wrap the 3-sphere of the internal space, extend along the holographic coordinate and intersect the D4-branes of the background at a single point, creating in this way a point-like codimension 4 defect.  Let us suppose that we write the  BO metric as in (\ref{10d_metric_string}) and let $\varphi$, $\theta$ and $\psi$ be three angles that parameterize the ${\mathbb S}^3$ inside the half ${\mathbb S}^4$. Then, the configuration we are going to study can be represented by the following array:
\beq
\label{wrappedD4_intersection}
\begin{array}{ccccccccccl}
 &x^1&x^2&x^3& x^4& r&\varphi &\theta&\psi&\alpha &  \\
D4: & \_ &\_ &\_ &\_  &\times & \times&\times &\times &\_ &      
\end{array}
\eeq
We will take the following set of worldvolume coordinates:
\beq
\zeta^{a}\,=\,(x^0, r,\varphi, \theta, \psi)\,\,,
\eeq
and we will adopt the following embedding ansatz:
\beq
\alpha\,=\,\alpha(r)\,\,,
\qquad\qquad
x^i\,=\,{\rm constant} \,\,,\qquad(i=1,\cdots, 4)\,\,.
\eeq
Inspecting the WZ term of the D4-brane worldvolume action one readily concludes that the flux of the RR 4-form  $F_4$  sources a non-trivial worldvolume gauge field  $F$ due to the term $\int C_3\wedge F$, where  $C_3$  is the 
potential for $F_4$. Since $C_3$ has only legs along the  ${\mathbb S}^3$, it follows that the $F_{0r}$ component of $F$ is the one that is induced by the RR flux. Accordingly,  we will switch on  in our ansatz the $F_{0r}$ component of $F$. The induced metric on the worldvolume in Einstein frame takes the form:
\beq
ds^2_{5,E}\,=\,K^2(\alpha)\,\Bigg[-{9\over 4}\,r^2\,(dx^0)^2\,+\,{9\over 4\,r^2}\,\Big(1+{4\over 9}\,r^2\,\alpha'^{\,2}\Big)\,dr^2\,+\,\sin^2\alpha\,d\Omega_3^2\Bigg]\,\,,
\label{induced_metric_wrapped_D4}
\eeq
where $d\Omega_3^2$ is the line element of the  ${\mathbb S}^3$, which we will represent as in 
(\ref{Omega_3_omegas}) in terms of the  three  $SU(2)$ invariant 3-forms that, in turn, can be parameterized  with the three angles  $\varphi$, $\theta$ and $\psi$  and their differentials as in (\ref{SU2_forms}).  Notice that, for fixed $\alpha$, the induced metric is of the form $AdS_2\times {\mathbb S}^3$. The worldvolume gauge field  $F$ lives in the  $AdS_2$ part of the metric. 

\subsection{Kappa symmetry analysis}
We will start our analysis by determining the conditions which make our configuration kappa-symmetric (and therefore SUSY-preserving). Since  our ansatz contains a worldvolume gauge field, the 
the kappa symmetry matrix $\Gamma_{\kappa}$ for the D4-brane is now:
\beq
\Gamma_{\kappa}\,=\,{1\over \sqrt{-\det(g+\hat F)}}\,\sum_{n=0}^{\infty}\,{1\over 2^n\,n!}\,
\gamma^{a_1\,b_1}\,\gamma^{a_2\,b_2}\,\cdots \gamma^{a_n\,b_n}\,
\hat F_{a_1\,b_1}\,\hat F_{a_2\,b_2}\,\cdots \hat F_{a_n\,b_n}\,{ J}^{(n)}\,\,,
\eeq
where $\hat F\,=\,e^{-{\Phi\over 2}}\,F$ and the matrix ${ J}^{(n)}$ is given by:
\beq
{ J}^{(n)}\,=\,{1\over 5!}\,\big(\Gamma_{11}\big)^{n+1}\,\epsilon^{a_1\cdots a_5}\,\gamma_{a_1\cdots a_5}\,\,.
\eeq
The $e^{-{\Phi\over 2}}$ factor multiplying $F$ is due to the fact that we are working in the Einstein frame. 
For our configuration, only the terms with $n=0,1$ contribute to $\Gamma_{\kappa}$:
\bear
&&\Gamma_{\kappa}\,=\,{1\over 5!\,\sqrt{-\det (g+\hat F)}}\,
\Big[\Gamma_{11}\,+\,{1\over 2}\,\gamma^{b_1\,b_2}\,\hat F_{b_1\,b_2}\,\Big]
\epsilon^{a_1\,\cdots a_5}\,\gamma_{a_1\,\cdots a_5}\,=\,\rc\rc
&&\qquad\qquad
=\,{1\over \sqrt{-\det (g+\hat F)}}\Big[\Gamma_{11}\,+\,\gamma^{x^0\,r}\,\hat F_{x^0\,r}\Big]\,
\gamma_{x^0\,r\,\varphi\,\theta\,\psi}\,\,.
\label{Gamma_k_wrapped}
\eear
We have to impose  (\ref{kappa_sym_cond})  for the matrix $\Gamma_{\kappa}$ written in (\ref{Gamma_k_wrapped}) and $\epsilon$ being a Killing spinor of the BO geometry. We will assume that the
$\epsilon$ are ordinary Killing spinors  which, in this $AdS_6$ coordinate system can be written as in 
(\ref{ordinary_spinors}) in terms of a constant spinor $\eta$. Actually, if we   define the modified matrix $\tilde\Gamma_{\kappa}$ as:
\beq
\tilde\Gamma_{\kappa}\,=\,e^{{\alpha\over 2}\,\Gamma^{789}}\,\Gamma_{\kappa}\,e^{-{\alpha\over 2}\,\Gamma^{789}}\,\,,
\label{tilde_Gamma_kappa}
\eeq
then,  the kappa symmetry condition (\ref{kappa_sym_cond}) takes the following form in terms of $\eta$:
\beq
\tilde\Gamma_{\kappa}\,\eta\,=\,\pm\eta\,\,.
\label{kappa_cond_eta}
\eeq
Let us now obtain $\tilde\Gamma_{\kappa}$ in terms of constant Dirac matrices corresponding to the 1-form basis 
written in (\ref{frame}). The induced $\gamma$-matrices for our ansatz are:

\bear
&&\gamma_{x^0}\,=\,{3\over 2}\,K(\alpha)\,r\,\Gamma_0\,\,,\qquad\qquad
\gamma_{r}\,=\,K(\alpha)\,\Big[{3\over 2r}\,\Gamma_5\,+\,\alpha'\,\Gamma_6\Big]\,\,,\rc\rc
&&\gamma_{\varphi}\,=\,{K(\alpha)\sin\alpha\over 2}\,\Big[\cos\theta\,\Gamma_9\,+\,\
\sin\theta\big(\cos\psi\,\Gamma_8+\sin\psi\Gamma_7\big)\Big]\,\,,\rc\rc
&&\gamma_{\theta}\,=\,{K(\alpha)\sin\alpha\over 2}\,\big[\cos\psi\,\Gamma_7\,-\,\sin\psi\,\Gamma_8\big]\,\,,
\qquad\qquad
\gamma_{\psi}\,=\,{K(\alpha)\sin\alpha\over 2}\,\Gamma_9\,\,.
\qquad\qquad
\eear
Therefore, their antisymmetrized product is:
\beq
\gamma_{x^0\,r\,\varphi\,\theta\,\psi}\,=\,-{9\over 32}\,K^5(\alpha)\,\sin^3\alpha\,\sin\theta\,\Big[
\Gamma_{05}\,+\,{2\over 3}\,r\alpha'\,\Gamma_{06}\Big]\,\Gamma_{789}\,\,.
\label{Induced_5_gammas}
\eeq
The indices of the induced $\gamma$-matrices are raised with the inverse of the induced metric. Thus, we have:
\beq
\gamma^{x^0\,r}\,=\,g^{x^0\,x^0}\,g^{r\,r}\,\gamma_{x^0\,r}\,=\,
\,=\,
{K^3(\alpha)\over 8}\,{\sin^3\alpha\,\sin\theta\over  1+\,{4\over 9}\,r^2\,\alpha'^{\,2}}\,
\Big[\,\Gamma_{05}\,+\,{2\,r\,\alpha'\over 3}\,\Gamma_{06}\,\Big]^2\,\Gamma_{789}\,\,.
\eeq
Since $\Gamma_{05}^2\,=\,\Gamma_{06}^2\,=\,1$ and $\{\Gamma_{05}, \Gamma_{06}\}=0$, it follows that:
\beq
\Big[\,\Gamma_{05}\,+\,{2\,r\,\alpha'\over 3}\,\Gamma_{06}\,\Big]^2\,=\,1\,+\,{4\over 9}\,r^2\,\alpha'^{\,2}\,\,,
\eeq
and then:
\beq
\gamma^{x^0\,r}\,\gamma_{x^0\,r\,\varphi\,\theta\,\psi}\,=\,
{K^3(\alpha)\over 8}\,\sin^3\alpha\,\sin\theta\,\Gamma_{789}\,\,.
\label{Induced_7_gammas}
\eeq
Using (\ref{Induced_5_gammas}) and (\ref{Induced_7_gammas}) to evaluate $\Gamma_{\kappa}$ in (\ref{Gamma_k_wrapped}), we get:
\beq
\sqrt{-\det (g+\hat F)}\,\Gamma_{\kappa}\,=\,{K^5\over 8}\,\sin^3\alpha\,\sin\theta
\Big[-{9\over 4}\,\Gamma_{11}\,\big(\Gamma_{05}\,+\,{2\,r\,\alpha'\over 3}\,\Gamma_{06}\,\big)\,+\,
{\hat F_{0r}\over K^2}\Big]\,\Gamma_{789}\,\,.
\eeq
We now compute $\tilde \Gamma_{\kappa}$, as defined in (\ref{tilde_Gamma_kappa}). We use:
\beq
e^{{\alpha\over 2}\,\Gamma^{789}}\,
\Gamma_{11}\,\big(\Gamma_{05}\,+\,{2\,r\,\alpha'\over 3}\,\Gamma_{06}\,\big)
\,e^{-{\alpha\over 2}\,\Gamma^{789}}\,=\,
\Gamma_{11}\,\big(\Gamma_{05}\,+\,{2\,r\,\alpha'\over 3}\,\Gamma_{06}\,\big)
(\cos\alpha\,-\,\sin\alpha\,\Gamma_{789})\,\,.
\eeq
Taking into account that $(\Gamma_{789})^2\,=\,-1$, we obtain:
\beq
\sqrt{-\det (g+\hat F)}\,\tilde\Gamma_{\kappa}\,={K^5\over 8}\,\sin^3\alpha\,\sin\theta\,\Big[-{9\over 4}\,
\Gamma_{11}\,\big(\Gamma_{05}\,+\,{2\,r\,\alpha'\over 3}\,\Gamma_{06}\,\big)
(\cos\alpha\,\Gamma_{789}\,+\,\sin\alpha)\,+\,
{\hat F_{0r}\over K^2}\,\Gamma_{789}\Big]\,\,.
\qquad\qquad\qquad\qquad
\eeq
Let us now study the action of $\tilde\Gamma_{\kappa}$ on the constant spinors $\eta$. After using the projections written in (\ref{ordinary_spinors}) we get:
\bear
&&\sqrt{-\det (g+\hat F)}\,\tilde\Gamma_{\kappa}\,\eta\,={K^5\over 8}\,\sin^3\alpha\,\sin\theta\,
\Bigg[-{3\over 2}\,\Big({3\over 2}\,\sin\alpha\,+\,\cos\alpha\,r\alpha'\Big)\Gamma_{11}\,\Gamma_{05}\,+\,\rc\rc
&&\qquad\qquad\qquad\qquad
+{3\over 2}\,\Big({3\over 2}\,\cos\alpha\,-\,\sin\alpha\,r\alpha'\Big)\Gamma_{6}\Gamma_{11}\,\Gamma_{05}\,+\,
{\hat F_{0r}\over K^2}\,\Gamma_6\,\Bigg]\eta\,\,.
\label{t_G_eta_1}
\eear
We now impose an extra projection which corresponds to having (anti)fundamental strings extended along the radial direction:
\beq
\Gamma_{11}\,\Gamma_{05}\,\eta\,=\,\sigma\,\eta\,\,,
\qquad\qquad
\sigma\,=\,\pm1\,\,.
\label{String_projection}
\eeq
Using (\ref{String_projection}) we can convert (\ref{t_G_eta_1}) into:
\bear
&&\sqrt{-\det (g+\hat F)}\,\tilde\Gamma_{\kappa}\,\eta\,={K^5\over 8}\,\sin^3\alpha\,\sin\theta\,
\Bigg[-{3\sigma\over 2}\,\Big({3\over 2}\,\sin\alpha\,+\,\cos\alpha\,r\alpha'\Big)\,+\,\rc\rc
&&\qquad\qquad\qquad\qquad\qquad\qquad
+\,\Bigg[{3\sigma\over 2}\,\Big({3\over 2}\,\cos\alpha\,-\,\sin\alpha\,r\alpha'\Big)\,+\,{\hat F_{0r}\over K^2}\Bigg]\,\Gamma_6\,
\Bigg]\,\eta\,\,.
\label{t_G_eta_2}
\eear
As $\tilde\Gamma_{\kappa}$ should act on $\eta$ as $\pm1$, the terms multiplying $\Gamma_6$ on the right-hand side of  (\ref{t_G_eta_2})
should vanish. This condition implies the following BPS relation between the worldvolume gauge field and the embedding function $\alpha(r)$:
\beq
\hat F_{0r}\,=\,{3\sigma\over 2}\,K^2(\alpha)\,\Big(\sin\alpha\,r\alpha'\,-\,{3\over 2}\,\cos\alpha\Big)\,\,.
\label{BPS_wrapped}
\eeq
If (\ref{BPS_wrapped}) holds we have:
\beq
\sqrt{-\det (g+\hat F)}\,\tilde\Gamma_{\kappa}\,\eta\,\Big|_{BPS}\,=-{9\sigma\over 32}\,
K^5(\alpha)\,\sin^3\alpha\,\sin\theta\,\Big(\sin\alpha\,+\,{2\over 3}\cos\alpha\,r\alpha'\Big)\eta\,\,.
\label{t_G_eta_3}
\eeq
In order to determine $\tilde\Gamma_{\kappa}\,\eta$ for the BPS configurations, 
let us now compute the DBI determinant for our ansatz. A short calculation shows that:
\beq
\sqrt{-\det (g+\hat F)}\,=\,{9\over 32}\,K^5(\alpha)\,\sin^3\alpha\sin\theta\,
\sqrt{1+{4\over 9}\,r^2\,\alpha'^{\,2}\,-\,{16\over 81\,K^4(\alpha)}\,\hat F_{0r}^2}\,\,.
\label{DBI_det_wrapped}
\eeq
When the BPS condition (\ref{BPS_wrapped})  is satisfied, the square root on the right-hand side of 
(\ref{DBI_det_wrapped}) becomes simply:
\beq
\sqrt{1+{4\over 9}\,r^2\,\alpha'^{\,2}\,-\,{16\over 81\,K^4(\alpha)}\,\hat F_{0r}^2}\,\Bigg|_{BPS}\,=\,
\sin\alpha\,+\,{2\over 3}\cos\alpha\,r\alpha'\,\,,
\label{BPS_sqrt}
\eeq
and, therefore:
\beq
\tilde\Gamma_{\kappa}\,\eta\,\big|_{BPS}\,=\,-\sigma\,\eta\,\,,
\eeq
which implies that our embeddings are kappa symmetric when (\ref{BPS_wrapped}) is satisfied. 

Let us look in more detail at the value of the gauge field we have found in (\ref{BPS_wrapped}) for the BPS configurations. Since:
\beq
K^2(\alpha)\,e^{{\Phi\over 2}}\,=\,{Q^{{1\over 2}}\over (3m)^{{1\over 3}}}\,
\big(\cos\alpha\big)^{-{1\over 3}}\,\,,
\eeq
we can rewrite $F_{0r}$ as:
\beq
F_{0r}\,=\,\sigma\,{3\over 2}\,{Q^{{1\over 2}}\over (3m)^{{1\over 3}}}\,
\Big[r\,\big(\cos\alpha\big)^{-{1\over 3}}\,\sin\alpha\,\alpha'\,-\,{3\over 2}\,\big(\cos\alpha\big)^{{2\over 3}}\Big]\,\,.
\label{F_BPS}
\eeq
In this form $F_{0r}$ can be written as a total radial derivative:
\beq
F_{0r}\,=\,-\sigma\,{9\over 4}\,{Q^{{1\over 2}}\over (3m)^{{1\over 3}}}\,
\partial_r\,\Big(r\big(\cos\alpha\big)^{{2\over 3}}\Big)\,\,.
\label{F_BPS_derivative}
\eeq
Let us represent $F_{0r}\,=\,-\partial_r\,A_0$ in the $A_r=0$ gauge. Clearly, the gauge potential $A_0$ is given by:
\beq
A_0\,=\,\sigma\,{9\over 4}\,{Q^{{1\over 2}}\over (3m)^{{1\over 3}}}\,
r\big(\cos\alpha\big)^{{2\over 3}}\,\,.
\eeq

\subsection{Equations of motion}
Let us now study the configuration of the wrapped D4-brane from the point of view of its worldvolume action. This action is the sum of a DBI and WZ term. The former is given by:
\beq
S_{DBI}\,=\,-T_4\,\int\,d^5\zeta\,e^{{\Phi\over 4}}\,
\sqrt{-\det\big(g_{5,E}+e^{-{\Phi\over 2}}\,F\big)}\,\,,
\eeq
where $T_4$ is the tension of the D4-brane and $g_{5,E}$ is the induced metric in Einstein frame written in (\ref{induced_metric_wrapped_D4}). 
After integrating over the angles of the worldvolume, we get:
\beq
S_{DBI}\,=\,\int dt\,dr\,{\cal L}_{DBI}\,\,,
\eeq
where the DBI lagrangian density ${\cal L}_{DBI}$ is given by:
\beq
{{\cal L}_{DBI}\over 16\,\pi^2\,T_4}\,=\,-{9\over 32}\,Q^{{3\over 2}}\,\sin^3\alpha\,
\sqrt{1+{4\over 9}\,r^2\,\alpha'^{\,2}\,-\,{16\over 81\,K^4(\alpha)}\,e^{-\Phi}\ F_{0r}^2}\,\,.
\eeq
Similarly, the WZ action can be written as:
\beq
S_{WZ}\,=\,\sigma\,T_4\,\int \hat C_3\wedge F\,\,,
\eeq
where $\sigma=\pm 1$ is the same sign as in  (\ref{String_projection}) and (\ref{BPS_wrapped}) and $\hat C_3$ is the pullback of the RR 3-form potential of $F_4$ (\ie\ $F_4=dC_3$). Taking into account the expression (\ref{F_4_expression}) of $F_4$:
\beq
F_4\,=\,-{5\over 12}\,Q\,(3m)^{{1\over 3}}\,\big(\cos\alpha\big)^{{1\over 3}}\,\sin^3\alpha\,\sin\theta\,d\alpha\wedge d\theta\wedge d\varphi\wedge d\psi\,\,,
\eeq
we write the 3-form potential $C_3$ as:
\beq
C_3\,=\,{Q\,(3m)^{{1\over 3}}\over 8}\,C(\alpha)\,\sin\theta\,\wedge d\theta\wedge d\varphi\wedge d\psi\,\,,
\eeq
where $C(\alpha)$ is the solution of the differential equation:
\beq
{d C\over d\alpha}\,=\,-{10\over 3}\,\big(\cos\alpha\big)^{{1\over 3}}\,\sin^3\alpha\,\,,
\qquad\qquad
C(\alpha=0)\,=\,0\,\,.
\label{eq_C}
\eeq
Notice that we have fixed our gauge freedom in such a way that we have no sources of worldvolume gauge field at $\alpha=0$, since the gauge potential vanishes at that point. This prescription is similar to the one used in \cite{Camino:2001at}.  The integration of (\ref{eq_C}) is straightforward and gives:
\beq
C(\alpha)\,=\,\big(\cos\alpha\big)^{{4\over 3}}\,\Big(\sin^2\alpha\,+\,{3\over 2}\Big)\,-\,{3\over 2}\,\,.
\label{C_alpha}
\eeq
If we  now write the WZ action as:
\beq
S_{WZ}\,=\,\int dt\,dr\,{\cal L}_{WZ}\,\,,
\eeq
then the WZ lagragian density is:
\beq
{{\cal L}_{WZ}\over 16\,\pi^2\,T_4}\,=\,\sigma\,{Q\,(3m)^{{1\over 3}}\over 8}\,C(\alpha)\,F_{0r}\,\,.
\eeq
Therefore, if ${\cal L}\,=\,{\cal L}_{DBI}+{\cal L}_{WZ}$ is the total lagrangian density:
\beq
S\,=\,S_{DBI}\,+\,S_{WZ}\,=\,\int dt\,dr\,{\cal L}\,\,,
\eeq
we have:
\beq
{{\cal L}\over 16\,\pi^2\,T_4}\,=\,-{9\over 32}\,Q^{{3\over 2}}\,\sin^3\alpha\,
\sqrt{1+{4\over 9}\,r^2\,\alpha'^{\,2}\,-\,{16\over 81\,K^4(\alpha)}\,e^{-\Phi}\ F_{0r}^2}\,+\,
\sigma\,{Q\,(3m)^{{1\over 3}}\over 8}\,C(\alpha)\,F_{0r}\,\,.
\label{total_Lag_density}
\eeq
The equation of motion for the gauge field derived from ${\cal L}$ implies Gauss' law:
\beq
{\partial {\cal L}\over \partial F_{0r}}\,=\,{\rm constant}\,\,.
\label{Gauss_law}
\eeq
Let us analyze in more detail this Gauss' law. 
The left-hand side of (\ref{Gauss_law}) is given by:
\beq
{\partial {\cal L}\over \partial F_{0r}}\,=\,16\pi^2\,T_4\,\Bigg[
{Q^{{3\over 2}}\over 18}\,{\sin^3\alpha\over K^4}\,
{e^{-\Phi}\,F_{0r}\over \sqrt{1+{4\over 9}\,r^2\,\alpha'^{\,2}\,-\,{16\over 81\,K^4}\,e^{-\Phi}\ F_{0r}^2}}\,+\,
\sigma\,{Q\,(3m)^{{1\over 3}}\over 8}\,C(\alpha)\Bigg]\,\,.
\eeq
Moreover, the right-hand side of (\ref{Gauss_law}) should be quantized. Following \cite{Camino:2001at}, we impose the following 
quantization condition:
\beq
{\partial {\cal L}\over \partial F_{0r}}\,=\,-\sigma\,n\,T_f\,\,,
\label{quantization_wrappedD4}
\eeq
where $T_f$ is the tension of the fundamental string and $n\in {\mathbb Z}$. Let us  next take into account that:
\beq
{n\,T_f\over 16\,\pi^2\,T_4}\,=\,{n\,\pi\over 2}\,\,,
\eeq
and let us define a new function ${\cal C}_{n}(\alpha)$ as:
\beq
{\cal C}_{n}(\alpha)\,\equiv\,C(\alpha)\,+\,{4\,n\,\pi\over Q\,(3m)^{{1\over 3}}}\,\,.
\eeq
Then, the quantization condition (\ref{quantization_wrappedD4})  can be written as
\beq
Q^{{3\over 2}}\,{\sin^3\alpha\over K^4}\,
{e^{-\Phi}\,F_{0r}\over \sqrt{1+{4\over 9}\,r^2\,\alpha'^{\,2}\,-\,{16\over 81\,K^4}\,e^{-\Phi}\ F_{0r}^2}}\,=\,-\,{9\sigma\over 4}\,
Q\,(3m)^{{1\over 3}}\,{\cal C}_{n}(\alpha)\,\,.
\label{quant_cond_cal_C}
\eeq
Moreover, since:
\beq
Q\,(3m)^{{1\over 3}}\,=\,{8\pi\over 3}\,N\,\,,
\label{Q_m_N}
\eeq
the function ${\cal C}_{n}(\alpha)$ can be written as:
\beq
{\cal C}_{n}(\alpha)\,=\,C(\alpha)+{3\over 2}\,{n\over N}\,=\,
\big(\cos\alpha\big)^{{4\over 3}}\,\Big(\sin^2\alpha\,+\,{3\over 2}\Big)\,+\,{3\over 2}\,\Big({n\over N}-1\Big)\,\,.
\eeq

Let us next find a first-order BPS differential equation for the embedding function  $\alpha\,=\,\alpha(r)$. Plugging (\ref{BPS_wrapped})  on the left-hand side of (\ref{quant_cond_cal_C}), we arrive at:
\beq
{Q^{{3\over 2}}\,e^{-{\Phi\over 2}}\over K^2}\,\sin^3\alpha\,\Big(\sin\alpha\,r\alpha'\,-\,{3\over 2}\,\cos\alpha\Big)\,=\,
-{3\over 2}\,Q\,(3m)^{{1\over 3}}\,\Big(
\sin\alpha\,+\,{2\over 3}\cos\alpha\,r\alpha'\Big)\,{\cal C}_{n}(\alpha)\,\,.
\eeq
Next, we use:
\beq
{Q^{{3\over 2}}\,e^{-{\Phi\over 2}}\over K^2}\,=\,Q\,(3m)^{{1\over 3}}\,
\big(\cos\alpha\big)^{{1\over 3}}\,\,,
\eeq
and solve for $\alpha'$. We get:
\beq
r\,\alpha'\,=\,{3\over 2}\,\sin\alpha\,
{\big(\cos\alpha\big)^{{4\over 3}}\,\sin^2\alpha\,-\,{\cal C}_{n}(\alpha)\over
\big(\cos\alpha\big)^{{1\over 3}}\,\sin^4\alpha\,+\,\cos\alpha\,{\cal C}_{n}(\alpha)}\,\,.
\label{BPS_alpha}
\eeq
Let us now define the function $\Lambda_n(\alpha)$ as:
\beq
\Lambda_n(\alpha)\,=\,
\big(\cos\alpha\big)^{{4\over 3}}\,\sin^2\alpha\,-\,{\cal C}_{n}(\alpha)\,\,.
\label{Lambda_n_def}
\eeq
This function appears in the numerator of the BPS equation (\ref{BPS_alpha}) for $\alpha(r)$.  As:
\beq
\big(\cos\alpha\big)^{{1\over 3}}\,\sin^4\alpha\,+\,\cos\alpha\,{\cal C}_{n}(\alpha)\,=\,
\big(\cos\alpha\big)^{{1\over 3}}\,\sin^2\alpha\,-\,\cos\alpha\,\Lambda_n(\alpha)\,\,,
\eeq
then, the  BPS equation (\ref{BPS_alpha}) can be rewritten as:
\beq
r\,\alpha'\,=\,{3\over 2}\,{\sin\alpha\,\Lambda_n(\alpha)\over 
\big(\cos\alpha\big)^{{1\over 3}}\,\sin^2\alpha\,-\,\cos\alpha\,\Lambda_n(\alpha)}\,\,.
\label{BPS_alpha_Lambda}
\eeq
The explicit form of $\Lambda_n(\alpha)$ is:
\beq
\Lambda_n(\alpha)\,=\,-{3\over 2}\Big[\big(\cos\alpha\big)^{{4\over 3}}\,+\,{n\over N}\,-\,1\Big]\,\,.
\label{Lambda_n}
\eeq
Its derivative is rather simple:
\beq
{d \,\Lambda_n(\alpha)\over d\alpha}\,=\,2\,\big(\cos\alpha\big)^{{1\over 3}}\,\sin\alpha\,\,.
\label{d_Lambda}
\eeq

In order to study the energy of our configurations let us define the energy density ${\cal H}$ by performing the Legendre transform of ${\cal L}$:
\beq
{\cal H}\,=\,F_{0r}\,{\partial\,{\cal L}\over \partial F_{0r}}\,-\,{\cal L}\,\,.
\eeq
Explicitly, we get:
\beq
{\cal H}\,=\,{9\pi^2\over 2}\,T_4\,Q^{{3\over 2}}\,\sin^3\alpha\,
{1+{4\over 9}\,r^2\,\alpha'^{\,2}\over  \sqrt{1+{4\over 9}\,r^2\,\alpha'^{\,2}\,-\,{16\over 81\,K^4}\,e^{-\Phi}\ F_{0r}^2}}\,\,.
\label{Legendre_H}
\eeq
Let us next eliminate $F_{0r}$  from the right-hand side of (\ref{Legendre_H})  by using Gauss' law. 
From (\ref{quant_cond_cal_C}) we obtain the relation between $F_{0r}$ and $\alpha(r)$:
\beq
F_{0r}\,=\,-{9\over 4}\sigma\,{Q^{{1\over 2}}\over (3m)^{{1\over 3}}}\,
{\sqrt{1+{4\over 9}\,r^2\,\alpha'^{\,2}}\over \big(\cos\alpha\big)^{{2\over 3}}\,
\sqrt{\sin^6\alpha\,+\,\big(\cos\alpha\big)^{-{2\over 3}} {\cal C}_n(\alpha)^2}}\,\,
 {\cal C}_n(\alpha)\,\,.
\eeq
Using this last expression we can easily demonstrate that:
\beq
\sqrt{1+{4\over 9}\,r^2\,\alpha'^{\,2}\,-\,{16\over 81\,K^4}\,e^{-\Phi}\ F_{0r}^2}\,=\,\sin^3\alpha\,\,
{\sqrt{1+{4\over 9}\,r^2\,\alpha'^{\,2}}\over \sqrt{\sin^6\alpha\,+\,\big(\cos\alpha\big)^{-{2\over 3}} {\cal C}_n(\alpha)^2}}\,\,.
\eeq
Plugging this result on the right-hand side of (\ref{Legendre_H}), we finally arrive at:
\beq
{\cal H}\,=\,{9\over 32\,\pi^2}\,Q^{{3\over 2}}\,
\sqrt{1+{4\over 9}\,r^2\,\alpha'^{\,2}}\,
\sqrt{\sin^6\alpha\,+\,\big(\cos\alpha\big)^{-{2\over 3}} {\cal C}_n(\alpha)^2}\,\,.
\label{cal_H}
\eeq
This function ${\cal H}$ is nothing but the Routhian density of the system,  obtained after eliminating the worldvolume gauge field, which is a cyclic variable. The equations of motion of the system are equivalent to the Euler-Lagrange equations derived from ${\cal H}$. One can directly show that any function $\alpha=\alpha(r)$ satisfying the first-order equation (\ref{BPS_alpha_Lambda}) also solves the second-order Euler-Lagrange equation. Instead of giving details of this computation, let us demonstrate that the solutions of (\ref{BPS_alpha_Lambda}) saturate an energy bound and, therefore, minimize $\int dr\, {\cal H}$. Indeed, let us  consider a general configuration with an arbitrary function $\alpha=\alpha(r)$. One can show that the hamiltonian density (\ref{cal_H}) can written as:
\beq
{\cal H}\,=\,{9\over 32\,\pi^2}\,Q^{{3\over 2}}\,
\sqrt{{\cal Z}^2\,+\,{\cal Y}^2}\,\,,
\eeq
where, for any function $\alpha=\alpha(r)$, ${\cal Z}$ is a total derivative:
\beq
{\cal Z}\,=\,{d\over dr}\,\Big[r\Big(\sin^2\alpha\,-\,(\cos\alpha)^{{2\over 3}}\,\Lambda_n(\alpha)\Big)\Big]\,\,,
\eeq
and ${\cal Y}$ is given by:
\beq
{\cal Y}\,=\,\sin\alpha\,(\cos\alpha)^{-{1\over 3}}\, \Lambda_n(\alpha)\,-\,{2\over 3}\,r\,\alpha'\,
\Big(\sin^2\alpha\,-\,(\cos\alpha)^{{2\over 3}}\,\Lambda_n(\alpha)\Big)\,\,.
\eeq
It immediately follows that ${\cal H}$ is bounded by:
\beq
{\cal H}\ge {9\over 32\pi^2}\,Q^{{3\over 2}}\,\big|{\cal Z}\big|\,\,.
\eeq
Since ${\cal Z}$ is a total derivative, the integrated energy $\int dr\, {\cal H}$ is bounded by a quantity that only depends on the boundary values of $\alpha(r)$. This implies that any $\alpha(r)$ saturating the bound also solves the Euler-Lagrange variational problem, \ie\ the equations of motion.

The saturation of the bound occurs when ${\cal Y}=0$, which is equivalent to our first-order BPS equation (\ref{BPS_alpha_Lambda}). This proves that the solutions of (\ref{BPS_alpha_Lambda}) also solve  the second-order equations of motion derived from the Routhian ${\cal H}$.

\subsection{Constant angle solutions}
\label{wrapped_D4_constant_angle}

Let us first recover the solutions of \cite{Assel:2012nf} by searching for constant angle solutions of the BPS equation (\ref{BPS_alpha_Lambda}). Clearly, a constant $\alpha$ (with $\alpha\not=0,\pi$) solves (\ref{BPS_alpha_Lambda})  if it is a zero of the function $\Lambda_n(\alpha)$. Thus, we define the angle $\alpha_n$ as the root of the equation:
\beq
\Lambda_n(\alpha_n)\,=\,0\,\,.
\label{BPS_cond_cosntant_angle}
\eeq
 The angles $\alpha_n$ are:
 \beq
 \cos\alpha_n\,=\,\Big(1-{n\over N}\Big)^{{3\over 4}}\,\,,
 \qquad\qquad
 0<n<N\,\,.
 \label{alpha_n}
 \eeq
They are a discrete set of ``latitudes" on the hemi-sphere, each of which defines a 3-sphere where we can wrap the D4-brane. We show below that these configurations behave as  bound states of $n$ fundamental strings stretched in the radial direction. Equivalently, this configuration should be dual to adding Kondo-like  impurities in the dual theory. In order to reach this conclusion, let us obtain the energy of  the BPS constant angle solutions (\ref{alpha_n}). One can check that they minimize ${\cal H}$  for constant $\alpha$:
\beq
{\partial {\cal H}\over \partial \alpha}\,\Big|_{\alpha\,=\,\alpha_n}\,=\,0\,\,.
\eeq
Moreover, we can compute the energy density for these configurations. Indeed, 
as $\Lambda_n(\alpha_n)=0$, we have:
\beq
{\cal C}_n(\alpha_n)\,=\,(\cos\alpha_n)^{{4\over 3}}\,\sin^2\alpha_n\,\,,
\eeq
and we can easily prove that:
\beq
{\cal E}_{n}\,=\,{\cal H}(\alpha=\alpha_n)\,=\,{9\over 32\,\pi^2}\,Q^{{3\over 2}}\,\sin^2\alpha_n\,\,.
\eeq
Using the value of the angles $\alpha_n$ written in (\ref{alpha_n}), we get:
\beq
{\cal E}_{n}\,=\,{9\over 32\,\pi^2}\,Q^{{3\over 2}}\,
\Bigg[1\,-\,\Big(1-{n\over N}\Big)^{{3\over 2}}
\Bigg]\,\,.
\eeq
Let us see that ${\cal E}_{n}$ can be interpreted as the energy density (or tension) of a bound state of $n$ fundamental strings. With this purpose,  let us consider the Nambu-Goto action of a fundamental string:
\beq
S_f\,=\,-T_f\,\int d^2\zeta\,e^{{\Phi\over 2}}\,\sqrt{-\det g_{2, E}}\,\,,
\eeq
where $g_{2, E}$ is the induced metric on the worldsheet and the dilaton factor appears because we are in the Einstein frame. 
We extend the string in $(t,r)$ at constant angle $\alpha$. The induced metric is:
\beq
ds_{2, E}^2\,=\,{9\over 4}\,K^2(\alpha)\,\Big[-r^2\,dt^2\,+\,{dr^2\over r^2}\,\Big]\,\,.
\eeq
Therefore, the energy density for the string is:
\beq
{\cal H}_f\,=\,{9\over 4}\,T_f\,e^{{\Phi\over 2}}\,K^2(\alpha)\,=\,
{9\over 8\pi}\,{Q^{{1\over 2}}\over (3m)^{{1\over 3}}\,\big(\cos\alpha\big)^{{1\over 3}} }\,\,.
\eeq
Using (\ref{Q_m_N}), we get:
\beq
{\cal H}_f\,=\,{27\over 64\pi^2}\,{Q^{{3\over 2}}\over N}\,\big(\cos\alpha\big)^{-{1\over 3}} \,\,.
\eeq
Let us define $\tau_f$ as the tension of the fundamental string  stretched along the radial direction at $\alpha=0$:
\beq
\tau_f\,\equiv\,{\cal H}_f(\alpha=0)\,\,,
\eeq
or equivalently:
\beq
\tau_f\,=\,{27\over 64\pi^2}\,{Q^{{3\over 2}}\over N}\,\,.
\eeq
It follows that the tension of the string at an angle $\alpha_n$ is related to the one at $\alpha=0$ as:
\beq
{\cal H}_f\,=\,{\tau_f\over \big(1-{n\over N}\big)^{{1\over 4}}}\,\,.
\label{f_tension_alpha_n}
\eeq
We can now write ${\cal E}_{n}$ in terms of $\tau_f$ as:
\beq
{\cal E}_{n}\,=\,{2\over 3}\,N\,\tau_f\,
\Bigg[1\,-\,\Big(1-{n\over N}\Big)^{{3\over 2}}
\Bigg]\,\,.
\eeq
This energy has the interpretation of the one corresponding to a bound state of $n$ strings, indeed corresponding to a Wilson loop in the antisymmetric representation as described in \cite{Assel:2012nf}. To check this interpretation, let us obtain the limit of ${\cal E}_{n}$ as $N\to\infty$:
\beq
\lim_{N\to\infty}\,{\cal E}_{n}\,=\,n\,\tau_f\,\,.
\eeq
Notice also  that $\alpha_n\to 0$ when $N\to\infty$. In this  $N\to\infty$ limit the strings become free. One can verify that the formation of the bound state is energetically favored with respect to having $n$ free strings at $\alpha=\alpha_n$. Indeed, the tension of one of such strings is given by (\ref{f_tension_alpha_n}) and one can show that: 
\beq
{\cal E}_{n}\,\le\,n\,{\tau_f\over 
\big(1-{n\over N}\big)^{{1\over 4}}}\,\,.
\eeq

 \subsection{General solution}
 Let us now find the general solution of the BPS  differential equation (\ref{BPS_alpha_Lambda}).  First of all, we use (\ref{d_Lambda}) to write (\ref{BPS_alpha_Lambda}) as:
 \beq
 r\,{d\alpha\over dr}\,=\,{3\sin\alpha\,\Lambda_n(\alpha)\over 
 \sin\alpha\,{d\Lambda_n(\alpha)\over d\alpha}\,-\,2\,\cos\alpha\,\Lambda_n(\alpha)}\,\,.
 \eeq
Written in this form the equation can be immediately integrated:
\beq
C\,r\,=\,{\Big[\Lambda_n(\alpha)\Big]^{{1\over 3}}\over \big(\sin\alpha\big)^{{2\over 3}}}\,\,,
\label{general_BPS_integral}
\eeq
where $C$ is a constant of integration. We should regard (\ref{general_BPS_integral}) as giving implicitly the function 
$\alpha=\alpha(r)$. The constant angle soutions correspond to take the constant $C=0$.  When $C\not=0$ and for generic values of $0<n<N$, the solution reaches $r=0$ with $\alpha\to\alpha_n$ and $r\to\infty$ at $\alpha=0$ or $\alpha=\pi$ (depending on the phase chosen for  the constant $C$).  Clearly, $n=0, N$ are special cases. When $n=0$, after choosing the phase of the constant $C$ appropriately,  the BPS solution is:
\beq
\Big({r\over r_0}\Big)^{{3\over 2}}\,=\,{
\Big[1-\big(\cos\alpha\big)^{{4\over 3}}\Big]^{{1\over 2}}\over \sin\alpha}\,\,,
\qquad\qquad (n=0).
\eeq
In this $n=0$ solution the coordinate $r$ remains finite as $\alpha\in [0, \pi/2]$ and
\beq
\lim_{\alpha\to 0}\,r\,=\,\Big({2\over 3}\Big)^{{1\over 3}}\,\,r_0
\qquad\qquad
\lim_{\alpha\to\pi/2}\,r\,=\,r_0\,\,,
\qquad\qquad (n=0).
\eeq
For $n=N$ the solution is:
\beq
\Big({r\over r_0}\Big)^{{3\over 2}}\,=\,{
\big(\cos\alpha\big)^{{2\over 3}}\over\sin\alpha}\,\,,
\qquad\qquad (n=N).
\eeq
In this case:
\beq
\lim_{\alpha\to 0}\,r\,=\,\infty\,\,,
\qquad\qquad
\lim_{\alpha\to\pi/2}\,r\,=\,0\,\,,
\qquad\qquad (n=N).
\eeq

\subsection{Fluctuations of the constant angle solutions}
In order to verify the stability of the constant angle configurations of the wrapped D4-branes we should analyze the fluctuations around the $\alpha=\alpha_n$ solutions of subsection \ref{wrapped_D4_constant_angle}. This study is performed in detail in appendix \ref{fluc_warpped_D4}, where we consider the most general fluctuation of the type:
\beq
\delta\alpha\,=\,\xi\,\,,
\qquad
\delta F_{\mu\nu}\,=\,f_{\mu\nu}\,\,,
\qquad
\delta x^{i}\,=\,\chi^i\,\,.
\eeq
These fluctuation modes are fields in $AdS_2\times {\mathbb S}^3$ that are generically coupled. Therefore, in some cases, one needs to redefine the fields in order to diagonalize the fluctuation equations. After separating the angular variables one ends up with reduced equations for massive fields in $AdS_2$. The masses of these $AdS_2$ fields depend on the angular momentum quantum number $l$  of the ${\mathbb S}^3$  harmonic. The spectrum of these $AdS_2$  masses and of the corresponding dual dimensions is worked out in detail in appendix \ref{fluc_warpped_D4} and will not be repeated here. It suffices to say that the dimensions are generically irrational but positive, which  ensures the stability that we wanted to check.

\section{Probe D6-branes}\label{D6}

Let us now address the problem of finding supersymmetric D6-brane probes in the BO background. We will consider the configuration in which the probe D6-brane creates a codimension 1 defect in the 5d gauge theory. In terms of the cartesian coordinates used in (\ref{10dmetric_cartesian})-(\ref{C5_cartesian}) our setup can be summarized in the array:
\beq
\label{D6intersection}
\begin{array}{ccccccccccl}
 &x^1&x^2&x^3& x^4& X^1&X^2 &X^3&X^4&z &  \\
D6: & \times &\times &\times &\_ &\times & \times&\_ &\_ &\times &      
\end{array}
\eeq
Actually, it will be more convenient to describe the embedding of the D6-brane to change from the  coordinates 
$(X^1, X^2, z)$ to $(\rho, \gamma, \phi)$, related as:
\beq
X^1\,=\,\rho^{{3\over 2}}\,\sin\gamma\,\sin\phi\,\,,
\qquad\qquad
X^2\,=\,\rho^{{3\over 2}}\,\sin\gamma\,\cos\phi\,\,,
\qquad\qquad
z\,=\,\rho^{{3\over 2}}\,\cos\gamma \ .
\eeq
Moreover, we will group the third and fourth components of $\vec X$  into a new vector $\vec y$:
\beq
\vec y\,=\,(y_1, y_2)\,=\,(X^3, X^4)\,\,.
\eeq
Let us choose the following set of worldvolume coordinates for the D6-brane probe:
\beq
\zeta^a\,=\,(x^0, x^1, x^2, x^3, \rho, \gamma, \phi)\,\,.
\label{wv_coordinates_D6}
\eeq
We embed the probe in such a way that:
\beq
x^4\,=\,x^4_0\,=\,{\rm constant}\,\,,
\qquad\qquad\qquad
\vec y\,=\,(y(\rho), 0)\,\,.
\eeq
The induced metric for this type of embeddings can be readily obtained from (\ref{10dmetric_cartesian}), namely:
\bear
&&ds_{7}^2\,=\,{Q^{{1\over 2}}\over (3m)^{{1\over 3}}}\,(\cos\gamma)^{-{1\over 3}}
\Bigg[{9\over 4}\,{(\rho^3+\vec y^{\,2})^{{5\over 6}}\over \rho^{{1\over 2}}}\,
dx^2_{1,3}\,+\,\Big({9\over 4}\,\rho\,+\,y'^{\,2}\Big){d\rho^2\over \rho^{{1\over 2}}(\rho^3+\vec y^{\,2})^{{5\over 6}}}\,+\rc\rc
&&\qquad\qquad\qquad\qquad\qquad
{\rho^{{5\over 2}}\over (\rho^3+\vec y^{\,2})^{{5\over 6}}}\,d\Omega_2^2
\Bigg]\,\equiv\,
{Q^{{1\over 2}}\over (3m)^{{1\over 3}}}\,(\cos\gamma)^{-{1\over 3}}\,
{\cal G}_{ab}\,\d\zeta^a\,d\zeta^b\,\,,
\label{7d_metric}
\eear
where $d\Omega_2^2=d\gamma^2+\sin^2\gamma\, d\phi^2\,\equiv\,h_{ij}\,d\zeta^i\,d\zeta^j$  is the metric of the 2-sphere and, in the last step, we have defined the effective 7d metric ${\cal G}_{ab}$.  Notice that, in the UV region $\rho\to\infty$, the metric ${\cal G}_{ab}$ is of the form:
\beq
{\cal G}_{ab}\,\d\zeta^a\,d\zeta^b\approx
{9\over 4}\,\Big(\rho^2\,dx^2_{1,3}\,+\,{d\rho^2\over \rho^2}\Big)\,+\,d\Omega_2^2\,\,,
\qquad\qquad (\rho\to\infty)\,\,,
\eeq
which corresponds to $AdS_5\times {\mathbb S}^2$.

Up to a constant, the DBI lagrangian density of the probe  is:
\beq
e^{-\Phi}\,\sqrt{-\det  g_7}\,=\,\Big({9\over 4}\Big)^2\,{Q^2\over (3m)^{{1\over 3}}}\,
{\sin\gamma\over (\cos\gamma)^{{1\over 3}}}\,\rho^{{5\over 2}}\,
\sqrt{{9\over 4}\,\rho\,+\,y'^{\,2}} \ .
\eeq
It is trivial to verify that the equation of motion of $y$  is satisfied if $y$ is constant. Let us take:
\beq
y\,=\,L\,\,,
\eeq
where $L$ is proportional to the mass of the quarks (\ie\ fields in the fundamental representation) introduced by the 
D6-brane.

We now consider a  more general configuration of the D6-brane with internal flux (corresponding to the Higgs branch of the theory).  We will choose  worldvolume coordinates as in (\ref{wv_coordinates_D6}) and will turn on a flux along $(\gamma, \phi)$. For consistency, the brane must be bent in the $x^4$ direction (see below). Therefore, our embedding is characterized by:
\beq
y\,=\,L\,\,,
\qquad\qquad
x^4\,=\,x(\rho)\,\,,
\qquad\qquad
F\,=\,F_{\gamma\phi}\,\,d\gamma\,\wedge\,d\phi\,\,,
\eeq
where $L$ is constant. The DBI lagrangian density is now proportional to:
\bear
&&e^{-\Phi}\,\sqrt{-\det(g_7+F)}\,=\,
\Big({3\over 2}\Big)^5\,{Q^2\over (3m)^{{1\over 3}}}\,{\sin\gamma\over (\cos\gamma)^{{1\over 3}}}\,
\rho^3\,
\sqrt{1+{(\rho^3+L^2)^{{5\over 3}}\over \rho}\,(x\,')^2}\times\qquad\rc\rc\rc
&&\qquad\qquad\qquad\qquad
\times\,
\sqrt{1+{(3m)^{{2\over 3}}\over Q}\,{(\rho^3+L^2)^{{5\over 3}}\over \rho^5}\,
{(\cos\gamma)^{{2\over 3}}\over \sin^2\gamma}\,F_{\gamma\phi}^2}\,\,.
\eear
In this configuration with  worldvolume flux the WZ term of the D6-brane is turned on and given by:
\beq
S_{WZ}\,=\,T_6\,\int\,\hat C_5\,\wedge\,F\,\,,
\eeq
where $\hat C_5$ is the pullback of the RR 5-form potential (\ref{C5_cartesian}), namely:
\beq
\hat C_5\,=\,-\Big({3\over 2}\Big)^5\,Q^{{3\over 2}}\,(\rho^3+L^2)^{{5\over 3}}\,x\,'\,
dx^0\wedge dx^1\wedge dx^2\wedge dx^3\wedge d\rho\,\,.
\eeq
Therefore, the WZ lagrangian density is:
\beq
{\cal L }_{WZ}\,=\,-T_6\,\Big({3\over 2}\Big)^5\,Q^{{3\over 2}}\,(\rho^3+L^2)^{{5\over 3}}\,x\,'\,F_{\gamma\phi}\,\,.
\eeq
Notice that the worldvolume flux sources a non-trivial bending $x=x(\rho)$, as stated above. The total lagrangian density takes the form:
\bear
&&{\cal L}\,=\,-T_6\,\Big({3\over 2}\Big)^5\,Q^2\Bigg[
{\sin\gamma\over (3m)^{{1\over 3}} (\cos\gamma)^{{1\over 3}}}\,\rho^3\,
\sqrt{1+{(\rho^3+L^2)^{{5\over 3}}\over \rho}\,(x\,')^2}\times\rc\rc\rc
&&\qquad\qquad\qquad\qquad
\times\,
\sqrt{1+{(3m)^{{2\over 3}}\over Q}\,{(\rho^3+L^2)^{{5\over 3}}\over \rho^5}\,
{(\cos\gamma)^{{2\over 3}}\over \sin^2\gamma}\,F_{\gamma\phi}^2}\,+\,
Q^{-{1\over 2}}\,(\rho^3+L^2)^{{5\over 3}}\,x\,'\,F_{\gamma\phi}\Bigg]\,\,.
\qquad\qquad
\eear
By inspecting ${\cal L}$ we notice that $x$ and the worldvolume gauge potential are cyclic coordinates. This means that the corresponding Euler-Lagrange equations can be integrated once as:
\beq
{\partial {\cal L}\over \partial F_{\gamma\phi}}\,=\,c_1\,\,,
\qquad\qquad
{\partial {\cal L}\over \partial x{\,'}}\,=\,c_2\,\,,
\label{first_integrals_Higgs_D6}
\eeq
with $c_1$ and $c_2$ constants. We will show below that the solution with $c_1=c_2=0$ is simple and preserves SUSY. Let us explore the equations for the gauge field. By computing the derivative of ${\cal L}$ with respect to 
$F_{\gamma\phi}$, we get:
\bear
&&{\partial {\cal L}\over \partial F_{\gamma\phi}}\,\sim\,
{(3m)^{{1\over 3}}\over Q}\,
{(\rho^3+L^2)^{{5\over 3}}\over \rho^2}\,
{(\cos\gamma)^{{1\over 3}}\over \sin\gamma}\,
\sqrt{1+{(\rho^3+L^2)^{{5\over 3}}\over \rho}\,(x\,')^2}\times\rc\rc\rc
&&\qquad\qquad\qquad\qquad
\times{F_{\gamma\phi}\over \sqrt{1+{(3m)^{{2\over 3}}\over Q}\,{(\rho^3+L^2)^{{5\over 3}}\over \rho^5}\,
{(\cos\gamma)^{{2\over 3}}\over \sin^2\gamma}\,F_{\gamma\phi}^2}}\,+\,
Q^{-{1\over 2}}\,(\rho^3+L^2)^{{5\over 3}}\,x'\,\,.
\label{partial_L_F_Higgs}
\eear
According to (\ref{first_integrals_Higgs_D6}) the right-hand side of (\ref{partial_L_F_Higgs}) has to be constant and, in particular, independent of the angle $\gamma$. Apart from the explicit dependence, only $F_{\gamma\phi}$ can contain functions of $\gamma$. Actually, in order to cancel the dependence on $\gamma$, it is straightforward to demonstrate that $F_{\gamma\phi}$ must be of the form:
\beq
F_{\gamma\phi}\,=\,f\,{\sin\gamma\over (\cos\gamma)^{{1\over 3}}}\,\,,
\label{F_gamma_Higgs}
\eeq
where $f$ is a constant. Notice that (\ref{F_gamma_Higgs}) is rather natural since the volume element of the $(\gamma, \phi)$ 2-sphere in the metric (\ref{7d_metric})  depends on $\gamma$ through $(\cos\gamma)^{-{1\over 3}}\,\sin\gamma$. 
Plugging (\ref{F_gamma_Higgs}) into (\ref{partial_L_F_Higgs}) we get:
\bear
&&{\partial {\cal L}\over \partial F_{\gamma\phi}}\,\sim\,
{(3m)^{{1\over 3}}\over Q}\,
{(\rho^3+L^2)^{{5\over 3}}\over \rho^2}\,
\sqrt{1+{(\rho^3+L^2)^{{5\over 3}}\over \rho}\,(x\,')^2}\times\rc\rc\rc
&&\qquad\qquad\qquad\qquad
\times{f\over \sqrt{1+{(3m)^{{2\over 3}}\over Q}\,{(\rho^3+L^2)^{{5\over 3}}\over \rho^5}\,
f^2}}\,+\,
Q^{-{1\over 2}}\,(\rho^3+L^2)^{{5\over 3}}\,x'\,\,.
\label{partial_L_F_Higgs_f}
\eear
Let us now impose the first equation in (\ref{first_integrals_Higgs_D6}) with $c_1=0$. It can readily be proved that this equation is equivalent to the following equation for $x\,'$:
\beq
{\sqrt{Q}\over (3m)^{{1\over 3}}}\,
{\rho^2\,x\,'\over \sqrt{1+{(\rho^3+L^2)^{{5\over 3}}\over \rho}\,(x\,')^2}}\,=\,-
{f\over \sqrt{1+{(3m)^{{2\over 3}}\over Q}\,{(\rho^3+L^2)^{{5\over 3}}\over \rho^5}\, f^2}}\,\,.
\label{D6_Higgs_c1eq0}
\eeq
After some computation, one can get  from (\ref{D6_Higgs_c1eq0}) the following simple expression for $x\,'$:
\beq
x\,'\,=\,-{(3m)^{{1\over 3}}\over \sqrt{Q}}\,\,{f\over \rho^2}\,\,,
\label{BPS_x_Higgs}
\eeq
which can be integrated as:
\beq
x(\rho)\,=\,{(3m)^{{1\over 3}}\over \sqrt{Q}}\,{f\over \rho}\,\,. 
\eeq
One can also show that this function $x(\rho)$ and the flux (\ref{F_gamma_Higgs}) satisfy the equation of motion of $x$ in (\ref{first_integrals_Higgs_D6}) with $c_2=0$. 

As mentioned above, this configuration is supersymmetric. We will verify this fact explicitly in the next subsection by analyzing the kappa symmetry of the probe D6-brane. 

\subsection{Kappa symmetry}
\label{kappa_D6_section}

Let us write the 10d metric in the Einstein frame in the coordinates used to describe the Higgs branch. We have:
\beq
ds^2_{10, E}\,=\,\Omega^2\Big[{9\over 4}\,dx_{1,4}^2+
{1\over (\rho^3+L^2)^{{5\over 3}}}\Big({9\over 4}\rho\, d\rho^2+\rho^3(d\gamma^2+\sin^2\gamma d\phi^2)+
dL^2+L^2d\beta^2\Big)\Big]\,\,,
\label{10d_metric_Einstein_for_D6}
\eeq
where $(dX^3)^2+(dX^4)^2=dL^2+L^2\,d\beta^2$ and $\Omega$ is the function:
\beq
\Omega\,=\,(3m)^{{1\over 24}}\, Q^{{5\over 16}}\,\rho^{{1\over 16}}(\rho^3+L^2)^{{5\over 16}}\,
(\cos\gamma)^{{1\over 24}}\,\,.
\eeq
Let us choose the following basis of 1-forms for the metric:
\bear
&& e^{\mu}\,=\,{3\over 2}\,\Omega\,dx^{\mu}\,\,,
\qquad (\mu=0,1,2,3,4)\,\,,\qquad\qquad\,\,
e^{\rho}\,=\,{3\over 2}\,\Omega\, {\rho^{{1\over 2}}\over  (\rho^3+L^2)^{{5\over 6}}}d\rho\,\,,\rc\rc
&&e^{\gamma}\,=\,\Omega\,{\rho^{{3\over 2}}\over  (\rho^3+L^2)^{{5\over 6}}}\,d\gamma\,\,,
\qquad\qquad\quad
e^{\phi}\,=\,\Omega\,{\rho^{{3\over 2}}\over  (\rho^3+L^2)^{{5\over 6}}}\sin\gamma \,d\phi\,\,,\rc\rc
&&e^{L}\,=\,{\Omega\over  (\rho^3+L^2)^{{5\over 6}}}\,dL\,\,,
\qquad\qquad\qquad
e^{\beta}\,=\,\Omega\,{L\over   (\rho^3+L^2)^{{5\over 6}}}\,d\beta\,\, ,
\eear
in terms of which we can write the ordinary (non-conformal) Killing spinors:
\beq
\epsilon\,=\,(\cos\gamma)^{{1\over 48}}\,
\rho^{{1\over 32}}\, (\rho^3+L^2)^{{5\over 32}}\,e^{{\gamma\over 2}\,\Gamma_{\rho\gamma}}\,
e^{{\beta-\phi\over 2}\,\Gamma_{\phi\gamma}}\,\eta\,\,,
\label{Killing_spinor_Higgs}
\eeq
where $\eta$ is a constant spinor satisfying the projections:
\beq
\Gamma_{11}\,\Gamma_{01234}\,\eta\,=\,\eta\,\,,
\qquad\qquad
\Gamma_{\rho}\,\eta\,=\,\eta\,\,.
\label{projections_eta_Higgs}
\eeq
The kappa symmetry matrix  for a D6-brane in the Higgs branch configuration is:
\beq
\Gamma_{\kappa}\,=\,{1\over 7!}\,{1\over \sqrt{-\det(g_{7,E}+\hat F)}}\,
\Big[1+{1\over 2}\,\gamma^{\nu_1\nu_2}\,\hat F_{\nu_1\nu_2}\,\Gamma_{11}\Big]
\epsilon^{\mu_1\,\cdots \mu_7}\,\gamma_{\mu_1\cdots \mu_7}\,\,,
\eeq
where $\hat F\,=\,e^{-{\Phi\over 2}}\,F$.  The induced $\gamma$-matrices appearing in $\Gamma_{\kappa}$, for the worldvolume coordinates (\ref{wv_coordinates_D6}), are:
\bear
&&\gamma_{\mu}\,=\,{3\over 2}\,\Omega\,\Gamma_{\mu}\,\,,
\qquad\qquad  (\mu=0,1,2,3)\,\,,
\qquad\qquad 
\gamma_{\rho}\,=\,{3\over 2}\,\Omega\,
\Bigg[{\rho^{{1\over 2}}\over  (\rho^3+L^2)^{{5\over 6}}}\,\Gamma_{\rho}\,+\,x\,'\,\Gamma_{4}\Bigg]\,\,,\rc\rc
&&\gamma_{\gamma}\,=\,\Omega\,{\rho^{{3\over 2}}\over  (\rho^3+L^2)^{{5\over 6}}}\,\Gamma_{\gamma}\,\,,
\qquad\qquad \qquad\qquad 
\gamma_{\phi}\,=\,\Omega\,{\rho^{{3\over 2}}\over  (\rho^3+L^2)^{{5\over 6}}}\,
\sin\gamma\,\Gamma_{\phi}\,\,,
\eear
where we have assumed that the D6-brane embedding is such that $L$ and $\beta$ are constant. 
From these expressions of the $\gamma$'s we get:
\bear
&&{1\over 7!}\,\epsilon^{\mu_1\,\cdots \mu_7}\,\gamma_{\mu_1\cdots \mu_7}\,=\,
\Big({3\over 2}\Big)^5\,\Omega^7\,\sin\gamma\,
{\rho^3\over  (\rho^3+L^2)^{{5\over 3}}}\,
\Bigg[{\rho^{{1\over 2}}\over  (\rho^3+L^2)^{{5\over 6}}}\,\Gamma_{\rho}\,+\,x\,'\,\Gamma_{4}\Bigg]
\Gamma_{\gamma\phi}\,\Gamma_{0123}
\,\,,\rc\rc
&&{1\over 2}\,\gamma^{\nu_1\nu_2}\,\hat F_{\nu_1\nu_2}\,=\,
{ (\rho^3+L^2)^{{5\over 3}}\over \Omega^2\,\sin\gamma\,\rho^3}\,\,
\hat F_{\gamma\phi}\,\Gamma_{\gamma\phi}\,\equiv\,
{\cal F}\,\Gamma_{\gamma\phi}\,\,.
\label{gammas_F_Higgs}
\eear
Let us assume that $F_{\gamma\phi}$ has the form (\ref{F_gamma_Higgs}).  Then, it is easy to check that the quantity ${\cal F}$ defined in the second equation in (\ref{gammas_F_Higgs}) takes the form:
\beq
{\cal F}\,=\,{(3m)^{{1\over 3}}\over \sqrt{Q}}\,
{(\rho^3+L^2)^{{5\over 6}}\over \rho^{{5\over 2}}}\,f\,\,.
\label{cal_F_Higgs}
\eeq
Moreover, if we define $\Lambda$ as:
\beq
\Lambda\equiv 
\Big({3\over 2}\Big)^5\,{\Omega^7\,\sin\gamma\,\over \sqrt{-\det(g_{7,E}+\hat F)}}
\,{\rho^{{7\over 2}}\over (\rho^3+L^2)^{{5\over 2}}}\,\,,
\eeq
then, the kappa symmetry matrix is:
\beq
\Gamma_{\kappa}\,=\,\Lambda\Big[1+{\cal F}\,\Gamma_{\gamma\phi}\,\Gamma_{11}\Big]
\Big[\Gamma_{\rho}\,+\,{(\rho^3+L^2)^{{5\over 6}}\over \rho^{{1\over 2}}}\,x'\,\Gamma_4\,\Big]\,
\Gamma_{0123}\,\Gamma_{\gamma\phi}\,\,.
\eeq
We have to impose the following  condition on the Killing spinor $\epsilon$:
\beq
\Gamma_{\kappa}\,\epsilon\,=\,\epsilon\,\,.
\eeq
Given the form of  the matrix $\Gamma_{\kappa}$ and the spinor $\epsilon$, it is straightforward to conclude that the above condition is equivalent to the following one on the constant spinor $\eta$ of (\ref{Killing_spinor_Higgs}):
\beq
\tilde \Gamma_{\kappa}\,\eta\,=\,\eta\,\,,
\label{Gamma_kappa_eta_Higgs}
\eeq
where $\tilde \Gamma_{\kappa}$ is defined as:
\beq
\tilde \Gamma_{\kappa}\,=\,e^{-{\gamma\over 2}\,\Gamma_{\rho\gamma}}\,\Gamma_{\kappa}\,
e^{{\gamma\over 2}\,\Gamma_{\rho\gamma}}\,\,.
\eeq
To compute $\tilde \Gamma_{\kappa}$ we use that:
\bear
&&
[\,\Gamma_{\rho\gamma}\,,\,\Gamma_{\rho}\,\Gamma_{\gamma\phi}\,\Gamma_{0123}\,]\,=\,
[\,\Gamma_{\rho\gamma}\,,\,\Gamma_{11}\,\Gamma_{4}\,\Gamma_{0123}\,]\,=\,0\,\,,\rc\rc
&&
\{\,\Gamma_{\rho\gamma}\,,\,\Gamma_{4}\,\Gamma_{\gamma\phi}\,\Gamma_{0123}\,\}\,=\,
\{\,\Gamma_{\rho\gamma}\,,\,\Gamma_{11}\,\Gamma_{\rho}\,\Gamma_{0123}\,\}\,=\,0\,\,.
\eear
Then:
\bear
&&\tilde\Gamma_{\kappa}\,=\,\Lambda\,
\Bigg[\Gamma_{\rho}\,-\,{\cal F}\,{(\rho^3+L^2)^{{5\over 6}}\over \rho^{{1\over 2}}}\,x\,'\,\Gamma_4\,
\Gamma_{\gamma\phi}\,\Gamma_{11}+\qquad\qquad\rc\rc
&&\qquad\qquad\qquad
+e^{-\gamma \Gamma_{\rho\gamma}}\Big({\cal F}\,\Gamma_{\gamma\phi}\,\Gamma_{11}\Gamma_{\rho}+
{(\rho^3+L^2)^{{5\over 6}}\over \rho^{{1\over 2}}}\,x'\,\Gamma_4\Big)\Bigg]\,
\Gamma_{0123}\,\Gamma_{\gamma\phi}\,\,.
\eear
Let us now impose the following extra projection  to $\eta$:
\beq
\Gamma_{0123}\,\Gamma_{\gamma\phi}\,\eta\,=\,\eta\,\,,
\eeq
which combined with the two conditions  (\ref{projections_eta_Higgs}) gives rise to:
\beq
\Gamma_{\gamma\phi}\,\Gamma_{11}\,\eta\,=\,\Gamma_4\,\eta\,\,.
\eeq
Using these projections, we can write the action of $\tilde\Gamma_{\kappa}$ on $\eta$ as:
\beq
\tilde\Gamma_{\kappa}\,\eta\,=\,\Lambda\,
\Bigg[1\,-\,{\cal F}\,{(\rho^3+L^2)^{{5\over 6}}\over \rho^{{1\over 2}}}\,x\,'\,+\,
\Bigg({\cal F}\,+\,{(\rho^3+L^2)^{{5\over 6}}\over \rho^{{1\over 2}}}\,x\,'\Bigg)
e^{-\gamma \Gamma_{\rho\gamma}}\Gamma_4\Bigg]\,\eta\,\,.
\label{Higgs_Gamma_kappa_eta}
\eeq
In order to satisfy (\ref{Gamma_kappa_eta_Higgs}), $\tilde\Gamma_{\kappa}$ should acts as the unity on $\eta$. Thus, the terms on the right-hand side of (\ref{Higgs_Gamma_kappa_eta}) containing matrices should vanish, which happens when the following BPS condition holds:
\beq
x\,'\,=\,-{\cal F}\,{\rho^{{1\over 2}}\over (\rho^3+L^2)^{{5\over 6}}}\,\,,
\eeq
which, after using (\ref{cal_F_Higgs}),  can be shown to be identical to (\ref{BPS_x_Higgs}). We can now evaluate 
$\tilde\Gamma_{\kappa}\,\eta$ on the BPS configuration:
\beq
\tilde\Gamma_{\kappa}\,\eta\big|_{BPS}\,=\,\Lambda_{BPS}\,(1+{\cal F}^2)\,\eta\big|_{BPS}\,\,.
\eeq
As:
\beq
\sqrt{-\det(g_{7,E}+\hat F)}\Big|_{BPS}\,=\,
\Big({3\over 2}\Big)^5\,\Omega^7\,
{\rho^{{7\over 2}}\,
\sin\gamma\over (\rho^3+L^2)^{{5\over 2}}}
\big(1+{\cal F}^2\big)\,\,,
\eeq
we have:
\beq
\Lambda_{BPS}\,=\,{1\over 1+{\cal F}^2}\,\,,
\eeq
and, indeed, one has:
\beq
\tilde\Gamma_{\kappa}\,\eta\big|_{BPS}\,=\,\eta\big|_{BPS}\,\,.
\eeq

\subsection{Fluctuations}

Let us now study the fluctuation spectrum of the D6-brane probe around the fluxless configuration with $y$ constant and $f=x'=0$. The detailed analysis of these configurations is carried out in appendix \ref{Fluct_D6} and summarized in this subsection. First of all, we notice that the unperturbed worldvolume metric in the UV is $AdS_5\times {\mathbb S}^2$, with an angular warping factor.  In our perturbed configuration the worldvolume gauge field is non-zero and we have two scalar fields $y$ and $\eta$ such that:
 \beq
 \delta y\,=\,\eta\,\,,
 \qquad\qquad
 \delta x^4\,=\,\chi\,\,.
 \eeq
 Therefore the scalar $\eta$ ($\chi$) represents the deformation of the D6-branes along the directions transverse (parallel) to the D4-branes of the background. Each of the independent modes is dual to a tower of operators 
whose  conformal dimensions have been determined and have the generic form:
 \beq
\Delta\,=\,\Delta_0\,+\,\alpha\,+{3n\over 2}\,+\,3k\,\,,
\eeq
where $\Delta_0$ is the dimension of the lowest-lying operator in the tower,   $n$ and $k$ are non-negative integers and, depending on the particular mode,  the number $\alpha$ can take the values $\alpha=0,1,2$ (see appendix 
\ref{Fluct_D6}  for details).

  The field theory dual to the D6-brane that we are fluctuating is a defect 4d CFT preserving ${\cal N}=1$ supersymmetry.  The fields of this theory are  the restriction of the rank $N$ $E_{N_f+1}$ theory on the D4-branes down to the (3+1)d defect,  in addition to $N_{D6}$ hypermultiplets. It is not difficult to match the different fluctuations with composite operators of the theory (see \cite{Arean:2006pk} for a similar analysis in the D3-D5 system). Let us denote by $(q\,,\,\tilde{q})$ the scalars of the quark hypermultiplet and by $\psi$ the fermionic components of this hypermultiplet. Let us consider first the fluctuation mode with the lowest dimension, which is the mode denoted by $I_+$ in subsection \ref{D6_fluct_I+modes} for $\alpha\,=\,k\,=\,n\,=\,0$. This mode has dimension $\Delta=2$ and is naturally identified with the operator $q^{\dagger}q-\tilde{q}^{\dagger}\tilde{q}$. On the other hand, the lowest-lying   vector mode (denoted by type II and type III in appendix \ref{Fluct_D6}) has dimension 
 $\Delta=3$ and is naturally identified with the vector current 
$\bar\psi\gamma_{\mu}\psi+q^{\dagger}D_{\mu}q-\tilde{q}^{\dagger}D_{\mu}\tilde{q}$.  Notice that these dimensions agree with the canonical ones since, in 4d,  a scalar field has dimension 1 and a spinor has dimension $3/2$. 
Moreover, these operators precisely correspond to the bosonic content of the $\mathcal{N}=1$ conserved flavor current multiplet.  In both towers of states, as it can be read-off from the form of the corresponding warped harmonics, the $\alpha+\frac{3n}{2}+3k$  contribution to $\Delta$ correspond to extra insertions of $\mathcal{O}_{n,\,k}^{\alpha} =\phi^{\alpha} \, A_1^n\,\mathcal{P}_k(|A_1|^2,\,\phi^3)$, where  $\mathcal{P}_k(x,\,y)$ is a degree $k$ homogeneous polynomial on its variables. The field $A_1\sim X^1+i X^2$ is a complex scalar of the 5d theory and 
$\phi\sim x^9\sim z^{{2\over 3}}$ is the real scalar of the 5d vector multiplet. The canonical dimension of $A_1$ (a scalar field in 5d) is $3/2$ and that of $\phi$ (a scalar in the 5d vector multiplet) is $1$, which matches with the previous expression of the operators $\mathcal{O}_{n,\,k}^{\alpha}$.  The precise expression of these homogeneous polynomials for the different modes can be found in appendix \ref{warped_harmonics}. 
  
The lowest-lying fluctuation mode of the transverse scalar  $\eta$  (studied in subsection \ref{D6_fluct_Smodes}) is naturally identified with an operator of the type $\bar{\psi}\psi+q^{\dagger}\phi q$. The dimension  for  this mode is
$\Delta=5/2$, which differs from the canonical value $\Delta=3$ for this type of bilinear operators.  However, these 
fluctuations correspond to non-holomorphic unprotected quantities in $\mathcal{N}=1$ for which holography gives a prediction. The same thing happens with the type $I_-$ modes of subsection \ref{D6_fluct_I-modes}.  Summarizing, we have the rough fluctuation/operator dictionary in this case:
 \begin{equation}
\begin{array}{c |  c ccc}
 {\rm Fluctuation} & \Delta &&& {\rm Dual\,\,operator}\\ \hline
 I_+ & 2+\alpha+\frac{3n}{2}+3k &&& \Big(q^{\dagger}q-\tilde{q}^{\dagger}\tilde{q}\Big)\,\mathcal{O}_{n,\,k}^{\alpha}  \\ 
 {\rm Vector} & 3+\alpha+\frac{3n}{2}+3k &&& \Big(\bar\psi\gamma_{\mu}\psi+q^{\dagger}D_{\mu}q-\tilde{q}^{\dagger}D_{\mu}\tilde{q}\Big)\,\mathcal{O}_{n,\,k}^{\alpha} \\ 
 {\rm Scalar} & \frac{5}{2}+\alpha+\frac{3n}{2}+3k &&& \Big(\bar{\psi}\psi+q^{\dagger}\phi q\Big)\,\mathcal{O}_{n,\,k}^{\alpha} 
 \end{array}
 \end{equation}

\section{Summary and conclusions}\label{conclusions}
  
  In this paper we have studied 5d QFT's by coupling them to lower-dimensional defects. On general grounds, such defects host, in addition to the restriction to the defect worldvolume of the ambient 5d fields, extra degrees of freedom in the fundamental representation of the gauge group, all combining into interesting defect QFT's. In particular, we have concentrated on a specific class of ambient 5d QFT's; namely high-rank generalizations of the $E_{N_f+1}$ theories of \cite{Seiberg:1996bd}. Besides exhibiting very interesting features, including $E_{N_f+1}$ global symmetries, these theories have a fully explicit gravity dual in type I' String Theory \cite{Brandhuber:1999np} which we have used to study holographically certain classes of defects. 

Technically, defects are constructed as probe branes in the (warped) $AdS_6$ background dual to the ambient QFT's. In this paper, in particular, we have concentrated on codimension 1 and codimension 2 defects, corresponding in the gravity dual respectively to probe D6- and D4-branes.

Starting with D4-branes, they give rise to codimension 2 defects, \textit{i.e.} 3d defect QFT's. We have found an infinite family of embeddings for the probe brane characterized by a holomorphic curve which encodes information about the vacuum of the defect QFT. Among all profiles, two cases, corresponding to the trivial vacuum and to the Higgs branch, are of special relevance, as the induced worldvolume metric on the defect brane approaches an $AdS_4\times \mathbb{S}^1$ with a radial-dependent warping. Interestingly, these defects can not be brought on top of the ``color branes" and must be separated a distance $a$. In turn, such distance governs the dynamics of the probe brane, effectively providing a hard-wall for the induced worldvolume metric. It would be very interesting to explore in more detail the features of such hard-wall, and their implications for the dynamics of the 3d defect QFT, including the precise operator-fluctuation dictionary which also seems to involve in a non-trivial way this scale $a$. This may be related to the fact that gauge fields in $AdS_4$ admit both quantizations  \cite{Witten:2003ya,Marolf:2006nd}, each leading to a different boundary theory where the gauge field is dual to either a gauge symmetry or to a global symmetry. Note also that the scalar $Y$ has dimension $\frac{d}{2}=\frac{3}{2}$, right at the boundary of the range for the two allowed quantizations. We leave for future work the study of the interplay between these particularities of $AdS_4$, the non-conformality (built in in the overall $\varrho$-dependent scaling of the metric) and the IR scale $a$.

We have also considered a different arrangement of probe D4-branes, this time creating codimension 4 defects in the ambient 5d QFT. Such defects host an effective $AdS_2$ worldvolume, and provide a generalization to the antisymmetric Wilson loop configurations in \cite{Assel:2012nf}.

We have also considered codimension 1 defects, engineered by probe D6-branes. In this case, in the asymptotic region, the worldvolume metric of the probe D6's is $AdS_5\times \mathbb{S}^2$. Moreover, upon setting the separation between the probe D6 and the ``color branes" to zero (\textit{i.e.} setting the mass of the 4d quarks to zero), the full worlvolume of the probe branes becomes $AdS_5$, thus suggesting that the defect QFT is indeed a 4d CFT (at least within the supergravity approximation). Note that such CFT would be very interesting, as it would inherit the $E_{N_f+1}$ global symmetry of the ambient 5d CFT in addition to the extra $U(F)$ associated to the $F$ extra D6-branes (for very small $F$, of course). It would be very interesting to further study the properties of this ``exotic" 5d CFT.

More generically, it would be very interesting to consider detailed tests of holography for our defect QFT's, including observables such as the Wilson loop for heavy quarks, entanglement entropy or the partition function itself as in \cite{Robinson:2017sup}, as well as a detailed analysis of the meson spectrum. It would also be very interesting to consider the more general backgrounds of  \cite{DHoker:2016ujz,DHoker:2017mds,DHoker:2017zwj} and study the corresponding defects, possibly allowing to construct a landscape of interesting 3d and 4d defect theories. One may for instance imagine taking an alternative approach to ours in this paper, and, rather than scanning for the possible defects in a fixed background, focus on some ``universal" defects present in a whole class. This approach may be also lead to interesting results in 4d and 6d theories upon using the backgrounds in \textit{e.g.}  \cite{Aharony:2012tz,Apruzzi:2013yva}.  We leave for future work these avenues, as well as the further and more detailed studies of the lower-dimensional theories hosted by the defects.

\vspace{2cm}
{\bf \large Acknowledgments}
We are grateful to C. N\'u\~nez for useful conversations. J.~M.~P. and A.~V.~R.   are funded by the Spanish grant FPA2017-84436-P, by Xunta de Galicia (GRC2013-024), by FEDER and by the Maria de Maeztu Unit of Excellence MDM-2016-0692.   J.~M.~P. is supported by the Spanish FPU fellowship FPU14/06300. D.R-G is partially supported by the Spanish government grant MINECO-16-FPA2015-63667-P, as well as by the Principado de Asturias through the grant FC-GRUPIN-IDI/2018/000174.

\appendix

\vskip 3cm
\renewcommand{\theequation}{\rm{A}.\arabic{equation}}
\setcounter{equation}{0}
%\medskip

\section{Killing spinors}\label{Killing}

In this appendix we review and adapt to our notation the determination of the Killing spinors for the BO background \cite{Passias:2012vp}.  Let us work in the $AdS_6$ coordinates of (\ref{10d_metric_string}) and (\ref{F_4_expression}). It is quite convenient to deal with the metric in the Einstein frame ($ds^2_{Ein}=e^{-{\Phi\over 2}}\,ds^2_{str}$), which is given by:
\beq
ds^2_{10\,,\,E}\,=\,K^2(\alpha)\Bigg[
{9\over 4}\,ds^2_{AdS_6}\,+\,d\alpha^2\,+\,{\sin^2\alpha\over 4}\,
d\Omega_3^2\Bigg]\,\,,
\eeq
where $K(\alpha)$ is the function written in (\ref{10d_metric_string}). Let us parameterize $d\Omega_3^2$ in terms of three left-invariant $SU(2)$ 1-forms $\omega^1$, $\omega^2$ and  $\omega^3$ satisfying 
$d\omega^i\,=\,-{1\over 2}\,\epsilon^{ijk}\,\omega^j\wedge \omega^k$. We have:
\beq
d\Omega_3^2\,=\,{1\over 4}\,\Big[(\omega^1)^2\,+\,(\omega^2)^2\,+\,(\omega^3)^2\Big]\,\,.
\label{Omega_3_omegas}
\eeq
The $\omega^i$'s can be represented in terms of three angles $(\varphi, \theta, \psi)$ as:
\beq
\omega^1=\cos\psi\,d\theta+\sin\psi\,\sin\theta\,d\varphi\,\,,\qquad
\omega^2=-\sin\psi\,d\theta+\cos\psi\,\sin\theta\,d\varphi\,\,,\qquad
\omega^3\,=\,d\psi+\cos\theta\,d\varphi\,\,.
\label{SU2_forms}
\eeq
In order to study the supersymmetry variations of the dilatino and gravitino, 
let us adopt the following basis of frame 1-forms:
\bear
&&e^{\mu}\,=\,{3\over 2}\,K(\alpha)\,r\,dx^{\mu}\,\,,
\qquad\qquad
(\mu=0,1,2,3,4)\,\,,\rc\rc
&&e^5\,=\,{3\over 2}\,K(\alpha)\,{dr\over r}\,\,,
\qquad\qquad
e^6\,=\,K(\alpha)\,d\alpha\,\,,\rc\rc
&&e^{6+a}\,=\,{1\over 2}\,\sin\alpha\,K(\alpha)\,\omega^a\,\,,
\qquad\qquad
(a=1,2,3)\,\,.
\label{frame}
\eear

The SUSY variations of the dilatino $\lambda$ and gravitino $\psi_M$ are:
\bear
&&\delta\,\lambda\,=\,\Big[-\,{1\over 2}\,\partial_M\,\Phi\,\Gamma^{M}\,-\,{5\over 4}\,m\,e^{{5\Phi\over 4}}\,-\,
{1\over 192}\,e^{{\Phi\over 4}}F^{(4)}_{MNPQ}\,\Gamma^{MNPQ}\,\Big]\epsilon\,\,,\rc\rc
&&\delta\psi_M\,=\,\Big[\nabla_{M}\,-\,{m\over 16}\,e^{{5\Phi\over 4}}\,\Gamma_M\,+\,
{e^{{\Phi\over 4}}\over 256}\,F^{(4)}_{NPQR}\big(\Gamma_{M}^{\,\,NPQR}\,-\,{20\over 3}\,\delta_M^{\,\,N}\,
\Gamma^{PQR}\,\big)\,\Big]\,\epsilon\,\,,
\label{SUSY_var}
\eear
where the Romans' mass $m$ for the BO background has been written  in (\ref{m_Q}) and the dilaton $\Phi$ and 4-form $F^{(4)}$ are those displayed in (\ref{10d_metric_string}) and (\ref{F_4_expression}) respectively. 
Let us first impose that the variation of the dilatino is zero. Using that:
\beq
m\,e^{{5\Phi\over 4}}\,=\,{1\over 3\,\cos\alpha\,K(\alpha)}\,\,,
\qquad\qquad
e^{{\Phi\over 4}}F_4\,=\,-{10\over 3}\,\big[K(\alpha)\big]^{-1}\,e^{6789}\,\,,
\eeq
we readily get that $\delta\lambda\,=\,0$ is equivalent to the following condition satisfied by the spinor $\epsilon$:
\beq
\Big[1\,+\,\sin\alpha\,\Gamma^6\,-\,\cos\alpha\,\Gamma^{6789}\,\big]\,\epsilon\,=\,0\,\,.
\label{dilatino_projection}
\eeq
We can rewrite this equation as:
\beq
\Gamma^{6789}\,e^{\alpha\,\Gamma^{789}}\,\epsilon\,=\,\epsilon\,\,.
\eeq
Taking into account that $\{\Gamma^{6789}\,,\,\Gamma^{789}\}\,=\,0$, we can solve this equation as:
\beq
\epsilon\,=\,e^{-{\alpha\over 2}\,\Gamma^{789}}\,\hat\eta\,\,,
\eeq
where the spinor $\eta$ satisfies the  following $\alpha$-independent projection:
\beq
\Gamma^{6789}\,\hat\eta\,=\,\hat\eta\,\,.
\label{rotated_spinor}
\eeq

Let us now impose the SUSY preserving conditions $\delta \psi_M=0$ for the gravitino. We need to compute the covariant derivative acting on spinors, which depends on the spin-connection. In the frame (\ref{frame})  the non-vanishing components of the spin-connection 1-form are:
\bear
&&\omega^{\mu\,5}\,=\,{2\over 3 K(\alpha)}\,e^{\mu}\,\,,
\qquad\qquad
(\mu\,=\,0,1,2,3,4)\,\,,\rc\rc
&&\omega^{p\,6}\,=\,-{1\over  20 K(\alpha)}\,\Phi'\,e^{p}\,\,,
\qquad\qquad
(p\,=\,0,1,2,3,4,5)\,\,,\rc\rc
&&\omega^{6\,n}\,=\,-{1\over 20\,K(\alpha)}\,(20\cot\alpha\,-\,\Phi')\,e^n\,\,,
\qquad\qquad
(n\,=\,7,8,9)\,\,,\rc\rc
&&\omega^{78}\,=\,-{\csc \alpha\over K(\alpha)}\,e^9\,\,,
\qquad
\omega^{89}\,=\,-{\csc \alpha\over K(\alpha)}\,e^7\,\,,
\qquad
\omega^{97}\,=\,-{\csc \alpha\over K(\alpha)}\,e^8\,\,,
\eear
where:
\beq
\Phi'\,=\,\partial_{\alpha}\Phi\,=\,{5\over6}\,\tan\alpha\,\,.
\eeq
When the index $M$ in the second equation in (\ref{SUSY_var}) is one of the $AdS_6$ directions, we find the following equations for $\epsilon$:
\bear
&&\Big({2\over r}\,\partial_{x^{\mu}}\,+\,\Gamma_{\mu\,5}\,-\,\Gamma_{\mu\,6789}\Big)\,\epsilon\,=\,0\,\,,
\qquad\qquad (\mu=0,1,2,3,4)\,\,,
\rc\rc
&&(2r\,\partial_r\,-\,\Gamma_{56789})\,\epsilon\,=\,0\,\,.
\label{gravitino_AdS}
\eear
If we take $M$ to be the polar angle $\alpha$, the gravitino equation becomes:
\beq
\Big(3\,\partial_{\alpha}\,-\,{1\over 16\,\cos\alpha}\,\Gamma^6\,+\,{25\over 16}\,\Gamma^{789}\,\Big)\,\epsilon\,=\,0\,\,.
\label{gravitino_alpha}
\eeq
Let us solve  this last equation  to determine the complete $\alpha$ dependence of $\epsilon$. We first write $\epsilon$ as in (\ref{rotated_spinor}), with  $\hat\eta$ being:
\beq
\hat\eta\,=\,f(\alpha)\,\eta\,\,,
\eeq
where $f(\alpha)$ is a function  proportional to the unit matrix to be determined. Plugging our ansatz into (\ref{gravitino_alpha}), we get:
\beq
\Big(\cos\alpha\,\Gamma^{789}\,-\,\Gamma^6\,+\,48\,\cos\alpha\,{f'\over f}\,\Big)\,\epsilon\,=\,0\,\,,
\eeq
where $f'=\partial_{\alpha} f$.  Taking into account that $\epsilon$ must satisfy (\ref{dilatino_projection}), this last equation reduces to:
\beq
\Big(\sin\alpha\,+\,48\,\cos\alpha\,{f'\over f}\,\Big)\,\Gamma^6\,\epsilon\,=\,0\,\,.
\label{f_Gamma6}
\eeq
As the eigenvalues of $\Gamma^6$ are $\pm 1$, $\Gamma^6$ can never annihilate $\epsilon$. Therefore, the only way to satisfy (\ref{f_Gamma6}) is by requiring that:
\beq
{f'\over f}\,=\,-{1\over 48}\,{\sin\alpha\over \cos\alpha}\,\,.
\eeq
This equation can be readily integrated, yielding:
\beq
f(\alpha)\,=\,\big[\cos\alpha\big]^{{1\over 48}}\,\,.
\eeq
Therefore, the Killing spinors are of the form:
\beq
\epsilon\,=\,\big[\cos\alpha\big]^{{1\over 48}}\,e^{-{\alpha\over 2}\,\Gamma^{789}}\,\eta\,\,,
\eeq
where $\eta$ is independent of $\alpha$ and should satisfy the projection (\ref{dilatino_projection}) (with $\hat\eta\to\eta$). 

Let us now look at (\ref{gravitino_AdS}). These equations can be easily recast in terms of $\eta$. Actually, if we define:
\beq
\Gamma_{*}\,\equiv\,\Gamma^{56789}\,\,,
\eeq
then, the system (\ref{gravitino_AdS}) becomes:
\bear
&&\Big(\partial_{\mu}\,+\,{r\over 2}\,\Gamma_{\mu}\,\Gamma_{5}\,\big(1-\Gamma_{*})\Big)\eta\,=\,0\,\,,\rc\rc
&&\Big(\partial_r\,-\,{1\over 2r}\,\Gamma_{*}\,\Big)\eta\,=\,0\,\,.
\label{spinors_AdS}
\eear
To solve this system we take into account that $\Gamma_*^2=1$ and consider constant spinors of the two posible eigenvalues of $\Gamma_*$:
\beq
\Gamma_{*}\,\eta_{\pm}\,=\,\pm\eta_{\pm}\,\,.
\eeq
There are two classes of solutions of (\ref{spinors_AdS}). The first type are the so-called ordinary spinors, which are independent of the cartesian coordinates $x^{\mu}$ and given by:
\beq
\eta_1\,=\,\sqrt{r}\,\eta_+\,\,,
\eeq
with $\eta_+$ being an arbitrary constant spinor with positive $\Gamma_*$ eigenvalue. The second class of solutions are the so-called conformal spinors, which do  depend on $x^{\mu}$ and are of the form:
\beq
\eta_2\,=\,\Big({1\over \sqrt{r}}\,\Gamma_5\,\Gamma_*\,+\,\sqrt{r}\,x^{\mu}\,\Gamma_{\mu}\,\Big)\,\eta_-\,\,,
\eeq
with $\eta_-$ being constant and satisfying $\Gamma_*\eta_-=-\eta_-$. Notice that the conformal spinors $\eta_2$ do not have well-defined 
$\Gamma_*$ eigenvalue since $\{\Gamma_*, \Gamma_{\mu}\}=0$.

One can check that the equations for the gravitino along the other angular directions are satisfied provided the spinors $\eta$ do not depend on the coordinates $(\theta, \varphi, \psi)$ used to parametrize the $SU(2)$ one-forms $\omega^1$, $\omega^2$ and $\omega^3$.

In our analysis of probes we will mostly deal with the ordinary spinors. Let us summarize its form:
\beq
\epsilon\,=\,\big[\cos\alpha\big]^{{1\over 48}}\,e^{-{\alpha\over 2}\,\Gamma^{789}}\,\eta\,\,,
\qquad\qquad
\Gamma^{6789}\,\eta\,=\,\eta\,\,,
\qquad
\Gamma^5\,\eta\,=\,\eta\,\,.
\label{ordinary_spinors}
\eeq
These two projections, in particular, imply that:
\beq
\Gamma^{01234}\,\Gamma_{11}\,\epsilon\,=\,\epsilon\,\,,
\eeq
with $\Gamma_{11}\,=\,\Gamma_{01234}\,\Gamma_{56789}$. 

In our studies of kappa symmetry for different probe branes we need to know the form of the ordinary Killing spinors in several coordinate systems and 1-form bases.  The form of the ordinary $\epsilon$'s in these coordinates can be easily obtained by following similar steps as those shown here. For example, in the cartesian coordinate system (\ref{10dmetric_cartesian})-(\ref{C5_cartesian}), the $\epsilon$'s can be written as:
\beq
\epsilon\,=\,\big(\vec X^{\,2}\,+\,z^2\big)^{{5\over 32}}\,z^{{1\over 48}}\,\eta\,\,,
\label{ordinary_spinors_cartesian}
\eeq
where $\eta$ are  constant spinors satisfying the projections:
\beq
\Gamma_{z}\,\,\eta\,=\,\eta\,\,,
\qquad\qquad
\Gamma_{11}\,\Gamma_{x^0\,x^1\,x^2\,x^3\,x^4}\,\eta\,=\,\eta\,\,,
\eeq
with the $\Gamma$'s being constant Dirac matrices for the different directions along the cartesian coordinates.  The representation (\ref{ordinary_spinors_cartesian}) is used in section \ref{kappa_D4_defects}. Similarly, in our analysis
of the SUSY D6-brane configurations we need the form of the spinors in the coordinate system (\ref{10d_metric_Einstein_for_D6}). This form has been written in section \ref{kappa_D6_section}.

\renewcommand{\theequation}{\rm{B}.\arabic{equation}}
\setcounter{equation}{0}
%\medskip

\section{Fluctuations of D4-brane defects}
\label{D4_defects_fluct}

In this appendix we will analyze the fluctuations of D4-brane probes around the static configuration studied in section 
\ref{D4_defects}. For simplicity we will restrict ourselves to the case in which the unperturbed probe lies at the origin 
of the $(x^3, x^4)$  and $(X^3, X^4)$ coordinates. Thus, at zero order  the embedding functions of the D4-brane probe are:
\beq
(X^3, X^4,z)^{(0)}\,=\,(0,0,a)\,\,,
\qquad\qquad
(x^3, x^4)^{(0)}\,=\,(0,0)\,\,.
\label{zero_order_emb_D4}
\eeq
Let us denote by $Y^i$ and $U^i$ the deviations from the values (\ref{zero_order_emb_D4}):
\beq
(X^3, X^4,z)\,=\,(Y^1,Y^2,a+Y^3)\,\,,
\qquad\qquad
(x^3, x^4)\,=\,(U^1, U^2)\,\,.
\label{D4_fluct_scalars}
\eeq
To find the equations that govern the $Y$ and $U$ fields we have to expand the probe action up to quadratic order in the fluctuations. To do so we first choose the following set of worldvolume coordinates:
\beq
\zeta^{a}\,=\,(x^0, x^1, x^2, \varrho, \vartheta)\,\,,
\eeq
where $\vartheta$   and $\varrho$  have been defined in (\ref{sigma_vartheta_def}) and (\ref{sigma_varrho}) respectively.  The induced metric in these coordinates can be obtained from (\ref{induced_metric_D4defects_BPS}) by taking $L=W=0$ and is given by:
\beq
ds^2_{5}\,=\,{Q^{{1\over 2}}\over (3ma)^{{1\over 3}}}\,\varrho^{{1\over 2}}
\Bigg[{9\over 4}\,\varrho^2\,dx^2_{1,2}\,+\,
{9\over 4}\,{d\varrho^2\over \varrho^2\big(1-{a^2\over \varrho^3}\big)}+
\Big(1-{a^2\over \varrho^3}\Big)\,d\vartheta^2\Bigg]\,\,.
\label{induced_metric_D4defects}
\eeq
Notice that $\varrho\ge a^{{2\over 3}}$ in (\ref{induced_metric_D4defects}). This minimal value of the holographic coordinate should be regarded as an IR scale of the problem. 

We will also excite in the probe its wordlvolume gauge field $A_a$. It can be readily shown that the equations of motion of the fluctuations can be derived from the following quadratic action:
\bear
&&S\,=\,\int d^5\zeta\Bigg[-{1\over 2}
\,{Q^{{1\over 2}}\over (3ma)^{{1\over 3}}}\,
\varrho^{-{1\over 2}}\,{\cal G}^{ab}\,\partial_a\,Y^i\,\partial_b\,Y^i\,-\,
{3\over 4}\,{Q^{{1\over 2}}\over (3ma)^{{1\over 3}}}\varrho^{{9\over 2}}
{\cal G}^{ab}\,\partial_a\,U^i\,\partial_b\,U^i\,+\rc\rc
&&\qquad\qquad\qquad\qquad
+\,\varrho^5\,\epsilon^{ij}\,
\partial_{\varrho} U^i\,\partial_{\vartheta} U^j\,-\,
{1\over 4}\,\varrho^2\,{\cal G}^{ab}\,{\cal G}^{cd}\,F_{ac}\,F_{bd}\,\Bigg]\,\,,
\label{D4_defect_fluct_action}
\eear
where $\epsilon^{12}=-\epsilon^{21}=1$,  $F_{ab}=\partial_a A_b-\partial_b A_a$ and ${\cal G}^{ab}$ is the inverse of the metric (\ref{induced_metric_D4defects}).  Notice that the $U^i$'s are coupled among themselves, due to the WZ term of the probe D4-brane.  The equations of motion derived from (\ref{D4_defect_fluct_action}) are:
\bear
&&\partial_a\,\Big[\varrho^{-{1\over 2}}\,{\cal G}^{ab}\,\partial_b\,Y^i\Big]\,=\,0\,\,,\rc\rc
&&\partial_a\,\Big[\varrho^{{9\over 2}}\,{\cal G}^{ab}\,\partial_b\,U^i\Big]\,-\,{10\over 3}\,\,
{(3ma)^{{1\over 3}}\over Q^{{1\over 2}}}\varrho^4\,\epsilon^{ij}\,\partial_{\vartheta}\,U^j\,=\,0\,\,,\rc\rc
&&\partial_a\,\Big[\varrho^{2}\,{\cal G}^{ab}\,{\cal G}^{cd}\,F_{bd}\Big]\,=\,0\,\,.
\label{eoms_D4_defect_fluct}
\eear
In the next subsections we will find solutions for the three types of fluctuations. 

\subsection{$Y$-fluctuations}

Let us write the equation in (\ref{eoms_D4_defect_fluct})  for $Y^i$ for the following ansatz in separated variables:
\beq
Y^i\,=\,y^i(\varrho)\,e^{i n\vartheta}\,\,,
\eeq
where $n\in {\mathbb Z}$. The equation for the radial function $y^i$ becomes:
\beq
\partial_{\varrho}\Big[\varrho\,\Big(1-{a^2\over \varrho^3}\Big)\,\partial_\varrho y^i\Big]\,-\,{9\over 4}\,
{n^2\over \varrho\,\Big(1-{a^2\over \varrho^3}\Big)}\,y^i\,=\,0\,\,.
\label{Y_fluct_eq}
\eeq
It turns out that this equation can be solved analytically in the form:
\beq
y^i(\varrho)\,=\,c_1^i\,\varrho^{{3n\over 2}}\Big(1-{a^2\over \varrho^3} \Big)^{{n\over 2}}\,+\,c_2^i\,
\varrho^{-{3n\over 2}}\,
\Big(1-{a^2\over \varrho^3} \Big)^{-{n\over 2}}\,\,,
\label{Y_fluct_sol}
\eeq
where $c_1^i$ and $c_2^i$ are constants. These fluctuations should be dual to operators in the defect theory whose dimensions can be obtained from the UV behavior of the fluctuation. Clearly, when $\varrho\to\infty$ the fluctuation behaves as:
\beq
y^i(\varrho)\,\sim\,c_1^i\,\varrho^{{3n\over 2}}\,+\,c_2^i\,\varrho^{-{3n\over 2}}\,\,,
\qquad\qquad (\varrho\to\infty)\,\,.
\label{UV_Y_fluct}
\eeq
As $\varrho$ is the canonical radial variable of the warped $AdS_4\times {\mathbb S}^1$ UV metric, the dimension $\Delta_Y$ can be easily extracted from the two possible UV behaviors displayed in (\ref{UV_Y_fluct}). One gets:
\beq
\Delta_Y\,=\,{3\over 2}\,+\,{3n\over 2}\,\,.
\label{Delta_Y}
\eeq

\subsection{$U$-fluctuations}
 For the $U^i$-fluctuations we adopt the ansatz:
\beq
U^i\,=\,w^i(\rho)\,e^{i\,n\,\vartheta}\,\,.
\eeq
The corresponding equations for the $w^1(\rho)$ and $w^2(\rho)$ are coupled and given by:
\beq
\partial_{\varrho}\,\Big[\varrho^6\,\Big(1-{a^2\over \varrho^3}\Big)\,\partial_\varrho w^i\Big]\,-\,
{9\over 4}\,n^2\,{\varrho^4\over 1-{a^2\over \varrho^3}}\,w^i\,-\,{15\,i\,n\over 2}\,\varrho^4\,\epsilon^{ij}\,w^j\,=\,0\,\,.
\label{wi_coupled_eqs}
\eeq
These equations can be easily decoupled by constructing the complex combination $w(\varrho)$ as:
\beq
w(\varrho)\,=\,w^1(\varrho)\,+\,i\,w^2(\varrho)\,\,.
\eeq
Indeed, the equation for $w(\varrho)$  that follows from (\ref{wi_coupled_eqs}) is:
\beq
\partial_{\varrho}\,\Big[\varrho^6\,\Big(1-{a^2\over \varrho^3}\Big)\,\partial_\varrho w\Big]\,-\,
{9\over 4}\,n^2\,{\varrho^4\over 1-{a^2\over \varrho^3}}\,w\,-\,{15\,n\over 2}\,\varrho^4\,w\,=\,0 \ .
\eeq
This equation can be solved in terms of a hypergeometric function as:
\bear
&&w(\varrho)\,=\,c_1\, \varrho^{{3n\over 2}}\,
\Big(1-{a^2\over \varrho^3}\Big)^{{n\over 2}}\,+\,\rc\rc
&&\qquad
+{c_2\over \varrho^{2+{3n\over 2}}}\,\Big(1-{a^2 \over \varrho^3 }\Big)^{{n\over 2}}
\Bigg[ F\Big({2\over 3}+n, n; {5\over 3}+n; {a^2\over \varrho^3}\Big)-
F\Big({2\over 3}+n, n+1; {5\over 3}+n; {a^2\over \varrho^3}\Big)
\Bigg]\,\,.\qquad\qquad
\eear
By looking at the two different types of behaviors at the UV of $w$, one gets:
\beq
w\,\sim\,\,c_1\,\varrho^{{3n\over 2}}\,+\,c_2\,\varrho^{-5-{3n\over 2}}\,\,,
\qquad\qquad (\varrho\to\infty)\,\,.
\eeq
It follows that the dimensions of the corresponding dual operators are:
\beq
\Delta_U\,=\,4\,+\,{3n\over 2}\,\,.
\label{Delta_U}
\eeq

\subsection{Vector fluctuations}

We finally look at the fluctuations of the worldvolume gauge field $A_a$, which we will refer to as $V_{\mu}$-fluctuations. The corresponding  Maxwell equation has been written in (\ref{eoms_D4_defect_fluct}) and are be solved by:
\beq
A_{\mu}\,=\,a_{\mu}(\varrho)\,e^{i n\vartheta}\,\,,
\qquad\qquad
A_{\varrho}\,=\,A_{\vartheta}\,=\,0\,\,,
\eeq
provided the radial function $a_{\mu}(\varrho)$ satisfies the differential equation:
\beq
\partial_{\varrho}\Big[\varrho\,\Big(1-{a^2\over \varrho^3}\Big)\,\partial_\varrho a_{\mu}\Big]\,-\,{9\over 4}\,
{n^2\over \varrho\,\Big(1-{a^2\over \varrho^3}\Big)}\,a_{\mu}\,=\,0\,\,.
\eeq
This is exactly the same equation as the one of the $Y$ scalars in (\ref{Y_fluct_eq}). Therefore, it follows that  $a_{\mu}(\varrho)$  can be solved as in (\ref{Y_fluct_sol}) and, therefore, the dimensions of the dual operators are:
\beq
\Delta_V\,=\,{3\over 2}\,+\,{3n\over 2}\,\,.
\label{Delta_V}
\eeq

\renewcommand{\theequation}{\rm{C}.\arabic{equation}}
\setcounter{equation}{0}
%\medskip

\section{Fluctuations of wrapped D4-branes}
\label{fluc_warpped_D4}

We now study fluctuations around the configurations of wrapped D4-branes at a constant angle $\alpha=\alpha_n$.
The DBI lagrangian density of the D4-brane is:
\beq
{\cal L}_{DBI}\,=\,-T_4\,e^{-\Phi}\,\sqrt{-\det(\hat g+F)}\,\,,
\eeq
where the $\hat g$ is the   string frame metric induced on the worldvolume.  In what follows it is quite convenient to put the dilaton inside the square root and to define an effective metric $G$ and a modified worldvolume gauge field ${\cal F}$ by:
\beq
G\,=\,e^{-{2\Phi\over 5}}\,g\,\,,
\qquad\qquad
{\cal F}\,=\,e^{-{2\Phi\over 5}}\,F\,\,,
\eeq
in terms of which ${\cal L}_{DBI}$ takes the form:
\beq
{\cal L}_{DBI}\,=\,-T_4\,\sqrt{-\det(\hat G+{\cal F})}\,\,.
\eeq
As:
\beq
e^{-{2\Phi\over 5}}\,=\,Q^{{1\over 10}}\,\big(3m\,\cos\alpha\big)^{{1\over 3}}\,=\,
{\big[K(\alpha)\big]^{8}\over Q^{{12\over 5}}}\,\,,
\eeq
the 10d effective metric is:
\beq
e^{-{2\Phi\over 5}}\,ds^2\,=\,Q^{{3\over 5}}\,
\big[{9\over 4}\,ds^2_{AdS_6}\,+\,d\Omega^2_4\Big]\,\,,
\eeq
and its pullback to the worldvolume becomes:
\beq
Q^{-{3\over 5}}\,\hat G_{\mu\nu}\,d\zeta^{\mu}\,d\zeta^{\nu}\,=\,
{9\over 4}\,ds^2_{AdS_2}\,+\,\sin^2\alpha\,d\Omega^2_3\,+\,\partial_{\mu}\alpha\,\partial_{\nu}\alpha\,
d\zeta^{\mu}\,d\zeta^{\nu}\,\,,
\eeq
where
\beq
ds^2_{AdS_2}\,=\,-r^2\,(dx^0)^2\,+\,{dr^2\over r^2}\,\,.
\eeq
Moreover, ${\cal F}$ is related to the original worldvolume gauge field $F$ as:
\beq
{\cal F}_{\mu\nu}\,=\,
Q^{{1\over 10}}\,\big(3m\,\cos\alpha\big)^{{1\over 3}}\,
F_{\mu\nu}\,\,.
\eeq
Then, the  non-vanishing components of the unperturbed  induced effective metric for the $\alpha=\alpha_n$ constant angle configuration are:
\beq
\bar G_{00}\,=\,-{9\over 4}\,Q^{{3\over 5}}\,r^2\,\,,
\qquad\qquad
\bar G_{rr}\,=\,{9\over 4}\,Q^{{3\over 5}}{1\over r^2}\,\,,
\qquad\qquad
\bar G_{ij}\,=\,Q^{{3\over 5}}\,\sin^2\alpha_n\,\tilde g_{ij}\,\,,
\eeq
where $\tilde g_{ij}$ is the metric on ${\mathbb S}^3$. Moreover,  the only non-zero component of the  unperturbed gauge field $ {\cal F}$ is:
\beq
\bar {\cal F}_{0r}\,=\,-{9\over 4}\,\sigma\,Q^{{3\over 5}}\,\cos\alpha_n\,\,.
\eeq
We now perturb the embedding angle and the gauge field as:
\beq
\alpha\,=\,\alpha_n\,+\,\xi\,\,,
\qquad\qquad
{\cal F}\,=\,\bar {\cal F}\,+\,\delta {\cal F}\,\,.
\eeq
The effective induced metric $G$ changes as:
\beq
\hat G\,=\,\bar G\,+\,\delta G\,\,.
\eeq
Let us write the different components of $\delta G$. 
When the indices $\mu,\nu$ are not both on ${\mathbb S}^3$, we have:
\beq
\delta G_{\mu\nu}\,=\,Q^{{3\over 5}}\,\partial_{\mu}\xi\,\partial_{\nu}\xi\,\,.
\eeq
If both indices  are on ${\mathbb S}^3$, we have:
\beq
\delta G_{ij}\,=\,Q^{{3\over 5}}\Big[
2\,\sin\alpha_n\,\cos\alpha_n\, \xi\,\tilde g_{ij}\,+\,
(\cos^2\alpha_n\,-\,\sin^2\alpha_n)\,\xi^2\,\tilde g_{ij}\,+\,
\partial_i\xi\,\partial_j\xi\Big]\\,.
\eeq
Let us assume that the worldvolume gauge field $F$ fluctuates as:
\beq
F_{\mu\nu}\,=\,\bar F_{\mu\nu}\,+\,f_{\mu\nu}\,\,,
\eeq
where the only non-vanishing components of the unperturbed gauge field $F$ are:
\beq
\bar F_{0r}\,=\,-{9\over 4}\,\sigma\,{Q^{{1\over 2}}\over (3m)^{{1\over 3}}}
\big(\cos\alpha_n\big)^{{2\over 3}}\,\,.
\eeq
It is now straightforward to relate $\delta {\cal F}$  to $f$ and $\xi$, namely:

\beq
\delta {\cal F}_{\mu\nu}\approx Q^{{1\over 10}}\,\big(3m\big)^{{1\over 3}}
\Big[(\cos\alpha_n)^{{1\over 3}}\,f_{\mu\nu}-
{\sin\alpha_n\over 3\,(\cos\alpha_n)^{{2\over 3}}}\,\bar F_{\mu\nu}\,\xi-
{\sin\alpha_n\over 3\,(\cos\alpha_n)^{{2\over 3}}}\,\xi\,f_{\mu\nu}-
{2+\cos^2\alpha_n\over 18\,(\cos\alpha_n)^{{5\over 3}}}\,\bar F_{\mu\nu}\,\xi^2
\Big]\,\,.\qquad
\eeq
When the indices  $(\mu,\nu)\not= (0,r), (r,0)$,  this expression gives:
\beq
\delta {\cal F}_{\mu\nu}\,=\,
Q^{{1\over 10}}\,\big(3m\big)^{{1\over 3}}
\Big[(\cos\alpha_n)^{{1\over 3}}\,f_{\mu\nu}\,-\,{\sin\alpha_n\over 3\,(\cos\alpha_n)^{{2\over 3}}}\,\xi\,f_{\mu\nu}\Big]\,\,,
\eeq
whereas $\delta {\cal F}_{0r}=-\delta {\cal F}_{r0}$ is:
\bear
&&\delta {\cal F}_{0r}\,=\,
{3\over 4}\,\sigma\,Q^{{3\over 5}}\,\sin\alpha_n\,\xi\,+\,
Q^{{1\over 10}}\,\big(3m\big)^{{1\over 3}}\,(\cos\alpha_n)^{{1\over 3}}\,f_{0r}\,+\rc\rc
&&\qquad\qquad\qquad\qquad
+{Q^{{3\over 5}}\over 8}\,\sigma\,{2+\cos^2\alpha_n\over \cos\alpha_n}\,\xi^2\,-\,
{1\over 3}\,Q^{{1\over 10}}\,\big(3m\big)^{{1\over 3}}\,
{\sin\alpha_n\over (\cos\alpha_n)^{{2\over 3}}}\,\xi\,f_{0r}\,\,.
\eear

The Born-Infeld  determinant can be written as:
\beq
\sqrt{-\det(\hat G+{\cal F})}\,=\,\sqrt{-\det \big(\,\bar G \,+\,{
\bar {\cal F}}\,\big)}\,
\sqrt{\det(1+X)}\,\,,
\label{detX}
\eeq
where the matrix $X$ is given by:
\beq
X\,\equiv\,\big(\,\bar G\,+\,{\bar {\cal F}}\,\big)^{-1}\,\,
\big(\,\delta G\,+\,\delta {\cal F}\,\big)\,\,.
\label{X_def}
\eeq
The unperturbed prefactor in (\ref{detX}) is given by:
\beq
\sqrt{-\det \big(\,\bar G \,+\,{
\bar {\cal F}}\,\big)}\,=\, {9\over 4}\,Q^{{3\over 2}}\,
\big(\sin\alpha_n\big)^4\,\sqrt{\tilde g}\,\, .
\eeq
Therefore,  the DBI lagrangian density is:
\beq
{\cal L}_{DBI}\,=\,-{9\over 4}\,T_4\,Q^{{3\over 2}}\, 
\big(\sin\alpha_n\big)^4\,\sqrt{\tilde g}\,
\sqrt{\det(1+X)}\,\,.
\label{L_DBI_X}
\eeq
To evaluate  the right-hand side of eq. (\ref{L_DBI_X}), we shall use 
the expansion:
\beq
\sqrt{\det(1+X)}\,=\,1\,+\,{1\over 2}\,\Tr X\,-\,{1\over 4}\,\Tr X^2\,+\,
{1\over 8}\,\big(\Tr X\big)^2\,+\,\mathcal{O}(X^3)\,\,.
\label{expansion}
\eeq
In the inverse  matrix $\big(\,\bar G\,+\,{ \bar {\cal F}}\,\big)^{-1}$ 
appearing in (\ref{X_def}) we will separate the symmetric and antisymmetric part:
\beq
\big(\,\bar G\,+\,{\bar {\cal F}}\,\big)^{-1}\,=\,{\cal G}^{-1}\,+\,
{\cal J}\,\,,
\label{openmetric}
\eeq
where ${\cal J}$ is the antisymmetric component. The symmetric matrix ${\cal G}$ is the so-called open string metric, and is the one that naturally shows up in  the fluctuations of the worldvolume when gauge fields are turned on. Notice that the matrix $X$ is the sum of four terms:
\beq
X\,=\,{\cal G}^{-1}\,\delta G\,+\,{\cal G}^{-1}\, \delta {\cal F}\,+\,
{\cal J}\,\delta G\,+\,{\cal J}\,\delta {\cal F}\,\,.
\eeq

The matrix $\bar G\,+\,{\bar {\cal F}}$ has a block structure. By computing  the inverse in the $0r$ sector, one gets:
\beq
\big(\,\bar G\,+\,{\bar {\cal F}}\,\big)^{-1}_{0r}\,=\,{4\over 9\,Q^{{3\over 5}}\,\sin^2\alpha_n}\,\,
\left(  \begin{array}{cc}-{1\over r^2}
&-\sigma\cos\alpha_n\\\\
\sigma\cos\alpha_n&r^2
\end{array}\right)\,\,.
\eeq
Then, the non-zero components of the antisymmetric tensor ${\cal J}$  are:
\beq
{\cal J}^{0r}\,=\,-{\cal J}^{r0}\,=\,-{4\,\sigma\over 9\,Q^{{3\over 5}}}\,{\cos\alpha_n\over \sin^2\alpha_n}\,\,,
\eeq
and the non-vanishing components of the open string metric are:
\beq
{\cal G}^{00}\,=\,-{4\over 9\,Q^{{3\over 5}}\,\sin^2\alpha_n}\,{1\over r^2}\,\,,
\qquad
{\cal G}^{rr}\,=\,{4\over 9\,Q^{{3\over 5}}\,\sin^2\alpha_n}\,r^2\,\,,
\qquad
{\cal G}^{ij}\,=\,{\tilde g^{ij}\over Q^{{3\over 5}}\,\sin^2\alpha_n}\,\,.
\eeq
Thus, the open string metric simply becomes:
\beq
{\cal G}_{\alpha\beta}\,d\zeta^{\alpha}\,d\zeta^{\beta}\,=\,
 Q^{{3\over 5}}\,\sin^2\alpha_n\Big[\,{9\over 4}\,ds^2_{AdS_2}\,
 +\,d\Omega_3^2\Big]\,\,.
 \eeq
Let us now evaluate the different values of  traces of the matrix $X$ up to second order in the fluctuations. First of all one can verify that  $\Tr X$ is given by:
\bear
&&\Tr X={20\over 3}\,\cot\alpha_n\,\xi\,+\,{8\,\sigma\over 9}\,
{(3m)^{{1\over 3}}\over \sqrt{Q}}\,
{(\cos\alpha_n)^{{4\over 3}}\over \sin^2\alpha_n}\,f_{0r}\,+\,
{5\over 9}\,{11\cos^2\alpha_n-5\over \sin^2\alpha_n}\,\xi^2\,-\,\rc\rc
&&\qquad\qquad\qquad\qquad
-{8\,\sigma\,(3m)^{{1\over 3}}\over 27 \sqrt{Q}}\,
{(\cos\alpha_n)^{{1\over 3}}\over \sin\alpha_n}\,\xi\, f_{0r}\,+\,
Q^{{3\over 5}}\,{\cal G}^{\alpha\beta}\,\partial_{\alpha}\xi\,\partial_{\beta}\xi\,\,,
\eear
whereas  $\Tr X^2$ is:
\bear
&&\Tr X^2={2\over 9}\,{55\cos^2\alpha_n+1\over \sin^2\alpha_n}\,\xi^2\,+\,
{32\over 81}\,{(3m)^{{2\over 3}}\over Q}\, 
{(\cos\alpha_n)^{{8\over 3}}\over \sin^4\alpha_n}\,f_{0r}^2\,+\,
\qquad\qquad\qquad\qquad
\rc\rc
&&\qquad
+{16\,\sigma\over 27}\,{(3m)^{{1\over 3}}\over \sqrt{Q}}\,
{(\cos^2\alpha_n+1)(\cos\alpha_n)^{{1\over 3}}\over \sin^3\alpha_n}\,
\xi\,f_{0r}\,-
(3m)^{{2\over 3}}\,Q^{{1\over 5}}\,(\cos\alpha_n)^{{2\over 3}}\,
{\cal G}^{\alpha\beta}\,{\cal G}^{\mu\nu}\,
f_{\alpha\mu}\,f_{\beta\nu}\,\,.
\qquad\qquad
\eear
Using these results in the expansion (\ref{expansion}) we get:
\bear
&&\sqrt{\det(1+X)}\,=\,1+{10\over 3}\,\cot\alpha_n\,\xi\,+\,
{4\sigma\over 9}\,{(3m)^{{1\over 3}}\over \sqrt{Q}}\,
{(\cos\alpha_n)^{{4\over 3}}\over \sin^2\alpha_n}\,f_{0r}+
{1\over 9}{50\cos^2\alpha_n-13\over \sin^2\alpha_n}\xi^2\,+\,\rc\rc
&&\qquad\qquad
+{8\sigma\over 27}\,{(3m)^{{1\over 3}}\over \sqrt{Q}}\,
{(\cos\alpha_n)^{{1\over 3}}\over \sin^3\alpha_n}
\big(5\cos^2\alpha_n-1\big)\,\xi\,f_{0r}\,+\,
{1\over 2}\,Q^{{3\over 5}}\,{\cal G}^{\alpha\beta}
\partial_{\alpha}\xi\partial_{\beta}\xi\,+\,\rc\rc
&&\qquad\qquad\qquad\qquad
+{(3m)^{{2\over 3}}\over 4}
\,Q^{{1\over 5}}\,(\cos\alpha_n)^{{2\over 3}}\,
{\cal G}^{\alpha\beta}\,{\cal G}^{\mu\nu}\,
f_{\alpha\mu}\,f_{\beta\nu}\,\,.
\eear
It is now straightforward to substitute into (\ref{L_DBI_X}) and get the DBI lagrangian density:
\bear
&&{\cal L}_{DBI}=T_4\,Q^{{3\over 2}}\,\sqrt{\tilde g}\Bigg[
-{15\over 2}\,(\sin\alpha_n)^3\cos\alpha_n\,\xi-
{1\over 4}\,(\sin\alpha_n)^2\,(50\cos^2\alpha_n-13)\,\xi^2\,-\,
\qquad\qquad\qquad\qquad
\rc\rc
&&\qquad\qquad
-{2\sigma\over 3}\,{(3m)^{{1\over 3}}\over \sqrt{Q}}\,\sin\alpha_n 
(\cos\alpha_n)^{{1\over 3}}(5\cos^2\alpha_n-1)\,\xi f_{0r}-
{9\over 8} Q^{{3\over 5}}(\sin\alpha_n)^4\,{\cal G}^{\alpha\beta}
\partial_{\alpha}\xi\partial_{\beta}\xi-\rc\rc
&&\qquad\qquad
-{9\over 16}\,Q^{{1\over 5}}\,(3m)^{{2\over 3}}
(\sin\alpha_n)^4\,(\cos\alpha_n)^{{2\over 3}}\,
{\cal G}^{\alpha\beta}\,{\cal G}^{\mu\nu}\,
f_{\alpha\mu}\,f_{\beta\nu}\Bigg]\,\,,
\label{cal_L_DBI}
\eear
where we have descarted the constant term and the term linear in $f_{0r}$, which do not contribute to the equations of motion.

Let  us now compute the WZ term of the action. The RR 3-form potential is given by:
\beq
C_3\,=\,Q\,(3m)^{{1\over 3}}\,C(\alpha)\,{\rm Vol}\big({\mathbb S}^3\big)\,\,,
\eeq
where $C(\alpha)$ is the function written in (\ref{C_alpha}). The WZ lagrangian density is:
\beq
{\cal L}_{WZ}\,=\,\sigma\,T_4\,\,Q\,(3m)^{{1\over 3}}\,\sqrt{\tilde g}\,C(\alpha)\,
F_{0r}\,\,.
\eeq
Let us expand the function $C(\alpha)$ and  the gauge field $F_{0r}$ up to second-order in the fluctuations:
\beq
C(\alpha_n+\xi)\,=\,C(\alpha_n)\,+\,C'(\alpha_n)\xi\,+\,
{1\over 2}\,C''(\alpha_n)\xi^2\,\,,
\qquad\qquad
F_{0r}\,=\,\bar F_{0r}\,+\,f_{0r}\,\,.
\eeq
Neglecting the constant term and the term linear in $f_{0r}$, which do not contribute to the equations of motion, we have at second-order in the fluctuations:
\beq
{\cal L}_{WZ}\,=\,\sigma\,T_4\,\,Q\,(3m)^{{1\over 3}}\,\sqrt{\tilde g}\,
\Big[\,C'(\alpha_n)\bar  F_{0r}\,\xi\,+\,C'(\alpha_n)\xi\,f_{0r}\,+\,
{1\over 2}\,C''(\alpha_n)\,\bar  F_{0r}\,\xi^2\Big]\,\,.
\eeq
The derivatives of the $C$ function appearing in ${\cal L}_{WZ}$ are:
\beq
C'(\alpha_n)=-{10\over 3}(\cos\alpha_n)^{{1\over 3}}\,\sin^3\alpha_n\,\,,
\qquad
C''(\alpha_n)=-{10\over 9}
{(\sin\alpha_n)^2\over (\cos\alpha_n)^{{2\over 3}}}
\Big[10\cos^2\alpha_n-1\Big]\,\,.
\eeq
Substituting these values we get the final form of the WZ lagrangian density:
\bear
&&{\cal L}_{WZ}\,=\,T_4\,Q^{{3\over 2}}\sqrt{\tilde g}\,
\Bigg[{15\over 2}\,(\sin\alpha_n)^3\,\cos\alpha_n\,\xi\,-\,
{10\,\sigma\over 3\sqrt{Q}}\,(3m)^{{1\over 3}}\,(\cos\alpha_n)^{{1\over 3}}\,
(\sin\alpha_n)^3\,\,\xi\,f_{0r}\,+\,\qquad\rc\rc
&&\qquad\qquad\qquad\qquad\qquad\qquad\qquad\qquad
+{5\over 4}\,(\sin\alpha_n)^2\big(10\cos^2\alpha_n\,-\,1\big)\,\xi^2\Bigg]\,\,.
\label{cal_L_WZ}
\eear
We can now add (\ref{cal_L_DBI}) and (\ref{cal_L_WZ}) to get the total lagrangian density for the fluctuations. Adjusting the global constant coefficient, we obtain that the equations of motion for the fluctuations are derived from the following lagrangian density:
\bear
&&{\cal L}=\sqrt{\tilde g}\,(\sin\alpha_n)^4\Bigg[
-{1\over 2}\,Q^{{3\over 5}}{\cal G}^{\alpha\beta}
\partial_{\alpha}\xi\partial_{\beta}\xi+{8\over 9}{\xi^2\over \sin^2\alpha_n}
-{1\over 4}\,Q^{{1\over 5}}\,(3m)^{{2\over 3}}
\,(\cos\alpha_n)^{{2\over 3}}\,
{\cal G}^{\alpha\beta}{\cal G}^{\mu\nu}
f_{\alpha\mu}\,f_{\beta\nu}\,-\,\rc\rc
&&\qquad\qquad\qquad
-{32\,\sigma\over 27}\,{(3m)^{{1\over 3}}\over \sqrt{Q}}\,
(\cos\alpha_n)^{{1\over 3}}\,{\xi\,f_{0r}\over (\sin\alpha_n)^3}\Bigg]\,\,.
\label{Lag_wrapped_D4_fluct}
\eear

Let us now write the equations of motion derived from the lagrangian (\ref{Lag_wrapped_D4_fluct}). The equation for the scalar $\xi$ is:
\beq
{1\over \sqrt{\tilde g}}\,\partial_{\alpha}
\Big[\sqrt{\tilde g}\,{\cal G}^{\alpha\beta}\,\partial_{\beta}\xi\Big]\,+\,
{16\over 9\,Q^{{3\over 5}}\,(\sin\alpha_n)^2}\,\xi\,-\,
{32\,\sigma\,(3m)^{{1\over 3}}\over 27\,Q^{{11\over 10}}}\,
{(\cos\alpha_n)^{{1\over 3}}\over  (\sin\alpha_n)^3}
\,f_{0r}\,=\,0\,\,,
\label{eom_scalar_wrapped_D4}
\eeq
while the equation for the gauge field becomes:
\beq
{1\over \sqrt{\tilde g}}\,\partial_{\alpha}
\Big[\sqrt{\tilde g}\,{\cal G}^{\alpha\beta}\,{\cal G}^{\mu\nu}\,f_{\beta\nu}\Big]\,+\,
{32\,\sigma\over 27\,(3m)^{{1\over 3}}\,Q^{{7\over 10}}}\,
{\delta_{0}^{\mu}\,\partial_{r}\xi\,-\,\delta_{r}^{\mu}\,\partial_{0}\xi
\over (\sin\alpha_n)^3\,(\cos\alpha_n)^{{1\over 3}}}\,=\,0\,\,.
\label{eom_vector_wrapped_D4}
\eeq
In what follows we study different solutions to these equations for the fluctuations. 

\subsection{Type  II modes}
Let $f_{\mu\nu}=\partial_{\mu}a_{\nu}-\partial_{\nu}a_{\mu}$ and consider the following ansatz for the solution of (\ref{eom_scalar_wrapped_D4}) and (\ref{eom_vector_wrapped_D4}):
\beq
\xi\,=\,e^{iE\,t}\,Y_l({\mathbb S}^3)\,\phi_1(r)\,\,,
\qquad
a_0\,=\,e^{iE\,t}\,Y_l({\mathbb S}^3)\,\phi_2(r)\,\,,
\qquad
a_r\,=\,a_i\,=\,0\,\,.
\eeq
where $Y_l({\mathbb S}^3)$ are spherical harmonics on the 3-sphere with angular momentum $l$. For this ansatz, let us now write down the equations of motion for the gauge field. Taking $\mu=i$ in (\ref{eom_vector_wrapped_D4}) we get:
\beq
\partial_0\,a_0\,=\,0\,\,,
\eeq
which means that $E=0$ and, thus, our fluctuations are static. Similarly, putting $\mu=r$ in (\ref{eom_vector_wrapped_D4}) one can verify that this equation is satisfied for $E=0$ (and, therefore, $\partial_0\,\xi=0$).  Taking $\mu=0$ in (\ref{eom_vector_wrapped_D4})  and using that:
\beq
\nabla^{2}_{{\mathbb S}^3}\,Y_l({\mathbb S}^3)\,=\,-l(l+2)\,Y_l({\mathbb S}^3)\,\,,
\qquad
l\,=\,0,1,\cdots\,\,,
\eeq
we get the following coupled equation for $\phi_2$ and $\phi_1$:
\beq
r^2\,\partial_r^2\,\phi_2\,-\,9\,l(l+2)\,\phi_2\,+\,6\,\sigma\,
{Q^{{1\over 2}}\over (3m)^{{1\over 3}}}\,
{\sin\alpha_n\over (\cos\alpha_n)^{{1\over 3}}}\,r^2\,\partial_r\,\phi_1\,=\,0\,\,.
\label{type_II_gauge_field_eq}
\eeq
The equation of motion of the scalar (\ref{eom_scalar_wrapped_D4})  for our ansatz becomes:
\beq
\partial_r\Big(r^2\,\partial_r\,\phi_1\Big)\,-\,
\Big(9\,l(l+2)-4\Big)\,\phi_1\,+\,
{8\sigma\over 3}\,{(3m)^{{1\over 3}}\over \sqrt{Q}}\,
{(\cos\alpha_n)^{{1\over 3}}\over \sin\alpha_n}\,\partial_r\,\phi_2\,=\,0\,\,.
\label{type_II_scalar_eq}
\eeq
Let us define a new function $\eta=\eta(r)$ as:
\beq
\eta\,=\,{(3m)^{{1\over 3}}\,\sigma\over 2\sqrt{Q}}\,
{(\cos\alpha_n)^{{1\over 3}}\over \sin\alpha_n}\,\partial_r\,\phi_2\,+\,3\,\phi_1\,\,.
\eeq
We can rewrite the equation for the scalar (\ref{type_II_scalar_eq}) in terms of $\eta$ as:
\beq
\partial_r\Big(r^2\,\partial_r\,\phi_1\Big)\,-\,
\Big(12+9\,l(l+2)\Big)\,\phi_1\,+\,{16\over 3}\,\eta\,=\,0\,\,,
\label{type_II_phi1_eta}
\eeq
whereas the equation (\ref{type_II_gauge_field_eq}) for the gauge field becomes:
\beq
r^2\,\partial_r\,\eta\,-\,{9\,\sigma\,(3m)^{{1\over 3}}\over 2\sqrt{Q}}\,l(l+2)\,
{(\cos\alpha_n)^{{1\over 3}}\over \sin\alpha_n}\,\phi_2\,=\,0 \ .
\label{type_II_gauge_eta}
\eeq
Let us compute the derivative with respect to $r$ of (\ref{type_II_gauge_eta}) and use in the result the relation between $\partial_r\,\phi_2$ and $\eta$. We get:
\beq
\partial_r\Big(r^2\,\partial_r\,\eta\Big)\,-\,9\,l(l+2)\,\eta\,+\,
27\,l(l+2)\,\phi_1\,=\,0\,\,.
\label{type_II_eta_phi1}
\eeq
Eqs. (\ref{type_II_phi1_eta}) and (\ref{type_II_eta_phi1}) constitute a system of ODEs for $\phi_1$ and $\eta$. Let us rewrite this system in a more convenient form. First of all, we define the operator ${\cal O}$ as the one that acts on any function $\psi$ as:
\beq
{\cal O}\,\,\psi\equiv \partial_r\Big(r^2\,\partial_r\,\psi\Big)\,\,.
\eeq
Moreover, let ${\cal M}$ be the following $2\times 2$ matrix:
\beq
{\cal M}\equiv\,
\begin{pmatrix}
12+9\,l(l+2)&&& -{16\over 3}\rc\rc
-27\,l(l+2)&&&9\,l(l+2)
\end{pmatrix}\,\,.
\eeq
Then, our system of ODEs can be written in matrix form as:
\beq
\Big({\cal O}\,-\,{\cal M}\Big)\,
\begin{pmatrix}
\phi_1\rc
\eta
\end{pmatrix}\,=\,0\,\,.
\eeq
In order to solve this system we first notice that the eigenvalues of ${\cal M}$ are:
\beq
\lambda_1=6+9\,l(l+2)-6\sqrt{1+4l(l+2)}\,\,,
\qquad
\lambda_2=6+9\,l(l+2)+6\sqrt{1+4l(l+2)}\,\,.
\qquad
\label{lambda_1_2}
\eeq
The corresponding eigenfunctions are:
\bear
&&\psi_{\lambda_1}\,=\,\eta\,-\,{9\over 8}
\Big[1-\sqrt{1+4l(l+2)}\Big]\phi_1\,\,,\rc\rc
&&\psi_{\lambda_2}\,=\,\eta\,-\,{9\over 8}\,
\Big[1+\sqrt{1+4l(l+2)}\Big]\phi_1 \ .
\eear
These functions satisfy:
\beq
\partial_r\Big(r^2\,\partial_r\,\psi_{\lambda_i}\Big)\,=\,
\lambda_i\,\psi_{\lambda_i}\,\,.
\label{type_II_ODE_psi_i}
\eeq
The solutions of (\ref{type_II_ODE_psi_i}) are:
\beq
\psi_{\lambda_i}\,=\,c_i\,r^{{1\over 2}\,(\sqrt{1+4\lambda_i}-1)}\,+\,
d_i\,r^{-{1\over 2}\,(\sqrt{1+4\lambda_i}+1)}\,\,,
\label{UV_behavior_psi_i}
\eeq
where $c_i$ and $d_i$ are constants. In general, in $AdS_2$, if a fluctuation behaves in the UV as:
\beq
\psi\sim c\,r^{-2a_1}\,+\,d\,r^{-2a_2}\,\,,
\qquad\qquad
(r\to\infty, a_2>a_1)\,\,,
\eeq
then, the dimension of the operator dual to the normalizable mode is:
\beq
\Delta={1\over 2}\,+\,a_2-a_1\,\,.
\label{Deltas_as}
\eeq
In our case 
\beq
a_1\,=\,-{1\over 4}\,(\sqrt{1+4\lambda_i}-1)\,\,,
\qquad\qquad
a_2\,=\,{1\over 4}\,(\sqrt{1+4\lambda_i}+1)\,\,,
\eeq
and, as a consequence the conformal dimension corresponding to the mode
$\psi_{\lambda_i}$ is:
\beq
\Delta_i\,=\,{1+\sqrt{1+4\,\lambda_i}\over 2}\,\,.
\label{Deltas_type_II_III}
\eeq
For $l>0$ these dimensions are irrational numbers. However, they are integers for 
$l=0$. Indeed, in this $l=0$ case $\lambda_1=0$ and $\lambda_2=12$ and we have:
\beq
\Delta_1\,=\,1\,\,,
\qquad\qquad
\Delta_2\,=\,4\,\,.
\eeq

\subsection{Type  III modes}
We now study a generalization of the type III modes in which the gauge field $a_{\mu}$ has only components along $AdS_2$ and the scalar $\xi$ is excited. The ansatz for $a_0$, $a_r$ and $\xi$ is:
\beq
a_0\,=\,e^{iE\,t}\,Y_l({\mathbb S}^3)\,A_0(r)\,\,,
\qquad
a_r\,=\,e^{iE\,t}\,Y_l({\mathbb S}^3)\,{V(r)\over  r^2}\,\,,
\qquad
\xi\,=\,e^{iE\,t}\,Y_l({\mathbb S}^3)\,\zeta(r)\,\,.
\eeq

The equation of motion (\ref{eom_vector_wrapped_D4}) for $\mu=i$ yields the following relation between the functions $A_0$ and $A_r$:
\beq
E\,A_0\,=\,-i\,r^2\,\partial_r\,V\,\,.
\label{A_0_V}
\eeq
Notice that now we are not forced to take $E=0$ and, therefore, these  fluctuation modes are not static. Moreover, 
eq. (\ref{eom_vector_wrapped_D4})  for $\mu=r$ becomes:
\beq
i\,E\,\partial_r\,A_0\,+\, {E^2\over r^2}\,V\,-\,9\,l(l+2)\,V\,+\,
{6\sigma\,\sqrt{Q}\over (3m)^{{1\over 3}}}\,{\sin\alpha_n\over (\cos\alpha_n)^{{1\over 3}}}\,i\,E\,\zeta\,=\,0\,\,.
\eeq
Eliminating $A_0$ in favor of $V$ by using (\ref{A_0_V}), we get:
\beq
{\cal O}\,V\,-\,9\,l\,(l+2)\,V\,+{6\sigma\,\sqrt{Q}\over (3m)^{{1\over 3}}}\,{\sin\alpha_n\over (\cos\alpha_n)^{{1\over 3}}}\,i\,E\,\zeta\,=\,0\,\,,
\label{type_II_eom_V}
\eeq
where now ${\cal O}$ is the differential operator which acts on any function $\psi$ as:
\beq
{\cal O}\,\psi\,\equiv\, \partial_r\Big(r^2\,\partial_r\,\psi\Big)\,+\,
{E^2\over r^2}\,\psi\,\,.
\eeq
Eq. (\ref{eom_vector_wrapped_D4}) for $\mu=0$, after eliminating $A_0$ by means of (\ref{A_0_V}), is just the derivative of (\ref{type_II_eom_V}). Moreover, the equation for the scalar $\zeta$ can be written as:
\beq
{\cal O}\,\zeta\,-\,{8\sigma\,(3m)^{{1\over 3}}\over 3 \sqrt{Q}}{
(\cos\alpha_n)^{{1\over 3}}\over \sin\alpha_n}\,{i\over E}\,{\cal O}\,V\,+\,
\Big(4-9\,l\,(l+2)\Big)\,\zeta\,=\,0\,\,.
\label{type_II_eom_zeta}
\eeq
Let us write (\ref{type_II_eom_V}) and (\ref{type_II_eom_zeta}) in a more convenient form. With this purpose we define a new function $S$ as:
\beq
S\equiv\,i\,E\,{6\sigma\,\sqrt{Q}\over (3m)^{{1\over 3}}}\,{\sin\alpha_n\over (\cos\alpha_n)^{{1\over 3}}}\,\zeta\,+\,
16\,V\,\,,
\eeq
and  the matrix ${\cal M}$ as:
\beq
{\cal M}\equiv\,
\begin{pmatrix}
9\,l(l+2)-4&&& 16\Big[4\,-\,9\,l(l+2)\big]\rc\rc
-1&&&9\,l(l+2)+16
\end{pmatrix}\,\,.
\eeq
Then, the system of ODEs for these modes can be written as:
\beq
\Big({\cal O}\,-\,{\cal M}\Big)\,
\begin{pmatrix}
S\rc
V
\end{pmatrix}\,=\,0\,\,.
\eeq
The eigenvalues of ${\cal M}$ are exactly the same as in (\ref{lambda_1_2}).  The corresponding eigenfunctions are:
\bear
&&\psi_{\lambda_1}\,=\,S\,-\,2
\Big[5-3\sqrt{1+4l(l+2)}\Big]V\,\,,\rc\rc
&&\psi_{\lambda_2}\,=\,S\,-\,2
\Big[5+3\sqrt{1+4l(l+2)}\Big]V\,\,.
\eear
They satisfy:
\beq
\partial_r\Big(r^2\,\partial_r\,\psi_{\lambda_i}\Big)\,+\,
{E^2\over r^2}\,\psi_{\lambda_i}\,=\,\lambda_i\,\psi_{\lambda_i}\,\,,
\eeq
and  have the same UV asymptotic behavior as in (\ref{UV_behavior_psi_i}). Thus the dimensions of the dual operators is just  given by (\ref{Deltas_type_II_III}). 

\subsection{Type I modes}

We now look at the fluctuation modes in which only the components of the gauge field along the internal 
${\mathbb S}^3$ are excited. When this is the case and we take $\mu=0,r$ in (\ref{eom_vector_wrapped_D4}), we get:
\beq
\partial_{0}\big(\,\nabla^{i}\,a_i\,\big)\,=\,0\,\,,\qquad\qquad
\partial_{r}\big(\,\nabla^{i}\,a_i\,\big)\,=\,0\,\,,
\eeq
where $\nabla^{i}$ is the covariant derivative on the 3-sphere:
\beq
\nabla^{i}\,a_i\,=\,
{1\over \sqrt{\tilde g}}\,\partial_i\,\big(\, \sqrt{\tilde g}\,
\tilde g^{ij}\,a_j\,\big)\,\,.
\eeq
Thus, clearly we should require that:
\beq
\nabla^{i}\,a_i\,=\,0\,\,.
\label{nabla-a}
\eeq
Taking $\mu=i$ in the equation of motion  (\ref{eom_vector_wrapped_D4}) for the gauge field, we get:
\beq
\tilde g^{ij}\,\Big[\partial_r\big(r^2\partial_r\,a_j)\,-\,{1\over r^2}\,\partial_0^2\,a_j\Big]\,+\,
{9\over \sqrt{\tilde g}}
\partial_k\,\big[\sqrt{\tilde g}\,\tilde g^{kl}\,\tilde g^{ij}\,
(\partial_l a_j-\partial_j a_l\,)\,\big]\,=\,0\,\,.
\eeq
The last term in this equation can be rewritten as:
\beq
{1\over \sqrt{\tilde g}}\,\partial_k\,\big[\sqrt{\tilde g}\,\tilde g^{kl}\,\tilde g^{ij}\,
(\partial_l a_j-\partial_j a_l\,)\,\big]\,=\,\nabla_k\nabla^k\,a^{i}\,-\,
R^{i}_{k}\,a^k\,-\,\nabla^{i}\,\nabla_k\,a^k\,\,,
\label{df}
\eeq
where $R^{i}_{k}\,=\,2\,\delta_{k}^{i}$ is the Ricci tensor of the ${\mathbb S}^{3}$-sphere and the $\nabla_k$ are covariant derivatives  on ${\mathbb S}^{3}$. Taking into account (\ref{nabla-a}) the last term in (\ref{df}) vanishes. Moreover, the first two terms in  (\ref{df}) can be written in terms of the Hodge-de Rham operator  $\Delta_1$ for vector fields (or 1-forms) on the ${\mathbb S}^{3}$-sphere, which acts on a vector field with components $f_i$ as:
\beq
\Delta_1\,f_i\,\equiv\, \nabla_k\nabla^k\,f_{i}\,-\,
R^{k}_{i}\,f_k\,\,.
\label{HdR}
\eeq
We get:
\beq
\partial_r\big(r^2\partial_r\,a_i)\,-\,{1\over r^2}\,\partial_0^2\,a_i\,+\,9\,\Delta_1\,a_i\,=\,0\,\,.
\eeq
Let us analyze this fluctuation equation by separating variables in the form:
\beq
a_i\,=\,e^{iEt}\,Y_{i}^{l}({\mathbb S}^{3})\,\phi(r)\,\,,
\label{typeI-sep-var}
\eeq
where $Y_{i}^{l}({\mathbb S}^{3})$ is a vector spherical harmonic on the 3-sphere satisfying:
\beq
\nabla^{i}\,Y_{i}^{l}({\mathbb S}^{3})\,=\,0\,\,.
\eeq
The Hodge-de Rham operator  $\Delta_1$ acts diagonally on the vector spherical harmonics and the eigenvalue depends on the number $l>1$ that characterizes the representation of $SO(4)$. One has:
\beq
\Delta_1\,Y_{i}^{l}({\mathbb S}^{3})\,=\,-(l+1)^2\,Y_{i}^{l}({\mathbb S}^{3})\,\,.
\eeq
The equation resulting for the radial function $\phi$ is:
\beq
\partial_r\big(r^2\partial_r\,\phi)\,+\,{E^2\over r^2}\,\phi\,-\,9\,(l+1)^2\,\phi\,=\,0\,\,.
\eeq
These modes behave asymptotically as:
\beq
\phi(r)\approx c\,r^{-2\,a_1}\,+\,d\,r^{-2\,a_2}\,\,,
\eeq
where 
\beq
a_1\,=\,-{1\over 4}\,\Big(\sqrt{1+36(l+1)^2}\,-\,1\Big)\,\,,
\qquad\qquad
a_2\,=\,{1\over 4}\,\Big(\sqrt{1+36(l+1)^2}\,+\,1\Big) \ .
\eeq
Applying (\ref{Deltas_as}) we get that the associated dimensions are:
\beq
\Delta\,=\,{1\over 2}\,\Big(1\,+\,\Big(\sqrt{1+36(l+1)^2}\Big)\,\,,
\qquad\qquad
(l\ge 1)\,\,.
\eeq
In these modes all $\Delta$'s are irrational. 

\subsection{Fluctuation of the cartesian coordinates}

Let us assume that the cartesian coordinates of the D4-brane embedding are not fixed but fluctuate. Let $\chi$ be such a fluctuation. The induced metric $G_{\mu\nu}$ gets an extra contribution given by:
\beq
\delta G_{\mu\nu}\,=\,{9\over 4}\,Q^{{3\over 5}}\,r^2\,\partial_{\mu}\chi\,\partial_{\nu}\chi\,\,.
\eeq
The corresponding contribution to ${\rm Tr}\,X$ at second-order is:
\beq
{9\over 4}\,Q^{{3\over 5}}\,r^2\,{\cal G}^{\mu\nu}\,\partial_{\mu}\chi\,\partial_{\nu}\chi\,\,,
\eeq
and the lagrangian for $\chi$ at quadratic order is:
\beq
{\cal L}_{\chi}\sim\,\sqrt{\tilde g}\,r^2\,{\cal G}^{\mu\nu}\,\partial_{\mu}\chi\,\partial_{\nu}\chi\,\,.
\eeq
The equations of motion derived from ${\cal L}_{\chi}$ are:
\beq
\partial_{\mu}\,\Big[r^2\,\sqrt{\tilde g}\,{\cal G}^{\mu\nu}\,\partial_{\nu}\chi\Big]\,=\,0\,\,,
\eeq
or equivalently:
\beq
\partial_r\big[r^4\,\partial_r\chi\big]\,-\,\partial_0^2\,\chi\,+\,9\,r^2\,\nabla_{{\mathbb S}^3}^2\,\chi\,=\,0\,\,.
\eeq
Let us separate variables and adopt the following ansatz:
\beq
\chi\,=\,e^{iE\,t}\,Y_l({\mathbb S}^3)\,\eta(r)\,\,.
\eeq
Then, the equation of motion for the radial function $\eta(r)$ is:
\beq
\partial_r\big[r^4\,\partial_r\eta\big]\,+\,E^2\,\eta\,-\,9\,r^2\,l(l+2)\,\eta\,=\,0\,\,.
\eeq
For $r\to\infty$:
\beq
\eta(r)\approx c\,r^{-2\,a_1}\,+\,d\,r^{-2\,a_2}\,\,,
\eeq
where 
\beq
a_1\,=\,-{3\over 4}\,\Big(\sqrt{1+4l\,(l+2)}\,-\,1\Big)\,\,,
\qquad\qquad
a_2\,=\,{3\over 4}\,\Big(\sqrt{1+4l\,(l+2)}\,+\,1\Big)\,\,.
\eeq
The corresponding dimensions, obtained from (\ref{Deltas_as}), are:
\beq
\Delta\,=\,{1\over 2}\,\Big(1+3\,\sqrt{1+4l\,(l+2)}\Big)\,\,,
\qquad\qquad
(l\ge 0)\,\,.
\eeq
This dimension is only integer for $l=0$ (it takes the value $\Delta=2$ in this case).

\renewcommand{\theequation}{\rm{D}.\arabic{equation}}
\setcounter{equation}{0}
%\medskip

\section{Fluctuations of the probe D6-branes}
\label{Fluct_D6}

Let us next study the fluctuations around the configuration with $x_4$ and $y$ constant. Accordingly, we write:
\beq
y\,=\,L\,+\,\eta\,\,,
\qquad\qquad
x^4\,=\,x^4_0\,+\,\chi\,\,.
\eeq
Moreover, we will assume that the D6-brane has a non-trivial worldvolume gauge field $F_{ab}$. The dynamics of these fluctuations is governed by a lagrangian density ${\cal L}$, which can be written in terms of the unperturbed effective metric ${\cal G}_{ab}$ defined in (\ref{7d_metric}).  Let us write its non-zero components:
\bear
&& {\cal G}_{\mu\, \nu}\,=\,
{9\over 4}\,{(\rho^3+L^{\,2})^{{5\over 6}}\over \rho^{{1\over 2}}}\,\,\eta_{\mu\,\nu}
\,\,,\qquad\qquad(\mu, \nu=0,1,2,3)\,\,,\rc\rc
&& {\cal G}_{\rho\, \rho}\,=\,
{9\over 4}\, {\rho^{{1\over 2}}\over (\rho^3+L^{\,2})^{{5\over 6}}}\,\,,\rc\rc
&&{\cal G}_{i\,j}\,=\, {\rho^{{5\over 2}}\over (\rho^3+L^{\,2})^{{5\over 6}}}\,h_{ij}\,\,,
\qquad\qquad\qquad\,(i,j=\gamma, \phi)\,\,.
\label{induced_metric_D6_fluct}
\eear
Notice that, in the UV region with $\rho\to\infty$, we can neglect the constant $L$ in (\ref{induced_metric_D6_fluct}) and the metric ${\cal G}_{ab}$ becomes $AdS_5\times {\mathbb S}^2$.  Recall also that the line element in (\ref{7d_metric}) contains an angular warp factor $(\cos\gamma)^{-{1\over 3}}$.

The equations of motion for the fluctuations  can be derived from the following lagrangian density:
\bear
&&{\cal L}\,=\,-{Q\over (3m)^{{2\over 3}}}\,{\sqrt{h}\over (\cos\gamma)^{{1\over 3}}}\,
\Bigg[{1\over 2}\, {\rho^{{5\over 2}}\over (\rho^3+L^{\,2})^{{5\over 6}}} {\cal G}^{ab}\,
\partial_a\,\eta\,\partial_b \eta\,+\,
{9\over 8}\,\rho^{{5\over 2}}\,(\rho^3+L^{\,2})^{{5\over 6}}\,{\cal G}^{ab}\,
\partial_a\,\chi\partial_b \chi\Bigg]\,+\,\rc\rc\rc
&&\qquad\qquad\qquad
-{1\over 4}\,(\cos\gamma)^{{1\over 3}}\,\sqrt{h}\,\rho^3\,F_{ab}\,F^{ab}\,+\,{5\over 2}\,
{Q^{{1\over 2}}\over (3m)^{{1\over 3}}}\,\rho^2\,(\rho^3+L^{\,2})^{{2\over 3}}\,\chi\,
\epsilon^{ij}\,F_{ij}\,\,,
\label{L_fluct_D6}
\eear
where $i,j$ are indices along the 2-sphere and $\epsilon^{ij}=\pm 1$.

Let us next write the equations of motion for the different fields in (\ref{L_fluct_D6}).
The equation of motion of the transverse scalars is:
\beq
\partial_a\,\Bigg[{\sqrt{h}\over (\cos\gamma)^{{1\over 3}}}\,
 {\rho^{{5\over 2}}\over (\rho^3+L^{\,2})^{{5\over 6}}} {\cal G}^{ab}\,
 \partial_b\,\eta\Bigg]\,=\,0\,\,.
 \label{transverse_scalar_eom}
 \eeq
The scalar $\chi$ is coupled to the worldvolume gauge field. Its equation of motion is:
\beq
\partial_a\,\Bigg[{\sqrt{h}\over (\cos\gamma)^{{1\over 3}}}\,
\rho^{{5\over 2}}\, (\rho^3+L^{\,2})^{{5\over 6}}\,{\cal G}^{ab}\,
\partial_b\,\chi\Bigg]\,+\,{10 \over 9}\,{(3m)^{{1\over 3}}\over Q^{{1\over 2}}}\,
\rho^2\, (\rho^3+L^{\,2})^{{2\over 3}}\,\epsilon^{ij}\,F_{ij}\,=\,0 \ .
 \label{chi_scalar_eom}
 \eeq
Finally, the equation of motion of the the worldvolume gauge field is:
\beq
\partial_a\Big[(\cos\gamma)^{{1\over 3}}\,\sqrt{h}\,\rho^3\,F^{ab}\Big]\,+\,5\,
{Q^{{1\over 2}}\over (3m)^{{1\over 3}}}\,
\rho^2\,(\rho^3+L^{\,2})^{{2\over 3}}\,\epsilon^{bj}\,\partial_j\,\chi\,=\,0\,\,,
\label{F_eom}
\eeq
where the antisymmetric tensor $\epsilon^{bj}=\epsilon_{bj}$ is zero unless $b$ is an index along the 2-sphere. 

\subsection{Scalar fluctuations}
\label{D6_fluct_Smodes}

Let us now solve (\ref{transverse_scalar_eom}) for the transverse scalar fluctuations $\eta$. First of all, we rewrite this equation as:
\beq
{\rho^3\over (\rho^3+L^{\,2})^{{5\over 3}}}\,\partial^{\mu}\partial_{\mu}\eta\,+\,
\partial_{\rho}\Big(\rho^2\,\partial_{\rho}\,\eta\Big)\,+\,
\nabla^2_{-{1\over 3}}\,\eta\,=\,0 \ ,
\eeq
where $\partial^{\mu}\partial_{\mu}$ is the laplacian in the flat Minkowski metric and $\nabla^2_{-{1\over 3}}$ is the  ${\mathcal S}^2$ warped laplacian operator defined in (\ref{harmonics_eom}) for $a=-{1\over 3}$. We will diagonalize these operators and, therefore, we will adopt  an ansatz of the type:
\beq
\eta\,=\,G^{(r)}(\rho)\,e^{i\,p x}\,Z^{(r)}(\gamma, \phi)\
\,\,,
\qquad\qquad
(r=1,2)\,\,,
\label{eta_ansatz}
\eeq
where $Z^{(1)}$  and  $Z^{(2)}$   are the warped harmonics defined in (\ref{Z_J_harmonics}), depending on two integers $n$ and $k$, and the 4-vector $p_{\mu}$ appearing in the plane wave factor 
$e^{i\,p x}$ in  (\ref{eta_ansatz}) is such that:
\beq
M^2\,=\,-p^2\,\,,
\eeq
with $M^2$ being the mass squared of the dual meson. The operator  $\nabla^2_{-{1\over 3}}$ acts on the functions $Z^{(r)}(\gamma, \phi)$  as:
\beq
\nabla^2_{-{1\over 3}}\,Z^{(r)}\,=\,-J_{Z}^{(r)}\Big(J_{Z}^{(r)}+{2\over 3}\Big)\,Z^{(r)}
\,\,,
\qquad\qquad\qquad
(r=1,2)\,\,,
\eeq
where $J_{Z}^{(1)}$ and  $J_{Z}^{(2)}$ are written in (\ref{Z_J_harmonics}). Then, the radial functions $G^{(r)}$ satisfy the equation:
\beq
\partial_{\rho}\Big(\rho^2\,\partial_{\rho}\,G^{(r)}\Big)\,+\,\Bigg[
M^2\,{\rho^3\over (\rho^3+L^{\,2})^{{5\over 3}}}\,-{9\over 4}\,J_{Z}^{(r)}\Big(J_{Z}^{(r)}+{2\over 3}\Big)\Bigg]
G^{(r)}\,=\,0\,\,,
\eeq
whose asymptotic  UV behavior for $\rho\to\infty$ is a superposition of the type:
\beq
G^{(r)}(\rho)\sim c_1\,\rho^{-1-{3\over 2}J_{Z}^{(r)}}\,+\,c_2\,\rho^{{3\over 2}J_{Z}^{(r)}}\,\,,
\eeq
which means that the corresponding dual operators have dimensions:
\beq
\Delta_S^{(r)}\,=\,{5\over 2}\,+\,{3\,J_{Z}^{(r)}\over 2}
\,\,,
\qquad\qquad
(r=1,2)\,\,.
\eeq
More specifically, we have:
\beq
\Delta_S^{(1)}\,=\,{9\over 2}\,+\,{3n\over 2}+2k\,\,,
\qquad\qquad
\Delta_S^{(2)}\,=\,{5\over 2}\,+\,{3n\over 2}+2k\,\,.
\eeq

\subsection{Type I modes}
\label{D6_fluct_Imodes}

We now study the fluctuation modes in which the worldvolume gauge field potential $A$ is involved ($F=dA$). First of all we consider the modes in which  only the components of $A$ along the ${\mathbb S}^2$ are non-zero, \ie\ such that:
\beq
A_{x^{\mu}}\,=\,A_{\rho}\,=\,0\,\,.
\eeq
Let us adopt an ansatz for the  ${\mathbb S}^2$ components of $A$ of the type:
\beq
A_i\,=\,{\cal F}(x,\rho)\, {\cal Z}_i\,(\gamma,\phi)\,\,,
\eeq
where $ {\cal Z}_i\,(\gamma,\phi)$ is a vector function of ${\mathbb S}^2$. It is straightforward to check that the gauge field equations (\ref{F_eom}) for $b=x^{\mu}, r$ are satisfied  for $A_{x^{\mu}}=A_{\rho}=0$  if the following equation for  $ {\cal Z}_i$ holds:
\beq
\partial_i\Big[(\cos\gamma)^{{1\over 3}}\,\sqrt{h}\,h^{ij}\, {\cal Z}_j\Big]\,=\,0\,\,.
\label{div_Z_i}
\eeq
Eq. (\ref{div_Z_i}) is automatically fulfilled if  $ {\cal Z}_i$ is represented in terms of the derivatives of a scalar function  ${\cal Z}$ as:
\beq
 {\cal Z}_i\,=\,{1\over (\cos\gamma)^{{1\over 3}}\,\sqrt{h}}\,h_{ik}\,\epsilon^{kl}\,
\nabla_l\,{\cal Z}\,\,.
\eeq
If follows from this representation that:
\beq
(\cos\gamma)^{{1\over 3}}\,\sqrt{h}\,h^{ij}\, {\cal Z}_j\,=\,\epsilon^{ij}\,\nabla_j\,{\cal Z}\,\,.
\eeq
Moreover, the field strength  $F_{ij}\,=\,\partial_i A_j\,-\,\partial_j A_i$
 on the  ${\mathbb S}^2$  takes the form:
\beq
F_{ij}\,=\,-\epsilon_{ij}\,{\sqrt{h}\over (\cos\gamma)^{{1\over 3}}}\,{\cal F}\,\,
\nabla^2_{-{1\over 3}}{\cal Z}\,\,.
\label{F_ij_type_I}
\eeq
Notice that $F_{ij}$ is coupled to the scalar fluctuation $\chi$ in the lagrangian ${\cal L}$ and, therefore, for consistency we have to include a non-zero $\chi$ is our ansatz. Moreover, from (\ref{F_ij_type_I}) it is clear that, in order to solve (\ref{chi_scalar_eom}) and (\ref{F_eom}) we must diagonalize  the differential operator $\nabla^2_{-{1\over 3}}$. Accordingly, we take ${\cal Z}$ to be one of the warped harmonics $Z^{(r)}$ introduced in 
(\ref{Z_J_harmonics}). We define the vector functions $ Z^{(r)}_{i}(\gamma,\phi)$ for $r=1,2$  as:
\beq
 Z^{(r)}_{i}(\gamma,\phi)\,\equiv\,{1\over (\cos\gamma)^{{1\over 3}}\,\sqrt{h}}\,h_{ik}\,\epsilon^{kl}\,
\nabla_l\, Z^{(r)}(\gamma,\phi)
\,\,,
\qquad\qquad
(r=1,2)\,\,,
\eeq
and we adopt the following ansatz for $A_i$ and $\chi$:
\beq
A_i\,=\,\Sigma^{(r)}(\rho)\,e^{ipx}\, Z^{(r)}_{i}(\gamma,\phi)\,\,,
\qquad\qquad
\chi\,=\,\Lambda^{(r)}(\rho)\,e^{ipx}\, Z^{(r)}(\gamma,\phi)\,\,.
\eeq
It follows from our ansatz that:
\beq
\epsilon^{ij}\,F_{ij}\,=\,2\,J_{Z}^{(r)}\Big(J_{Z}^{(r)}+{2\over 3}\Big)\,
{\sqrt{h}\over (\cos\gamma)^{{1\over 3}}}
\Sigma^{(r)}(\rho)\,e^{ipx}\, Z^{(r)}\,\,.
\eeq
Using these results it is easy to verify that (\ref{F_eom}) becomes:
\bear
&&\partial_{\rho}\,\Big[(\rho^3+L^2)^{{5\over 3}}\partial_{\rho}\Sigma^{(r)}\Big]\,+\,
\Bigg(
M^2\,\rho\,-\,{9\over 4}\,\,J_{Z}^{(r)}\Big(J_{Z}^{(r)}+{2\over 3}\Big)\,
{(\rho^3+L^2)^{{5\over 3}}\over \rho^2}\Bigg)\Sigma^{(r)}\,+\,\rc\rc
&&\qquad\qquad\qquad\qquad\qquad\qquad
+{45\over 4}\,{Q^{{1\over 2}}\over (3m)^{{1\over 3}}}\,\rho^2\,(\rho^3+L^2)^{{2\over 3}}\,
\Lambda^{(r)}\,=\,0\,\,,
\eear
while (\ref{chi_scalar_eom}) is equivalent to:
\bear
&&\partial_{\rho}\,\Big[\rho^2\,(\rho^3+L^2)^{{5\over 3}}\partial_{\rho}\Lambda^{(r)}\Big]\,+\,
\Bigg(
M^2\,\rho^3\,-\,{9\over 4}\,J_{Z}^{(r)}\Big(J_{Z}^{(r)}+{2\over 3}\Big)\,
(\rho^3+L^2)^{{5\over 3}}\,\Bigg)\Lambda^{(r)}\,+\,\rc\rc
&&\qquad\qquad\qquad\qquad
+5\,{ (3m)^{{1\over 3}}\over Q^{{1\over 2}}}\,
J_{Z}^{(r)}\Big(J_{Z}^{(r)}+{2\over 3}\Big)\,\rho^2\,
\,(\rho^3+L^2)^{{2\over 3}}\,\Sigma^{(r)}\,=\,0\,\,.
\eear
Therefore, we have a system of two coupled differential equations. In order to decouple them let us define a new function $V^{(r}(\rho)$ by:
\beq
V^{(r)}(\rho)\,\equiv\,{Q^{{1\over 2}}\over (3m)^{{1\over 3}}}\,
\rho\, \Lambda^{(r)}(\rho)\,\,,
\eeq
and the  second-order differential operator ${\cal O}$ as the one that acts on any function of $F(\rho)$ by:
\beq
{\cal O}\,F\,\equiv\,{4\over 45}\,
\partial_{\rho}\,\Big[(\rho^3+L^2)^{{5\over 3}}\partial_{\rho}F\Big]\,+\,
{4\over 45}\,\Bigg(
M^2\,\rho\,-\,{9\over 4}\,J_{Z}^{(r)}\Big(J_{Z}^{(r)}+{2\over 3}\Big)\,
{(\rho^3+L^2)^{{5\over 3}}\over \rho^2}\,\Bigg)\,F \ .
\eeq
Then, the system of equations becomes simply
\bear
&&{\cal O}\,\Sigma^{(r)}\,+\,\rho\,(\rho^3+L^2)^{{2\over 3}}\,V^{(r)}\,=\,0\,\,,\rc\rc
&&{\cal O}\,V^{(r)}\,\,+\,{4\over 9}\,\rho\,(\rho^3+L^2)^{{2\over 3}}\,\Bigg(J_{Z}^{(r)}\Big(J_{Z}^{(r)}+{2\over 3}\Big)\,
\Sigma^{(r)}\,-\,V^{(r)}\Bigg)\,=\,0 \ .
\eear
To decouple this system we look for combinations of the type $V^{(r)}+\alpha\,\Sigma^{(r)}$, where $\alpha$ is a constant coefficient  that must satisfy the following quadratic equation:
\beq
{9\over 4}\,\alpha^2\,-\,\alpha\,-\,J_{Z}^{(r)}\Big(J_{Z}^{(r)}+{2\over 3}\Big)\,=\,0\,\,,
\eeq
whose solutions are:
\beq
\alpha_+\,=\,{2\over 3}\,\Big(J_{Z}^{(r)}+{2\over 3}\Big)\,\,,
\qquad\qquad
\alpha_-\,=\,-{2\over 3}\,J_{Z}^{(r)}\,\,.
\eeq
We now treat these two cases separately. 

\subsubsection{Type ${\rm I}_+$ modes}
\label{D6_fluct_I+modes}

Let us consider $\alpha=\alpha_+$ and define $\eta_+$ as:
\beq
\eta_+\,=\,V\,+\,{2\over 3}\,\Big(J_{Z}^{(r)}+{2\over 3}\Big)\Sigma\,\,.
\eeq
The equation satisfied by  $\eta_+$ is:
\beq
\partial_{\rho}\,\Big[(\rho^3+L^2)^{{5\over 3}}\partial_{\rho}\eta_+\Big]+
\Bigg[
M^2\,\rho-{9\over 4}J_{Z}^{(r)}\Big(J_{Z}^{(r)}+{2\over 3}\Big)
{(\rho^3+L^2)^{{5\over 3}}\over \rho^2}+{15\over 2}\,J_{Z}^{(r)}\,\rho
(\rho^3+L^2)^{{2\over 3}}\Bigg]\eta_+=0\,\,.
\eeq
At the UV ($\rho\to\infty$) the solutions of this equation behave as:
\beq
\eta_+\sim c_1\,\rho^{-{3\over 2}\,J_{Z}^{(r)}}\,+\,c_2\,
\rho^{{3\over 2}\,J_{Z}^{(r)}-4}\,\,.
\eeq
The corresponding conformal dimensions are:
\beq
\Delta_{I_+}^{(r)}\,=\,2\,+\Big|{3\over 2}\,J_{Z}^{(r)}-2\Big|
\,\,,
\qquad\qquad
(r=1,2)\,\,.
\eeq
More specifically:
\beq
\Delta_{I_+}^{(1)}\,=\,2\,+\,{3n\over 2}\,+\,3k\,\,,
\qquad\qquad
\Delta_{I_+}^{(2)}\,=\,{3n\over 2}\,+\,3k\,\,.
\eeq
(The branch 2 requires that $k\ge 1$ or $n\ge2$). 

\subsubsection{Type ${\rm I}_-$ modes}
\label{D6_fluct_I-modes}

When  $\alpha=\alpha_-$ we define:
\beq
\eta_-\,=\,V\,-\,{2\over 3}\,J_{Z}^{(r)}\,\Sigma\,\,,
\eeq
which obeys the following differential equation:
\beq
\partial_{\rho}\,\Big[(\rho^3+L^2)^{{5\over 3}}\partial_{\rho}\eta_-\Big]+
\Bigg[
M^2\,\rho-{9\over 4}J_{Z}^{(r)}\Big(J_{Z}^{(r)}+{2\over 3}\Big)
{(\rho^3+L^2)^{{5\over 3}}\over \rho^2}-
{15\over 2}\,\Big(J_{Z}^{(r)}+{2\over 3}\Big)\,\rho
(\rho^3+L^2)^{{2\over 3}}\Bigg]\eta_-=0\,\,.
\eeq
At the UV the fluctuation $\eta_-$ behaves as:
\beq
\eta_-\sim c_1\,\rho^{-5-{3\over 2}\,J_{Z}^{(r)}}\,+\,c_2\,
\rho^{1+{3\over 2}\,J_{Z}^{(r)}}\,\,.
\eeq
The associated conformal dimension is
\beq
\Delta_{I_-}^{(r)}\,=\,5\,+{3\over 2}\,J_{Z}^{(r)}
\,\,,
\qquad\qquad
(r=1,2)\,\,.
\eeq
For the two branches, these dimensions are:
\beq
\Delta_{I_-}^{(1)}\,=\,7\,+\,{3n\over 2}\,+\,3k\,\,,
\qquad\qquad
\Delta_{I_-}^{(2)}\,=\,5+{3n\over 2}\,+\,3k\,\,.
\eeq

\subsection{Type II modes}
\label{D6_fluct_IImodes}

For these modes $\chi=0$ and the only non-vanishing components of $A$ are those along the Minkowski directions:
\beq
A_{\mu}\,=\,\Phi_{\mu}(x,\rho)\,{\cal Y}(\gamma, \phi)\,\,,
\qquad\qquad
A_{\rho}=A_i\,=\,0\,\,.
\eeq
The non-trivial components of $F_{ab}$ are:
\beq
F_{\mu\,\nu}\,=\,(\partial_{\mu}\Phi_{\nu}-\partial_{\nu}\Phi_{\mu})\,{\cal Y}\,\,,
\qquad\qquad
F_{\mu\,\rho}\,=\,-\partial_{\rho}\Phi_{\mu}\,{\cal Y}\,\,,
\qquad\qquad
F_{\mu\,i}\,=\,-\Phi_{\mu}\nabla_i{\cal Y}\,\,.
\eeq
The equations of motion (\ref{F_eom}) for $b=\rho, i$ are satisfied if $\Phi_{\mu}$ is transverse, namely when:
\beq
\partial^{\mu}\Phi_{\mu}\,=\,0\,\,.
\label{transv_A_mu}
\eeq
Moreover, when $b$ is a Minkowski direction, we get:
\beq
{\cal Y}\,\partial_{\rho}\big[\rho^3\,\partial^{\mu}\,\Phi_{\mu}\big]\,+\,
{\rho^4\over (\rho^3+L^2)^{{5\over 3}}}\,{\cal Y}\,\partial^{\nu}\partial_{\nu}\,\Phi_{\mu}\,+\,
{9\over 4}\,\rho\,\Phi_{\mu}\,\nabla^2_{{1\over 3}}\,{\cal Y}\,=\,0 \ .
\label{eom_type_II_A_mu}
\eeq
To separate variables in this equation,  we take ${\cal Y}=Y^{(r)}$ for $r=1,2$ and
\beq
\Phi_{\mu}\,=\,\xi_{\mu}\,U^{(r)}(\rho)\,e^{ipx}\,\,,
\eeq
where $\xi_{\mu}$ is a constant vector  transverse to the momentum, \ie\ satisfying
\beq
p^{\mu}\,\xi_{\mu}\,=\,0\,\,,
\eeq
which ensures that (\ref{transv_A_mu}) is fulfilled. 
The total ansatz for  $A_{\mu}$ is thus:
\beq
A_{\mu}\,=\,\xi_{\mu}\,U^{(r)}(\rho)\,e^{ipx}\,Y^{(r)}(\gamma, \phi)\,\,,
\qquad\qquad
(r=1,2)\,\,.
\eeq
For this ansatz (\ref{eom_type_II_A_mu}) becomes:
\beq
\partial_{\rho}\big[\rho^3\,\partial_{\mu}\,U^{(r)}\big]\,+\,
\Bigg[M^2\,{\rho^4\over (\rho^3+L^2)^{{5\over 3}}}\,-\,{9\over 4}\,\rho\,J_Y^{(r)}\Big(J_Y^{(r)}+{4\over 3}\Big)\Bigg]
U^{(r)}\,=\,0.
\eeq
For large $\rho$ the solutions of this equation behave as:
\beq
U^{(r)}\,\sim\,c_1\,\rho^{{3\over 2}J_Y^{(r)}}\,+\,c_2\,
\rho^{-{2-{3\over 2}J_Y^{(r)}}}.
\eeq
The conformal dimensions of these fluctuations are
\beq
\Delta_{II}^{(r)}\,=\,3\,+{3\over 2}\,J_{Y}^{(r)}
\,\,,
\qquad\qquad
(r=1,2)\,\,.
\eeq
Using the values of the $J_{Y}^{(r)}$ for the two branches we get:
\beq
\Delta_{II}^{(1)}\,=\,4\,+\,{3n\over 2}\,+\,3k\,\,,
\qquad\qquad
\Delta_{II}^{(2)}\,=\,3+{3n\over 2}\,+\,3k\,\,.
\eeq

\subsection{Type III modes}
\label{D6_fluct_IIImodes}

We now consider the ansatz:
\beq
A_{\mu}\,=\,0\,\,,
\qquad\qquad
A_{\rho}\,=\,{\cal F}(x,\rho)\,{\cal Y}(\gamma, \phi)\,\,,
\qquad\qquad
A_{i}\,=\,{\cal H}(x,\rho)\,\nabla_i\,{\cal Y}(\gamma, \phi)\,\,.
\eeq
The non-vanishing components of the field strength are:
\beq
F_{\mu i}\,=\,\partial_{\mu}\,{\cal H} \,\nabla_i\,{\cal Y}\,\,,
\qquad\qquad
F_{\mu \rho}\,=\,\partial_{\mu}\,{\cal F} \,{\cal Y}\,\,,
\qquad\qquad
F_{\rho i}\,=\,(\partial_{\mu}\,{\cal H}\,-\,{\cal F})\,\nabla_i\,{\cal Y}\,\,.
\eeq
Notice that $F_{ij}=0$  and, therefore, we can put consistently to zero the scalar fluctuation $\chi$
in (\ref{chi_scalar_eom}). Let us now study (\ref{F_eom}) for different values of the index $b$. 
The equation for $A_{\rho}$ is:
\beq
\rho^3\,\partial^{\mu}\partial_{\mu}\,{\cal F}\,{\cal Y}\,+\,
{9\over 4}\,(\rho^3+L^2)^{{5\over 3}}\,\big(
{\cal F}\,-\,\partial_{\mu}\,{\cal H}\big)\,
\nabla^2_{{1\over 3}}\,{\cal Y}\,=\,0\,\,.
\label{eom_III_rho}
\eeq
The equation for $A_{i}$ is:
\beq
\rho\,\partial^{\mu}\partial_{\mu}\,{\cal H}\,-\,
\partial_{\rho}\Big[(\rho^3+L^2)^{{5\over 3}}\,({\cal F}\,-\,\partial_{\mu}\,{\cal H}\big)\Big]\,=\,0\,\,.
\label{eom_III_i}
\eeq
Finally, the equation for $A_{\mu}$ becomes:
\beq
\partial_{\mu}\Big[
{\cal H}\,\nabla^2_{{1\over 3}}\,{\cal Y}\,+\,{4\over 9}\,
\partial_{\rho}\big(\rho^3\,{\cal F}\big)\,{\cal Y}\Big]\,=\,0\,\,.
\label{eom_III_mu}
\eeq
By inspecting (\ref{eom_III_rho}), (\ref{eom_III_i}) and (\ref{eom_III_mu}) it follows that the separation of variables will require that we take ${\cal Y}(\gamma, \phi)$ to be an eigenfunction of the operator $\nabla^2_{{1\over 3}}$. Therefore,  let us take the functions ${\cal Y(\gamma, \phi)}$ as the harmonic functions $Y^{(r)}(\gamma, \phi)$ for 
$r=1,2$. Moreover, we consider  functions ${\cal F}$ and ${\cal H}$ that depend on $x$ as a plane  wave:
\beq
{\cal F}(x,\rho)\,=\,e^{ipx}\,\Gamma^{(r)}(\rho)\,\,,
\qquad\qquad
{\cal H}(x,\rho)\,=\,e^{ipx}\,G^{(r)}(\rho)\,\,,
\eeq
which means that our ansatz for $A_{\rho}$ and $A_{i}$ is:
\beq
A_{\rho}\,=\,\Gamma^{(r)}(\rho)\,e^{ipx}\,Y^{(r)}(\gamma, \phi)\,\,,
\qquad\qquad
A_{i}\,=\,G^{(r)}(\rho)\,e^{ipx}\,\nabla_i\,Y^{(r)}(\gamma, \phi)\,\,,
\qquad\qquad
(r=1,2)\,\,.
\eeq
Then, (\ref{eom_III_mu}) allows us to write $G^{(r)}$ in terms of $\Gamma^{(r)}$ as:
\beq
G^{(r)}\,=\,{4\over 9\rho}\,{\partial_{\rho}\big(\rho^3\,\Gamma^{(r)}\big)\over 
J_Y^{(r)}(J_Y^{(r)}+{4\over 3})}\,\,.
\label{type_III_G_vs_Gamma}
\eeq
We can now use (\ref{type_III_G_vs_Gamma}) to eliminate $G^{(r)}$ in (\ref{eom_III_rho}) and (\ref{eom_III_i}).  It turns out that these two equations reduce to a single one for $\Gamma^{(r)}$ given by:
\beq
\partial^2_{\rho}\,\Gamma^{(r)}\,+\,{5\over \rho}\,\partial_{\rho}\,\Gamma^{(r)}\,+\,
\Bigg[M^2\,{\rho\over (\rho^3+L^2)^{{5\over 3}}}\,+\,{3\over \rho^2}\,-\,
{9\over 4\rho^2}\,J_Y^{(r)}\Big(J_Y^{(r)}+{4\over 3}\Big)\Bigg]\Gamma^{(r)}\,=\,0\,\,.
\eeq
The UV behavior of the solutions of this equations is of the form:
\beq
\Gamma^{(r)}\,\sim\,c_1\,\rho^{-{1+{3\over 2}J_Y^{(r)}}}\,+\,c_2\,
\rho^{-{3-{3\over 2}J_Y^{(r)}}}.
\eeq
It follows that the conformal dimensions of these fluctuations are
\beq
\Delta_{III}^{(r)}\,=\,3\,+{3\over 2}\,J_{Y}^{(r)}
\,\,,
\qquad\qquad
(r=1,2)\,\,.
\eeq
Therefore:
\beq
\Delta_{III}^{(1)}\,=\,4\,+\,{3n\over 2}\,+\,3k\,\,,
\qquad\qquad
\Delta_{III}^{(2)}\,=\,3+{3n\over 2}\,+\,3k\,\,.
\eeq

\vskip 3cm
\renewcommand{\theequation}{\rm{E}.\arabic{equation}}
\setcounter{equation}{0}
%\medskip

\section{Warped harmonics}
\label{warped_harmonics}

Let us consider a 2d space  with coordinates $(\gamma, \phi)$ and metric:
\beq
ds_2^2\,=\,h_{ij}\, d\zeta^i\,d\zeta^j\,=\,(d\gamma)^2\,+\,\sin^2\gamma\,(d\phi)^2\,\,.
\eeq
The coordinate $\phi$ takes values in $[0, 2\pi]$, whereas $\gamma\in [0,\pi/2]$. We now consider a real scalar field $\lambda$ with lagrangian density:
\beq
{\cal L}\sim (\cos\gamma)^{a}\,\sqrt{h}\,\Big( h^{ij}\,\partial_i\,\lambda\,\partial_j\,\lambda\,+\,{\cal M}^2\,\lambda^2\Big)\,\,,
\label{armonic_action}
\eeq
where $a$ is a real number  ($|a|\le 1$) that  parameterizes the warp factor. It can be understood as originated in some dimensional reduction. The equation of motion derived from (\ref{armonic_action}) is:
\beq
\nabla_{a}^2\,\,\lambda\,+\,{\cal M}^2\,\lambda\,=\,0\,\,,
\eeq
where $\nabla_{a}^2$ is the following laplacian-like operator:
\beq
\nabla_{a}^2\,\lambda\,\equiv\,{1\over (\cos\gamma)^{a}\,\sqrt{h}}\,
\partial_i\,\Big((\cos\gamma)^{a}\,\sqrt{h}\,h^{ij}\,\partial_j\,\lambda\Big)\,\,.
\label{harmonics_eom}
\eeq
Let us parameterize the mass ${\cal M}$ as:
\beq
{\cal M}^2\,=\,J\,(J\,+1\,+a)\,\,,
\eeq
where $J\ge 0$. The equation of motion (\ref{harmonics_eom}) takes the form:
\beq
\partial_{\gamma}^2\,\lambda\,+\,\big(\cot \gamma\,-\,a\,\tan\gamma\big)\,\partial_{\gamma}\,\lambda\,+\,
{\partial_{\phi}^2\,\lambda\over \sin^2\gamma}\,=\,-J(J+1+a)\,\lambda\,\,.
\eeq
Let us separate variables as:
\beq
\lambda(\gamma, \phi)\,=\,e^{i n\phi}\,\Lambda(\gamma)\,\,,
\eeq
where $n\in {\mathbb Z}$ due to the periodicity condition $\lambda(\gamma, \phi+2\pi)\,=\,\lambda(\gamma, \phi)$. The equation satisfied by $\Lambda$ is:
\beq
\partial_{\gamma}^2\,\Lambda\,+\,\big(\cot \gamma\,-\,a\,\tan\gamma\big)\,\partial_{\gamma}\,\Lambda\,-\,
{n^2\over \sin^2\gamma}\,\Lambda\,=\,-J(J+1+a)\,\Lambda\,\,.
\eeq
There are two independent solutions of this differential equation, which we will denote by $\Lambda^{(1)}(\gamma)$ and $\Lambda^{(2)}(\gamma)$, and are given by:
\bear
&&\Lambda^{(1)}(\gamma)\,=\,{(\cos\gamma)^{1-a}\over (\sin\gamma)^{|n|}}\,F\Big(
1-{|n|-J\over 2}\,,\, {1-a-|n|-J\over 2}\,;\,{3-a\over 2}\,;\, \cos^2\gamma\Big)\,\,,\rc\rc
&&\Lambda^{(2)}(\gamma)\,=\,{1\over (\sin\gamma)^{|n|}}\,F\Big(
-{|n|+J\over 2}\,,\,{1+a-|n|+J\over 2}\,;\,{1+a\over 2}\,;\, \cos^2\gamma\Big)\,\,.
\label{Lambdas_General}
\eear
We now impose regularity conditions to these solutions. Due to the $(\sin\gamma)^{-|n|}$ factor on the right-hand side of (\ref{Lambdas_General}), it is clear that we must impose that the hypergeometric functions vanish  at $\gamma=0$. To evaluate them at this point, we use the general formula:
\beq
F(A,B; C;1)\,=\,{\Gamma(C)\,\Gamma(C-A-B)\over \Gamma(C-A)\,\Gamma(C-B)}\,\,.
\label{hypergeometric_z_1}
\eeq
Then, the hypergeometric functions in (\ref{Lambdas_General}) vanish when one of the gamma functions in the denominator of (\ref{hypergeometric_z_1}) have a pole or, equivalently, when either $C-A$ or $C-B$ in 
(\ref{hypergeometric_z_1}) is zero or a negative integer. Let us apply this procedure to $\Lambda^{(1)}$ first.  One can easily show that the number $J$ must take the values:
\beq
J^{(1)}\,=\,1\,-\,a\,+\,|n|\,+\,2k\,\,,
\qquad\qquad
k\ge 0\,\,,\,\,k\in {\mathbb Z}\,\,.
\eeq
The corresponding function $\Lambda^{(1)}$ is:
\beq
\Lambda^{(1)}(\gamma)\,=\,{(\cos\gamma)^{1-a}\over (\sin\gamma)^{|n|}}\,F\Big(
{3-a\over 2}\,+\,k\,,\,-k\,-\,|n|; {3-a\over 2};\cos^2\gamma\Big)\,\,.
\eeq
Let us now rewrite  $\Lambda^{(1)}$ using the identity:
\beq
F(A,B; C;z)\,=\,(1-z)^{C-A-B}\,F\big(C-A\,,\, C-B\,;\,C\,;\,z)\,\,.
\label{hypergeomeytric_identity}
\eeq
We get:
\beq
\Lambda^{(1)}(\gamma)\,=\,(\cos\gamma)^{1-a}\, (\sin\gamma)^{|n|}\,
F\Big(-k\,,\,{3-a\over 2}\,+\,k\,+\,|n|\,;\,{3-a\over 2}\,;\,\cos^2\gamma\Big)\,\,.
\label{Lambda_1_regular}
\eeq
Similarly, the regular $\Lambda^{(2)}$ functions are those for which $J$ takes the values:
\beq
J^{(2)}\,=\,|n|\,+\,2k\,\,.
\eeq
They are given by:
\beq
\Lambda^{(2)}(\gamma)\,=\,{1\over (\sin\gamma)^{|n|}}\,F\Big(
{1+a\over 2}\,+\,k\,,\,-k-\,|n|\,;\,{1+a\over 2}\,;\,\cos^2\gamma\Big)\,\,.
\eeq
Using again (\ref{hypergeomeytric_identity}) we can rewrite these functions as:
\beq
\Lambda^{(2)}(\gamma)\,=\, (\sin\gamma)^{|n|}\,F\Big(-k\,,\,{1+a\over 2}\,+\,k\,+\,|n|\,;\,{1+a\over 2}\,;\,
\cos^2\gamma\Big)\,\,.
\label{Lambda_2_regular}
\eeq
Notice that the hypergeometric functions of (\ref{Lambda_1_regular}) and (\ref{Lambda_2_regular}) are polynomials of degree $2k$ in $\cos\gamma$.

\subsection{Y-polynomials}
 We now consider the harmonic functions corresponding to the values of the constant $a$ that appear in our study probe D6-branes. First we take $a=1/3$ and define the $Y$-functions by:
\beq
Y^{(r)}(\gamma, \phi)\,\equiv\,e^{i\,n\,\phi}\,\Lambda^{(r)}(\gamma)\Big|_{a={1\over 3}}\,\,,
\qquad\qquad
(r=1,2)\,\,.
\label{Y_functions_def}
\eeq
Notice that $Y^{(r)}$ depends on the integers $k$ and $n$. This dependence is not written explicitly in our notation in order to make it simpler.  In what follows $n$ is supposed to be greater or equal to zero. For negative $n$ the harmonics are defined by complex conjugation:
\beq
Y^{(r)}_{|_{k,-n}}\,=\,\big[Y^{(r)}\big]^*_{|_{k,n}}\,\,.
\label{negative_n_harmonic}
\eeq
Explicitly, the $Y$-functions and the corresponding values of $J$ are:
\bear
&&Y^{(1)}\,=\,(\cos\gamma)^{{2\over 3}}\,(\sin\gamma\,e^{i\,\phi})^{n}\,
F\Big(-k\,,\,{4\over 3}\,+\,k\,+n\,;\,{4\over 3}\,;\,\cos^2\gamma\Big)\,\,,\rc\rc
&&Y^{(2)}\,=\,(\sin\gamma\,e^{i\,\phi})^{n}\,
F\Big(-k\,,\,{2\over 3}\,+\,k\,+n\,;\,{2\over 3}\,;\,\cos^2\gamma\Big)\,\,,\rc\rc
&&J_{Y}^{(1)}\,=\,{2\over 3}\,+\,2k\,+\,n\,\,,
\qquad\qquad\qquad
J_{Y}^{(2)}\,=\,2k\,+\,n\ .
\eear
In order to write the $Y$-functions as homogeneous polynomials, let us introduce the cartesian-like coordinates 
$X_1$, $X_2$ and $z$ by:
\beq
X_1\,=\,\rho^{{3\over 2}}\,\sin\gamma\,\cos\phi\,\,,
\qquad
X_2\,=\,\rho^{{3\over 2}}\,\sin\gamma\,\sin\phi\,\,,
\qquad
z\,=\,\rho^{{3\over 2}}\,\cos\gamma\,\,,
\eeq
where $r$ is a radial variable. Using that:
\beq
\sin\gamma\,e^{i\phi}\,=\,\rho^{-{3\over 2}}\,(X_1+iX_2)\,\,,
\qquad
\cos\gamma\,=\,\rho^{-{3\over 2}}\,z\,\,,
\qquad
\rho^3\,=\,(X_1)^2+(X_2)^2+z^2\,\,,
\eeq
we can write $Y^{(1)}$ and $Y^{(2)}$ as:
\bear
&&Y^{(1)}\,=\,\rho^{-{3\over 2}\,J_Y^{(1)}}\,z^{{2\over 3}}\,
(X_1+iX_2)^n\,{\cal P}_{2k}^{(1)}(X_1, X_2, z)\,\,,\rc\rc
&&Y^{(2)}\,=\,\rho^{-{3\over 2}\,J_Y^{(2)}}\,
(X_1+iX_2)^n\,{\cal P}_{2k}^{(2)}(X_1, X_2, z)\,\,,
\eear
where ${\cal P}_{2k}^{(1)}$ and ${\cal P}_{2k}^{(2)}$ are homogeneous polynomials of $X_1$, $X_2$ and $z$ of degree 
$2k$, given by:
\bear
&&{\cal P}_{2k}^{(1)}(X_1, X_2, z)\,=\,r^{3k}\,
F\Big(-k\,,\,{4\over 3}\,+\,k\,+n\,;\,{4\over 3}\,;\,\cos^2\gamma\Big)\,\,,\rc\rc
&&{\cal P}_{2k}^{(2)}(X_1, X_2, z)\,=\,r^{3k}\,
F\Big(-k\,,\,{2\over 3}\,+\,k\,+n\,;\,{2\over 3}\,;\,\cos^2\gamma\Big).
\eear
The first non-trivial ${\cal P}$ polynomials are:
\bear
&&{\cal P}_{2}^{(1)}(X_1, X_2, z)\,=\,(X_1)^2+(X_2)^2\,-\,{3\over 4}\,(n+1)\,z^2\,\,,\rc\rc
&&{\cal P}_{2}^{(2)}(X_1, X_2, z)\,=\,(X_1)^2+(X_2)^2\,-\,{3\over 2}\,(n+1)\,z^2\,\,.
\eear

\subsection{Z-polynomials}
The $Z$-harmonics are obtained by taking $a=-1/3$ in our general equations. For $n\ge 0$ they are defined by:
\beq
Z^{(r)}(\gamma, \phi)\,\equiv\,e^{i\,n\,\phi}\,\Lambda^{(r)}(\gamma)\Big|_{a=-{1\over 3}}\,\,,
\qquad\qquad
i=(1,2)\,\,.
\eeq
This definition is extended to $n<0$ by complex conjugation as in (\ref{negative_n_harmonic}). The $Z$ functions and the corresponding values of $J$ are:
\bear
&&Z^{(1)}_{k,n}\,=\,(\cos\gamma)^{{4\over 3}}\,(\sin\gamma\,e^{i\,\phi})^{n}\,
F\Big(-k\,,\,{5\over 3}\,+\,k\,+n\,;\,{5\over 3}\,;\,\cos^2\gamma\Big)\,\,,\rc\rc
&&Z^{(2)}_{k,n}\,=\,(\sin\gamma\,e^{i\,\phi})^{n}\,
F\Big(-k\,,\,{1\over 3}\,+\,k\,+n\,;\,{1\over 3}\,;\,\cos^2\gamma\Big)\,\,,\rc\rc
&&J_{Z}^{(1)}\,=\,{4\over 3}\,+\,2k\,+\,n\,\,,
\qquad\qquad\qquad
J_{Z}^{(2)}\,=\,2k\,+\,n\,.
\label{Z_J_harmonics}
\eear
In terms of $X_1$, $X_2$ and $z$ we have:
\bear
&&Z^{(1)}\,=\,\rho^{-{3\over 2}\,J_Z^{(1)}}\,z^{{4\over 3}}\,
(X_1+iX_2)^n\,{\cal Q}_{2k}^{(1)}(X_1, X_2, z)\,\,,\rc\rc
&&Z^{(2)}\,=\,\rho^{-{3\over 2}\,J_Z^{(2)}}\,
(X_1+iX_2)^n\,{\cal Q}_{2k}^{(2)}(X_1, X_2, z)\,\,,
\eear
where ${\cal Q}_{2k}^{(1)}$ and ${\cal Q}_{2k}^{(2)}$ are homogeneous polynomials  of the cartesian coordinates of degree $2k$, given by:
\bear
&&{\cal Q}_{2k}^{(1)}(X_1, X_2, z)\,=\,\rho^{3k}\,
F\Big(-k\,,\,{5\over 3}\,+\,k\,+n\,;\,{5\over 3}\,;\,\cos^2\gamma\Big)\,\,,\rc\rc
&&{\cal Q}_{2k}^{(2)}(X_1, X_2, z)\,=\,\rho^{3k}\,
F\Big(-k\,,\,{1\over 3}\,+\,k\,+n\,;\,{1\over 3}\,;\,\cos^2\gamma\Big)\,\,.
\eear
Let us write the first non-trivial $Q$ polynomials:
\bear
&&{\cal Q}_{2}^{(1)}(X_1, X_2, z)\,=\,(X_1)^2+(X_2)^2\,-\,{3\over 5}\,(n+1)\,z^2\,\,,\rc\rc
&&{\cal Q}_{2}^{(2)}(X_1, X_2, z)\,=\,(X_1)^2+(X_2)^2\,-\,3\,(n+1)\,z^2\,\,.
\eear

\end{document}